\documentclass[a4paper,twocolumn,showpacs,superscriptaddress,floatfix]{revtex4-1}

\usepackage{graphicx}
\usepackage{epsf}
\usepackage{epsfig}
\usepackage{amssymb,amsmath,mathrsfs,amsthm}
\usepackage[usenames]{color}
\usepackage{amssymb}
\usepackage{times}

\newcommand{\be}{\begin{equation}}
\newcommand{\ee}{\end{equation}}
\newcommand{\bea}{\begin{eqnarray}}
\newcommand{\eea}{\end{eqnarray}}
\newcommand{\nn}{\nonumber}

\font\tenscr=rsfs10 scaled1100
\font\sevenscr=rsfs7 % scaled \magstep1
\font\fivescr=rsfs5 % scaled \magstep1
\skewchar\tenscr='177
\skewchar\sevenscr='177
\skewchar\fivescr='177
\newfam\scrfam
\textfont\scrfam=\tenscr
\scriptfont\scrfam=\sevenscr
\scriptscriptfont\scrfam=\fivescr

\def\scri{{\fam\scrfam I}}

\theoremstyle{plain}

\begin{document}

\title{Pseudospectrum and black hole quasi-normal mode (in)stability}

\author{Jos\'e Luis Jaramillo}
\affiliation{Institut de Math\'ematiques de Bourgogne (IMB), UMR 5584, CNRS, Universit\'e de
Bourgogne Franche-Comt\'e, F-21000 Dijon, France}
\author{Rodrigo Panosso Macedo}
\affiliation{School of Mathematical Sciences, Queen Mary, University of
  London, \\ Mile End Road, London E1 4NS, United Kingdom}
\author{Lamis Al Sheikh}
\affiliation{Institut de Math\'ematiques de Bourgogne (IMB), UMR 5584, CNRS, Universit\'e de
Bourgogne Franche-Comt\'e, F-21000 Dijon, France}

\begin{abstract}
 We study the stability of quasi-normal modes (QNM) in asymptotically flat black hole spacetimes
by means of a pseudospectrum analysis. The construction of the
  Schwarzschild QNM pseudospectrum reveals: i) the  stability of the slowest decaying QNM
  under perturbations respecting the asymptotic structure,  reassessing the
  instability of the fundamental QNM discussed by Nollert \cite{Nollert:1996rf} as an ``infrared'' effect;
  ii) the instability of all overtones under small scale (``ultraviolet'') perturbations of sufficiently 
  high frequency, that migrate towards universal QNM branches along pseudospectra boundaries,
  shedding light on  Nollert \& Price's analysis \cite{Nollert:1996rf,Nollert:1998ys}.
  Methodologically, a compactified hyperboloidal approach to QNMs is adopted to cast QNMs in terms of
  the spectral problem of a non-selfadjoint operator.
  In this setting, spectral (in)stability is naturally addressed through the
  pseudospectrum notion, that we construct numerically via Chebyshev spectral methods and foster in gravitational physics.
  After illustrating the approach with the P\"oschl-Teller potential,
  we address the Schwarzschild black hole case, where QNM (in)stabilities are physically relevant in the
  context of black hole spectroscopy in gravitational wave physics
  and, conceivably, as probes into fundamental high-frequency spacetime fluctuations at the Planck scale.
\end{abstract}


\maketitle

\section{Introduction: QNMS and (in)stability}

\subsection{The black hole QNM stability problem and the pseudospectrum}
Structural stability is essential in the modelling and understanding of physical phenomena.
In the context of spectral problems pervading physics, often related to wave phenomena
in both classical and quantum theories, this  concerns in particular the basic question
about the stability of the physical spectrum of the system. Thereupon, one  needs to assess the following questions:
how does the spectrum react to small changes of the underlying system? Is the spectrum stable,
i.e., do small perturbations lead to tiny deviations? Or is it unstable, with small changes in the system
leading to drastic  modifications of the spectrum?
In the present work, we study such kind of
spectral stability question in the setting of black hole (BH) spacetimes.
Specifically, the problem we address here is the spectral robustness of
BH QNMs, namely the stability of the resonant frequencies of
BHs under perturbations. From a methodological perspective, our
spectral (in)stability analysis is built upon the notion of the so-called pseudospectrum.

\subsubsection{Spectral instability and pseudospectrum}
The physical  status of spectral stability depends crucially on whether the underlying system is
conservative or not. In particular, conservative systems do have stable spectra and therefore the
spectral instability question, being solved from scratch, is not relevant. Such spectral stability
is familiar in (standard) quantum mechanics, where (time-independent) perturbation theory 
precisely relies on it. It is the selfadjoint nature of the relevant operators (namely ``Hermitian
matrices'' in the finite dimensional case) 
that accounts for such spectral stability. More systematically, this is a consequence
of the so-called 'spectral theorem' for selfadjoint operators: eigenvectors form an orthogonal and complete set,
whereas eigenvalues are real and stable. This provides the mathematical background for the key physical
notion of normal mode, associated with the characteristic (real) vibrating frequencies
of a conservative system and its natural oscillation modes.

The situation is more delicate for non-conservative systems, modelled in terms of
non-selfadjoint operators (non-Hermitian matrices). Such systems occur naturally
whenever there exist flows (e.g. energy, particle, information) into external degrees of freedom
that are out of the (Hilbert) space under consideration (see \cite{Ashida:2020dkc} for
a recent and extensive review on non-Hermitian physics;  cf. e.g. its Table 1 
for a list of several classical and quantum systems described by non-Hermitian operators).
In this setting the 'spectral theorem' is lost: eigenvectors are in general neither complete
nor orthogonal, and eigenvalues (now in general complex) are potentially unstable.
We focus here on this latter point, namely the potential spectral instability
of a class of non-selfadjoint operators associated with the non-conservative system
defined by the scattering of fields by BHs where, critically, the field leaks away from the system
at far distances and through the BH horizon.

The notion of pseudospectrum
\cite{TreTreRed93,Trefe97,Davie00,Sjost03,trefethen2005spectra,Davie07,KreSieTat15,Sjostrand2019,EmbTre_webpage} provides
a powerful tool for the analysis of the properties of non-selfadjoint operators.
In particular, its use is well spread  whenever stability issues of non-conservative systems are addressed,
from pioneering applications in hydrodynamics \cite{TreTreRed93} to recent advances \cite{ColRomHan19}
covering a wide range in physics. Broadly speaking in order to gain some first intuition, the pseudospectrum
provides a visualization (and actually a characterization) of the spectral instability of our operator in terms
of a kind of 'topographic map' on the complex plane, where the 'peaks' (actually end points
of infinitely-high throats) lay at the complex eigenvalues of the non-perturbed operator. With this picture in mind,
spectral stability is assessed in terms of the ``thickness'' of the throats: very thin throats decreasing fast
from the eigenvalues correspond to spectral stability, whereas broad slowly decreasing throats
indicate spectral instability. Expressing this in terms of 'level-sets', contour lines
corresponding to 'heights'
$1/\epsilon$ define a family of nested sets around eigenvalues, referred to as $\epsilon$-pseudospectra,
that determine the  regions in which eigenvalues can potentially 'migrate' under a system perturbation of size $\epsilon$. 
The (non-perturbed) spectrum corresponds to the set defined by $\epsilon\to 0$.
Therefore, tightly packed  contour lines around eigenvalues corresponding to strong gradients indicate
spectral stability, whereas contour lines  with low gradients extending far from the eigenvalues
signal spectral instability.

\subsubsection{Black hole QNMs in gravitational physics}
BH QNMs encode the resonant response to (linear) perturbations of the BH spacetime.
In spite of being triggered by perturbations, QNMs constitute an intrinsic property
of the background and, therefore, QNM frequencies encode crucial geometric information
about BHs and their environment. Thus, they have become a fundamental tool in astrophysics,
fundamental gravitational physics, and mathematical relativity in their attempts to
probe spacetime geometry through perturbation theory and scattering methods (see
e.g. \cite{Chandrasekhar:579245,Kokkotas:1999bd,Nollert:1999ji,Berti:2009kk,Konoplya:2011qq}
for systematic presentations and reviews).

Upon perturbation, and after an initial transient, the perturbative
field propagating on the background spacetime shows an exponentially-damped,
oscillatory behavior. QNM frequencies are the set of complex numbers encoding the
oscillatory frequencies and decaying time scales of the propagating linear (scattered) field.
To fix ideas, this is illutrated by the BH formed after the merger of a compact binary,
in the emerging setting of gravitational wave (GW) astronomy.
After the transient merger phase, the resulting perturbed BH evolves towards
stationarity in a linear ringdown phase dominated by QNMs.  In particular,
the late time behaviour of the GW signal is controlled by the fundamental
or slowest decaying QNM mode, namely the QNM frequency with smallest (in absolute value)
imaginary part and therefore closest to the real axis. Nonetheless, QNMs with larger
imaginary parts and referred to as overtones  --- with different oscillatory frequencies and faster
decaying time scales than the fundamental QNM--- are also present in the GW signal, its analysis
being at the basis of the BH spectroscopy research
program~\cite{Berti:2005ys,Dreyer:2003bv,Baibhav_2018,Isi:2019aib,Giesler:2019uxc,Cabero:2019zyt,Maggio:2020jml}.
Beyond GW physics, QNMs play a key role in gravitational physics as a crossroads
among different limits and regimes of the theory, 
encompassing problems in the evoked GW astrophysical
setting, in semiclassical gravity (e.g. \cite{York83}) and gravity-fluid (AdS/CFT) dualities (e.g. \cite{Horowitz:1999jd}),
in analogue gravity \cite{Barcelo:2005fc} or in foundational questions in mathematical relativity
(e.g. \cite{Dyatlov:2013wzt,Warnick:2013hba,HinVas17}), among other problems ranging from the classical
to the quantum regime. In this latter setting, and given the importance they will
have in our present discussion, it is worthwhile to signal that QNM overtones have been
proposed as a possible probe into the quantum aspects of spacetime
\cite{York83,Nollert:1993zz,Hod98,Maggi08,BabDagKun11,OlmDie20,Carneiro:2020uww}.

To be more specific, the discussion of BH QNMs is set in terms of the wave equations arising when General Relativity
is considered at linear order for fields propagating over a fixed BH background. We will focus in this
work on the asymptotically flat case, where geometry approaches Minkoswki (in an appropriate sense) at large
distances.
One must then impose appropriate boundary conditions to the underlying wave equations: as one moves
far way for the source, the waves must propagate out to infinity, whereas they must propagate to the interior
of the black hole at the horizon. The resulting outgoing boundary conditions define a leaky system.
QNMs are obtained from the spectral problem associated with this system.
A crucial point in the present discussion is that such spectral problem can be cast
as a proper 'eigenvalue problem' for a certain non-selfadjoint operator.
At this point we make contact with the potential spectral instability associated with
non-selfadjoint operators that we have discussed above.
The ultimate significance of QNM frequencies depends directly on the understanding
and control of their spectral stability. 

\subsection{BH QNM instability: Nollert \& Price's pioneer work}
Nollert's groundbreaking work in BH QNM spectral stability \cite{Nollert:1996rf},
complemented by the analysis in Nollert \& Price's \cite{Nollert:1998ys}, showed evidence of an overall instability
of the Schwarzschild QNM spectrum --- affecting both the fundamental QNM and the overtones ---
under a class of small scale perturbations
(see also the recent works \cite{Daghigh:2020jyk,Qian:2020cnz}).
The analysis by Nollert and Price, both numerical and analytic in an asymptotic treatment,
demonstrated the robustness of such QNM instability. However, it remained to be elucidated if
such an instability corresponds to the specific form of the considered perturbations,
and therefore could be an artifact of the employed approach, or if it rather
responds to a structural feature within  the theory with actual implications on the
physics of BH QNMs.

More specifically, these works considered a family of step-like approximations to
the Schwarzschild curvature potential. In a first step, the authors calculated the QNMs corresponding
to the step-like approximation for the potential (perturbed QNMs) finding a strong deviation
from the original values (non-perturbed QNMs), with a clear and systematic pattern: perturbed
QNMs distribute along new QNM branches with a qualitative structure dramatically distinct
from that of non-perturbed QNMs. In a second step, they performed time evolutions
of the wave equation under the step-like approximated potential in a bid to identify and
extract the perturbed QNMs from the wave signal. In contrast with the
spectral problem, time evolutions presented an overall stable behaviour under perturbations
of the potential. Specifically, Nollert \& Price's work demonstrates that, for the studied class of perturbations:
\begin{itemize}
\item[i)] QNM overtones are strongly unstable, their instability increasing with their damping.
\item[ii)] The fundamental, slowest decaying, QNM is unstable. 
\item[iii)] The black hole ringdowns, at intermediate late times,
  according to the non-perturbed fundamental mode. Only at very late times
  the ringdown frequency is controlled by the  perturbed fundamental QNM mode.
\end{itemize}
These results have been confirmed and expanded in \cite{Daghigh:2020jyk,Qian:2020cnz} to
perturbations of the scattering potential extending the step-like approximation, but still sharing
the feature of presenting a discontinuity at the potential or some of its derivatives.

Beyond Nollert \& Price's works, research in BH QNM spectral (in)stability has been further pursued in different
gravitational physics settings.
In astrophysics, the understanding of possible environmental observational signatures in
``dirty'' BH scenarios has prompted a research line \cite{Leung:1999iq,Barausse:2014tra} 
that has been significantly intensified recently~\cite{Cardoso:2019mqo,McManus:2019ulj,Hui:2019aox,Daghigh:2020jyk,Qian:2020cnz}.
On the other hand, regarding investigations on the fundamental structure of spacetime, the perspective of accessing
quantum scales through high-frequency instabilities of QNM overtones has also 
tantalized a systematic research  \cite{Nollert:1993zz,Hod98,Maggi08,BabDagKun11,OlmDie20,Carneiro:2020uww}.

In spite of these efforts, a comprehensive picture of BH QNM (in)stability seems to be lacking.
At this point it is worthwhile to explicitly distinguish between the instability in QNM frequencies
  and the instability in late ringdown frequencies. 
  The former refers to the spectral instability in the 'frequency domain' approach, 
  when solving the spectral problem associated with the wave equation. The latter would refer to
  a dynamical instability in the 'time domain' approach, when solving the initial data dynamical problem.
  Both problems are intimately related, but are indeed different.
  In particular, it is known that the two sets of frequencies can indeed decouple
  (e.g.  \cite{Nollert:1996rf,Nollert:1998ys,Khanna:2016yow,Cardoso:2016rao,Konoplya:2016hmd,Daghigh:2020jyk,Konoplya:2020fwg,Qian:2020cnz} in the gravitational context).
  Still, the separation between QNM and ringdown frequencies signals an 'anomaly' and, therefore,
  pinpoints a structural feature in the physical system requiring specific study.
  In the present work, we focus on QNM (in)stability in the spectral sense.

In this context, the stability status of the slowest decaying QNM --- presenting precisely the tension described
  above between calculated spectral instability and observed robustness in the ringdown signal ---
remains unclear, whereas
the elucidation of the lowest overtone subject to high-frequency instability is an open problem.
Under the light of the discussion above on the fundamental role of BH QNMs in different
settings of gravitational physics, the clarification of these two points
is a first-order problem from a strictly physical perspective. Moreover,
if establishing the stability status of BH QNMs is key in general BH physics,
the problem is actually urgent in gravitational wave astrophysics.
Indeed, in the era of gravitational-wave astronomy, the stability of the fundamental QNM and the overtones
is paramount for BH spectroscopy.

The implementation of an analysis based on the pseudospectrum permits to address systematically
these questions and to provide sound answers to
points i) and ii) above. In short, and anticipating the results later discussed in detail,
  such analysis confirms the instability behaviour of QNM overtones -- point i) -- and provides
  a framework for its systematic study, whereas
  it disproves the instability of the fundamental QNM -- point ii) -- if asymptotic properties
  of the spacetime are respected, its unstable behaviour in \cite{{Nollert:1996rf}}
  resulting an artifact consequence of ``cutting'' the effective potential at a finite distance.
Regarding point iii), from the stability of the fundamental QNM we conclude that
  the late time ringdown is indeed dominated by the unperturbed slowest decaying QNM,
  (without any very late transition to a 'perturbed ringdown' frequency),
  but the systematic analysis
  of the detailed relation between QNM frequencies and BH ringdown frequencies lays beyond
  the scope of the present work and will be the subject of a specifically targeted research
   focused on the potential implications of BH QNM instability on GW astrophysics \cite{JarMacShe21}.

  \subsection{The present approach}

  \subsubsection{The basic ingredients: hyperboloidal approach and pseudospectrum}
  The calculation of BH QNMs has been the subject of systematic study
  in gravitational physics and there exists a variety of standard approaches
  to address this problem (cf. e.g. \cite{Kokkotas:1999bd,Nollert:1999ji,Berti:2009kk,Konoplya:2011qq}).
From a methodological perspective, our discussion relies on two
main ingredients at a conceptual level:
\begin{itemize}
\item[i)] A hyperboloidal approach to QNMs: this geometric approach casts the QNM calculation as a proper
  eigenvalue problem of a particular non-selfadjoint operator.
\item[ii)] Pseudospectrum: together with related spectral tools,
  it provides the key instrument 
  to study the potential spectral instability of the relevant non-selfadjoint operator.
\end{itemize}
The combination of these two elements
permits to develop a systematic treatment of the problem. To the best of our knowledge,
no systematic treatment of BH QNM (in)stability based on the pseudospectrum exists in the
literature. At a first exploratory stage, prior to a full analytical study,
the present work addresses pseudospectra in a numerical approach. This sets a
challenging numerical problem demanding high accuracy, which is here addressed by introducing
a third key ingredient in our approach: the use of spectral numerical methods.

\subsubsection{Beyond gravity: QNMs, pseudospectrum and interdisciplinary physics}
Before entering into the detailed discussion of BH QNM stability, let us stress that both QNMs
and the pseudospectrum provide independent, but indeed complementary, arenas for
interdisciplinary research in physics and related disciplines.

Regarding QNMs, beyond the present
gravitational context, the notion of QNM spreads in physics, e.g. in electromagnetism and
optics, acoustics,  or --- under the related notion of resonance in quantum mechanics --- in atomic,
nuclear and molecular physics.
Beyond physics, QNMs enter in the discussion of scattering problems in geometry \cite{Zworski99} and chaotic dynamics
(see \cite{zworski2017mathematical,dyatlov2019mathematical}
for a systematic review of scattering resonances or QNMs from a mathematical perspective).
Together with the extent of the applicability of the QNM notion, an important aspect concerns
timing. Indeed, the synergy observed in this sense in recent years among different
subdisciplines in the gravitational setting (namely GW astrophysics, AdS/CFT dualities and mathematical
relativity) extends remarkably to other fields in physics, as perfectly illustrated by recent breakthroughs
in optical nanoresonator QNMs, namely photonic and plasmonic resonances \cite{sauvan2013theory,LalYanVyn17}.

Regarding the pseudospectrum, its use in physics naturally extends over the study of stability and
spectral problems in non-conservative systems, from which we highlight its applications in hydrodynamics 
\cite{TreTreRed93} and in non-Hermitian quantum mechanics \cite{KreSieTat15}.
Beyond physics, systematic applications are found in numerical analysis,
the original context where the notion was formulated. This wide 
range of applications become  intertwined  methodologically by the pseudospectrum.
The present approach 
to   BH QNM stability, that introduces (to the best of our knowledge) the
pseudospectrum into gravity, incorporates gravitational physics to
this multifaceted research scheme. When combined with the large
range of applicability of the QNM notion in physics, it outlines a robust and potentially rich
frame for interdisciplinary research in physics.

\bigskip

The article is structured as follows. Section \ref{s:Hyper_approach}  presents a qualitative description of
the hyperboloidal approach to scattering problems and reviews the literature on this geometrical framework.
Beyond reviewing the main concepts, with a focus on QNMs, this section identifies and constructs 
the appropriate scalar product in the problem.
Section \ref{s:spectral_stability_pseudospectrum} introduces the
basic elements to study spectral instability of non-selfadjoint operators,
in particular the notion of pseudospectrum. Section \ref{s:numerical_approach}
presents the numerical spectral tools to be employed in the present approach.
Then, section \ref{s:PT} illustrates all the previous elements in
the toy-model provided by the P\"oschl-Teller potential, that also
anticipates some of the main results in the BH setting. Section
\ref{s:QNM_Schwarzschild} contains the main contribution in the
present work, namely the construction of the Schwarzschild QNM pseudospectrum
and the consequent analysis of BH QNM (in)stability. Conclusions and perspectives
are finally presented in section \ref{s:conclusions_perspectives}. A series of four appendices
complete some points in  the technical discussion of the main text.
Throughout the work, we adopt units in which the speed of light and the gravitational constant are unit ($c=G=1$).

\section{Hyperboloidal approach to QNMs}
\label{s:Hyper_approach}

\subsection{Hyperboloidal approach: a heuristic introduction }
\label{s:hyper_approach_primer}
Our approach to QNMs strongly relies on casting the discussion
in terms of the spectral problem of a (non-selfadjoint) operator.
In our scheme, this is achieved by means of a so-called hyperboloidal approach
to wave propagation, that provides a systematic framework
exploiting the geometric asymptotics of the spacetime, in particular
enforcing the relevant outgoing boundary conditions in a
geometric way. We start with a heuristic discussion
of the basics, aiming at providing an intuitive picture
and explicitly sacrificing rigor.

The notion of wave zone is a familiar concept in physics. It describes a region far away from a source where the degrees of a freedom of a given field (non-necessarily linear) propagate as a free wave, independently of their interior sources and obeying the superposition principle. Roughly speaking, this region is characterised by $r/R \gg 1$, where $r$ is the location of a distant
observer and $R$ is a typical length scale of the source. This concept is addressed formally by taking appropriate limits $r \rightarrow \infty$ or $1/r \rightarrow 0$. From a spacetime perspective, however, such a limit must be carefully understood. 

\begin{figure}[h!]
\centering
\includegraphics[width=8.5cm]{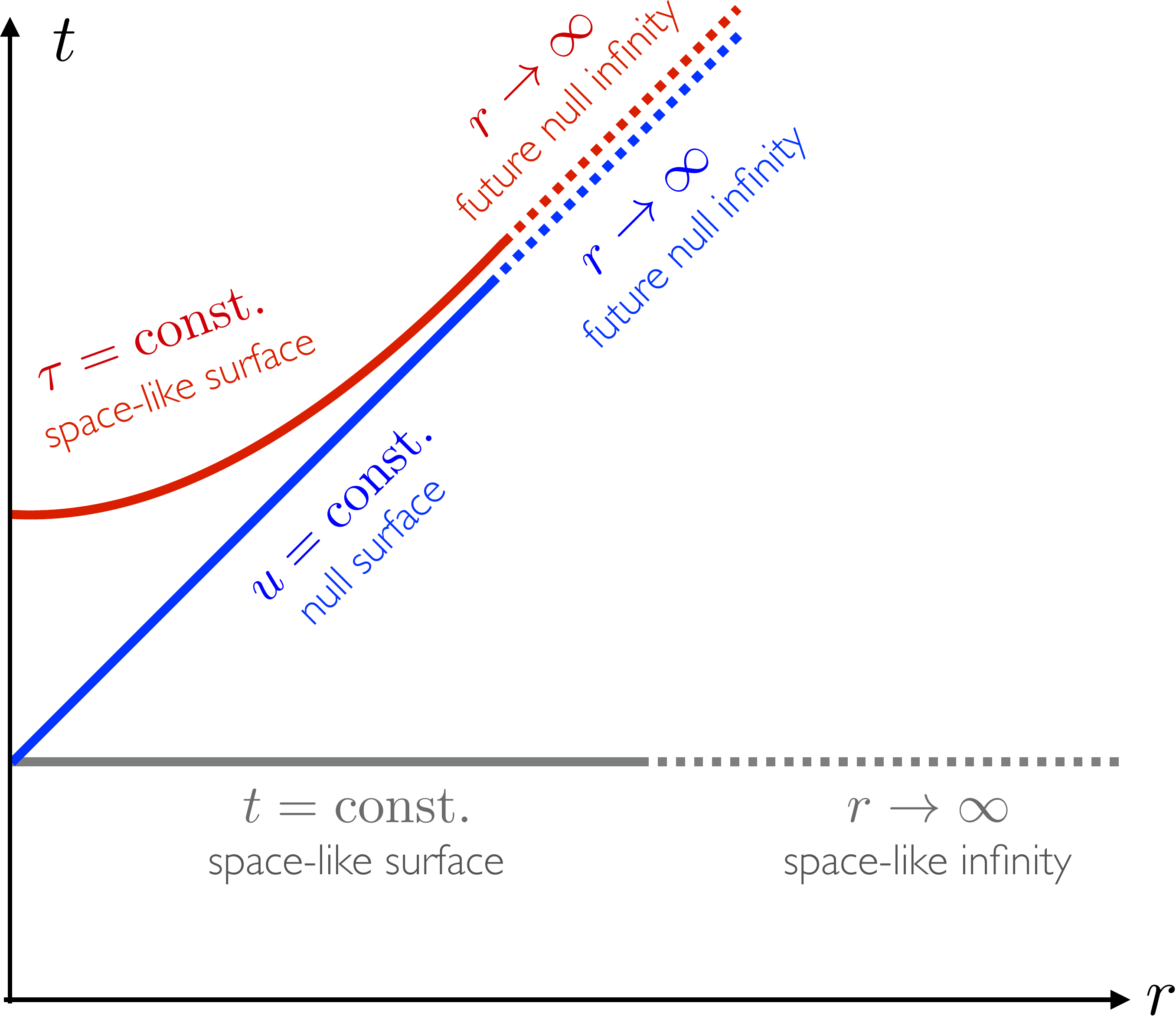}
\caption{Schematic representation of the different ``$r\rightarrow \infty$'' limits 
  along curves within different types of spacetime hypersurfaces.
  Cauchy hypersurfaces, of spacelike character and represented by the ``$t=\mathrm{const}$'' condition
  in the figure, are such that this limit attains the so-called spatial infinity $i^0$, whereas in null
  hypersurfaces satisfying rather ``$u=\mathrm{const}$'' (with $u=t-r$ a null 'retarded time')
  the limit attains the outgoing wave zone modelled by future null infinity  $\scri^+$.
  The hyperboloidal approach offers an intermediate possibility, where the limit is taken
  along spacelike hypersurfaces,
  formally represented by the ``$\tau=\mathrm{const}$'', but still reaching  $\scri^+$ asymptotics.}
\label{fig:SpaceTimeCartoon}
\end{figure}

To fix ideas, let us consider a physical scenario in spherical symmetry, where
a wave propagating at finite speed is described in a standard spherical coordinate system $(t,r,\theta, \varphi)$
(for simplicity, let us consider momentarily a flat spacetime where we ignore gravity effects).
The retarded time coordinate  $u = t - r$ corresponds to the time at which an outgoing wave, passing by the observer at $r$
at time $t$, was emitted by a source located at the origin. Crucially, ``light rays'' propagate
along (characteristic) curves satisfying $u=\mathrm{const}$. In this setting,
and as illustrated in Fig.~\ref{fig:SpaceTimeCartoon}, taking the limit $r \rightarrow \infty$
corresponds to completely different geometric statements depending on whether 
one stays at the hypersurface $t=\mathrm{const}$ or rather on $u=\mathrm{const}$.
The limit attained by 'spacelike' (geodesic) curves satisfying the former condition ($t=\mathrm{const}$)
is referred to as 'space-like infinity' and denoted $i^0$, whereas lightlike or null (future geodesic) curves
satisfying the latter condition ($u=\mathrm{const}$) attain a limit referred to as future null infinity,
denoted as $\scri^+$. It is future null infinity $\scri^+$ that formally captures the intuitive notion
of outgoing wave zone. 

Other alternatives to the $t=\mathrm{const}$ and $u=\mathrm{const}$ hypersurfaces are possible,
something natural in a general relativistic context implementing coordinate choice freedom.
A particularly convenient possibility in our present problem consists in choosing 
a third alternative: to keep space-like hypersurfaces defined as level sets of
an appropriate time function $\tau$, while reaching future null infinity 
as $r\rightarrow\infty$ so as to enforce the outgoing character of the radiation. 
Such a third option is displayed in Fig.~\ref{fig:SpaceTimeCartoon} as a $\tau=\mathrm{const}$ 
hypersurface. The asymptotic geometry of such hypersurfaces is that of a hyperboloid,
a feature giving name to the resulting hyperboloidal approach.

The previous heuristic picture of spacetime asymptotics is formalised in the geometric notion of conformal
infinity~\cite{Pen63,Ger76,Ashtekar80,Ash84a,Wald84,Fri03a,Kroon2016}, that provides 
a rigorous and geometrically well-defined strategy to deal with radiation problems of compact isolated bodies. A
conformal compactification maps the infinities of the physical spacetime into a finite region
delimited by the boundaries of a conformal manifold. Specifically,
$\scri^+$ corresponds to the future endpoints of null geodesics, whereas a time function $\tau$
will be referred to as  hyperboloidal if hypersurfaces $\tau=\mathrm{const}$ intersect  $\scri^+$,
being therefore adapted to the geometrical structure at the infinitely far away wave zone.

The hyperboloidal formulation has proved to be a powerful tool in mathematical
and numerical relativity, permitting to obtain
existence results in the non-linear treatment of Einstein equations, as
illustrated in the semiglobal result in \cite{Fri86b}, or providing
a natural framework for the extraction of the GW waveform in numerical
dynamical evolutions of GW sources. Together with those fully non-linear
studies, over the last decade the hyperboloidal approach has been
successfully applied to problems defined on fixed spacetime
backgrounds (see e.g. \cite{PanossoMacedo:2020biw} and references therein).
In particular, \cite{Zenginoglu:2011jz}
proposes a hyperboloidal approach to BH perturbation theory.

This is our setting for QNMs, where the hyperboloidal framework
permits to implement geometrically the outgoing boundary conditions at $\scri^+$,
in a strategy first proposed by Schmidt in~\cite{Schmi93}.
The adopted (compactified) hyperboloidal approach 
provides a geometric framework to study QNMs, that characterizes resonant frequencies in terms of an eigenvalue problem
~\cite{Zenginoglu:2011jz,Schmi93,Dyatlov_2011,Warnick:2013hba,Ansorg:2016ztf,PanossoMacedo:2018hab,PanossoMacedo:2020biw,Hafner:2019kov,Gajic:2019qdd,Gajic:2019oem,galkowski2020outgoing,Bizon:2020qnd}.
As explained above, the scheme geometrically imposes QNM outgoing boundary conditions
by adopting a spacetime slicing that intersects
future null infinity $\scri^+$ and, in the BH setting, penetrates the horizon.
Since light cones point outwards at the boundary of the domain, outgoing boundary conditions
are automatically imposed for propagating physical degrees of freedom.
Along these lines, our scheme to address
the BH QNM (in)stability problem strongly relies on the  hyperboloidal approach, since
it provides the rationale to define the non-selfadjoint operator on which
a pseudospectrum analysis is then performed.

\subsection{Wave equation in the compactified hyperboloidal approach}
\label{s:wave_eq_hyper_approach}
We focus on the propagation and, more generally, the scattering problem  
of (massless) linear fields
on stationary spherically symmetric BH backgrounds.
For concreteness, let us first consider a scalar field $\Phi$, satisfying the wave
equation
\bea
\label{e:box_Phi}
\square \Phi = \nabla^a \nabla_a\Phi = 0 \ . 
\eea
We adopt standard Schwarzschild coordinates
\bea
\label{e:spherically_symmetric_metric}
ds^2 = -f(r) dt^2 + f(r)^{-1} dr^2 + r^2 \left({d \theta^2} + \sin^2\theta d\varphi^2\right) \ ,
\eea
and emphasize that $t\!=\!\mathrm{const}$ slices correspond to Cauchy
hypersurfaces intersecting both the horizon bifurcation sphere and spatial infinity $i^0$. 
If we consider the rescaling 
\bea
\Phi = \frac{1}{r} \phi \ ,
\eea
then Eq. (\ref{e:box_Phi}) rewrites, expanding $\phi$ in spherical harmonics with $\phi_{\ell m}$ modes
and using the tortoise coordinate defined by $\displaystyle \frac{dr}{dr_*} = f(r)$
(with the appropriate integration constant), as
\bea
\label{e:wave_equation_tortoise}
\left(\frac{\partial^2}{\partial t^2} -\frac{\partial^2}{\partial r_*^2} + V_{\ell} \right)\phi_{\ell m} = 0 \ ,
\eea
where now $r_*\!\in \;]-\infty,\infty[$. 
    Remarkably, when considering electromagnetic and (linearized) gravitational fields, the respective
    geometric wave equations corresponding to Eq. (\ref{e:box_Phi}) can be cast in the form
    of Eq. (\ref{e:wave_equation_tortoise}) for appropriate effective scalar potentials. Specifically,
    two scalar fields with different parity can be introduced, satisfying Eq. (\ref{e:wave_equation_tortoise})
    with suitable potentials $V_{\ell}$. In the gravitational case, the axial parity
    is subject to the so-called Regge-Wheeler potential, whereas the polar one is controlled by the Zerilli potential
    (cf. e.g \cite{Chandrasekhar:579245,Kokkotas:1999bd,maggiore2018gravitational}).

\begin{figure}[h!]
\centering
\includegraphics[width=8.5cm]{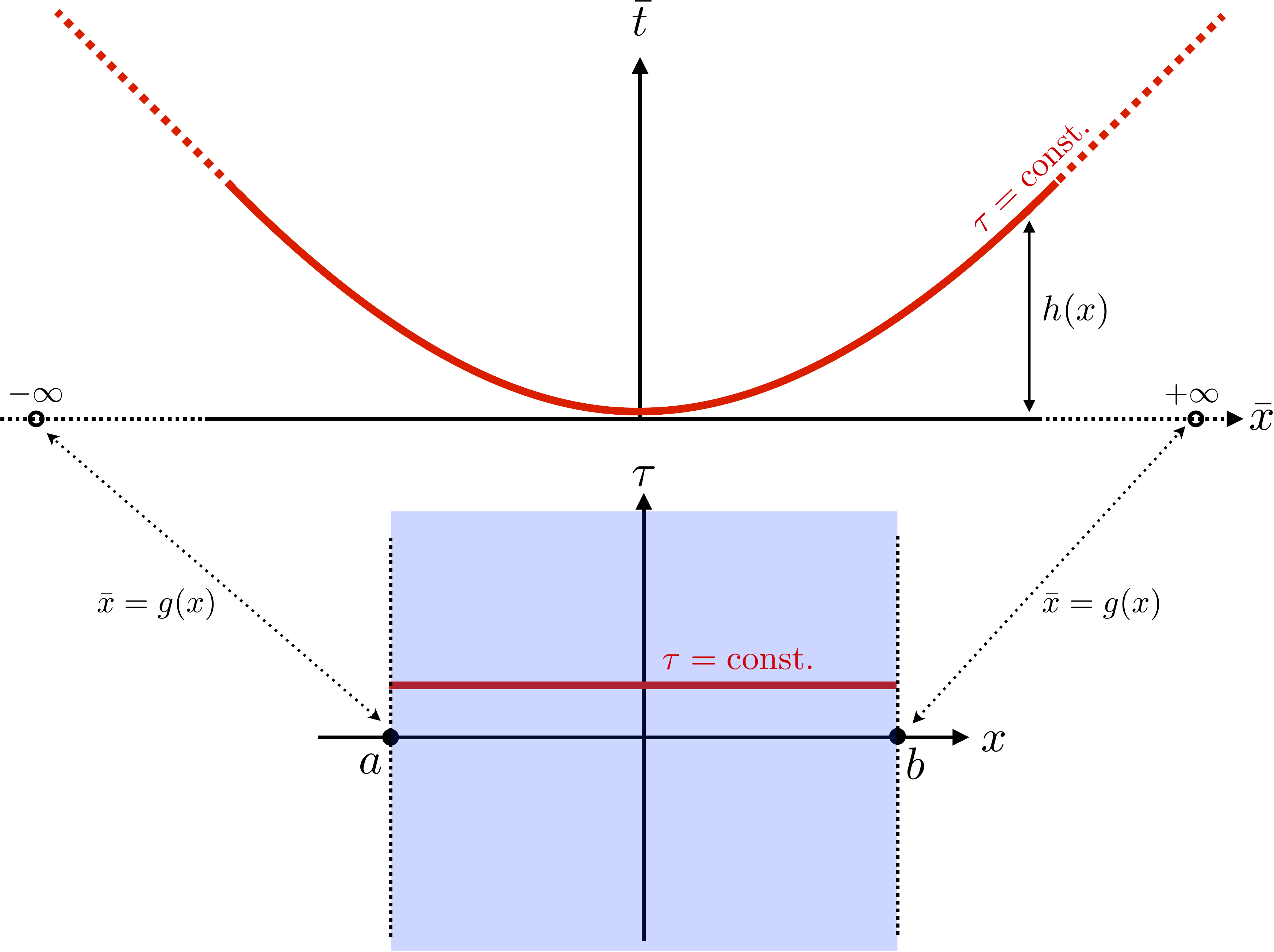}
\caption{Schematic representation of the hyperboloidal coordinate transformation in Eq. \eqref{e:change_variables}. {\em Top panel:} Dimensionless Schwarzschild time and tortoise coordinates $(\bar t, \bar x)$. The height function $h(x)$ bends the time slices so that future null infinity (and/or black-hole horizon) is reached in the limit $\bar x \rightarrow \pm \infty$. {\em Bottom panel:} Spatially compactified
  hyperboloidal coordinates $(\tau, x)$. The compactification function $g(x)$ maps the infinite domain $\bar x \in ]-\infty, \infty[$ onto the finite interval $x\in ]a,b[$. Points $b$ and $a$ are added  at the boundary, representing null infinity $\scri^+$ and/or the BH horizon.
  The blue stripe shows the domain of integration of the wave equation in Eq. (\ref{e:wave_equation_hg}) in these compactified
  hyperboloidal coordinates,
  namely $(\tau, x)\in ]-\infty, +\infty[\times [a,b]$, corresponding to the full original domain $(\bar t, \bar x) \in \mathbb{R}^2$
  of Eq. (\ref{e:wave_equation_tortoise}).
  }
\label{fig:HypCoordTransf}
\end{figure}

    The BH event horizon and (spatial) infinity correspond, respectively, to $r_*\!\!\to\!\! -\infty$
    and $r_*\!\!\to\!\!+\infty$. We extend the domain of $r_*$ to $[-\infty,\infty]$
    and introduce the dimensionless quantities
    \bea
    \label{e:dimensional_rescaling}
   \bar{t}=\frac{t}{\lambda} \ \ , \ \ \bar{x}=\frac{r_*}{\lambda} \ \ , \ \
    \hat{V}_\ell=\lambda^2 V_\ell \ ,
    \eea
    for an appropriate length scale $\lambda$ to be chosen in each specific setting.
    More importantly, we consider  coordinates $(\tau,x)$ that implement the compactified hyperboloidal approach 
\bea
\label{e:change_variables}
\left\{
\begin{array}{rcl}
\bar{t} &=& \tau - h(x)  \\ 
\bar{x} &=& g(x)
\end{array}
\right. \ .
\eea
Specifically (see Fig.~\ref{fig:HypCoordTransf}):
\begin{itemize}
\item[i)] The height function $h(x)$ implements the hyperboloidal slicing, i.e.
$\tau = \mathrm{const}$ is a horizon-penetrating hyperboloidal
slice $\Sigma_\tau$ intersecting future $\scri^+$.
\item[ii)] The function $g(x)$ introduces a spatial compactification from
$\bar{x}\in[-\infty,\infty]$ to a compact interval $[a,b]$.
\end{itemize}

We note that the compactification is performed only in the spatial direction
along the hyperboloidal slice, and not in time, so that the latter can be Fourier transformed
in a unbounded domain.
The relevant compactification
here is a partial one, and not the total spacetime compactification leading
to Carter-Penrose diagrams.
The choice of $h(x)$ and $g(x)$ is, as we
comment below, subject to certain restrictions.
Under transformation  (\ref{e:change_variables}), the wave equation (\ref{e:wave_equation_tortoise})
writes 
\bea
\label{e:wave_equation_hg}
&&\Bigg[\left(1 -\left(\frac{h'}{g'}\right)^2\right)\partial_\tau^2 -\frac{2}{g'} \left(\frac{h'}{g'}\right)
  \partial_\tau\partial_x  
  -\frac{1}{g'} \left(\frac{h'}{g'}\right)'\partial_\tau  \nn \\
&&  -\frac{1}{g'} \partial_x\left(\frac{1}{g'}\partial_x\right)
  + \hat{V}_{\ell} \Bigg]\phi_{\ell m} = 0 \, ,
\eea
where the prime denotes derivative with respect to $x$. Admittedly, expression (\ref{e:wave_equation_hg})
appears more intricate than Eq. (\ref{e:wave_equation_tortoise}). However,  this change encodes a neat
geometric structure and, as we shall argue, it plays a crucial role in our construction and discussion of the
relevant spectral problem.

\subsection{First-order reduction in time and spectral problem}
The structure in Eq. (\ref{e:wave_equation_hg}) is made more apparent by performing
a first-order reduction in time, by introducing 
\bea
\label{e:psi_u_def}
\psi_{\ell m}=\partial_\tau \phi_{\ell m} \ \ , \ \ u_{\ell m} =
\begin{pmatrix}
  \phi_{\ell m} \\
  \psi_{\ell m}
\end{pmatrix}  \ .
\eea
Then, Eq. (\ref{e:wave_equation_hg}) becomes
\bea
\label{e:wave_eq_1storder}
\partial_\tau u_{\ell m}  = i L  u_{\ell m} \ ,
\eea
where the operator $L$ is defined as
\bea
\label{e:L_operator}
L =\frac{1}{i}\!
\left(
  \begin{array}{c|c}
    0 & 1 \\
    \hline 
   L_1 & L_2
  \end{array}
  \right) \ ,
  \eea  
with
\bea
\label{e:L_1-L_2}
L_1 &=& \frac{1}{w(x)}\big(\partial_x\left(p(x)\partial_x\right) - q_\ell(x)\big) \nn \\
L_2 &=& \frac{1}{w(x)}\big(2\gamma(x)\partial_x + \partial_x\gamma(x)\big) \ ,
\eea
and
\bea
\label{e:functions_L1_L2}
&& w(x)=\frac{g'^2-h'^2}{|g'|} \ \ , \ \ p(x) = \frac{1}{|g'|} \ \ , \ \
q_\ell(x)=|g'|\;\hat{V}_\ell \nn \\
&& \gamma(x)=\frac{h'}{|g'|} \ .
\eea
The structure of $L_1$ is that of a  Sturm-Liouville operator. In particular,
functions $h(x)$ and $g(x)$ are chosen such that they guarantee the positivity
of the weight function $w(x)$, namely $w(x)>0$. The operator
$L_2$ has also a neat geometric/analytic structure adapted to the integration by parts,
being symmetric in the following form: $L_2 = \displaystyle \frac{1}{w(x)}\Big(\gamma(x)\partial_x
+ \partial_x(\gamma(x) \cdot) \Big)$.

A key property of coordinate transformation (\ref{e:change_variables}) is that
it preserves, up to the overall constant $\lambda$, the
timelike Killing vector $t^a$  controlling stationarity
\bea
\label{e:Timelike_Killing}
t^a = \partial_t = \frac{1}{\lambda}\partial_{\bar{t}} = \frac{1}{\lambda}\partial_{\tau} \ .
\eea
In this sense functions $t$ and $\lambda \tau$ ``tick'' at the same pace, namely they
are natural parameters of $t^a$, i.e. $t^a(t)=t^a(\lambda \tau) =1$ (the role of the constant
$\lambda$ being just that of keeping proper dimensions).
This is crucial for the consistent definition of QNM frequencies by Fourier (or Laplace)
transformation from Eqs. (\ref{e:wave_equation_tortoise}) and (\ref{e:wave_equation_hg}),
since variables $\omega$ respectively conjugate to $t$ and $\tau$ then coincide (up to the
constant $1/\lambda$). In other words: the change of time coordinate
in Eq. (\ref{e:change_variables}) does not affect the values of the obtained QNM frequencies.

Performing then the Fourier transform in $\tau$ in the first-order (in time) form (\ref{e:wave_eq_1storder})
of the wave equation (with standard sign convention for the Fourier modes,
$u_{\ell m}(\tau ,x)\sim u_{\ell m}(x)e^{i\omega \tau }$)
we arrive at the spectral problem for the operator $L$
\bea
\label{e:QNMspectralproblem}
L \, u_{n, \ell m} = \omega_{n,\ell m} \; u_{n, \ell m} \ ,
\eea
or, more explicitly
\bea
\label{e:QNMspectralproblem_boxes}
\left(
  \begin{array}{c|c}
    0 & 1 \\
    \hline 
   L_1 & L_2
  \end{array}
  \right)
  \begin{pmatrix}
  \phi_{\ell m} \\
  \psi_{\ell m}
  \end{pmatrix}
  =
   i\omega_{n,\ell m}
  \begin{pmatrix}
  \phi_{n,\ell m} \\
  \psi_{n,\ell m}
\end{pmatrix} \ .
\eea

\subsubsection{Regularity and outgoing boundary conditions}
\label{s:regularity_BC}
As emphasized at the beginning of this section, a major motivation for the
adopted hyperboloidal approach
is the geometric imposition of outgoing boundary conditions at future null infinity and
at the event horizon: being null hypersurfaces with light cones pointing outwards
from the integration domain, the physical causally propagating degrees of
freedom (as the scalar fields we consider here) should not admit
boundary conditions, as long as they satisfy the appropriate
regularity conditions.
%In other words, no external boundary conditions are needed.
How does this translate into the analytic scheme resulting from
the change of variables (\ref{e:change_variables})?

The key point is that transformation (\ref{e:change_variables}) must be such that
$p(x)$ in the Sturm-Liouville operator $L_1$ in Eq. (\ref{e:L_1-L_2})
vanishes at the boundaries of the compactified spatial domain $[a,b]$
\bea
\label{e:vanishing_p_borders}
p(a)=p(b)=0 \ .
\eea
This will be illustrated explicitly in the study cases discussed later.
Then the elliptic operator $L_1$ is a `singular' Sturm-Liouville operator,
this impacting directly on the boundary conditions it admits. Specifically,
if (appropriate) regularity is enforced on eigenfunctions, then $L_1$ does not admit boundary conditions. 
Moreover, such absence of boundary conditions extends to the full operator $L$ in
the hyperbolic problem. In brief: if sufficient regularity is imposed
on the space of functions $u_{n,\ell m}$, then  wave equations (\ref{e:wave_equation_hg}),
$(\ref{e:wave_eq_1storder})$ and the spectral problem
(\ref{e:QNMspectralproblem}) do not admit boundary conditions, as
a consequence of the  vanishing of  $p(x)$ at the boundaries of $[a,b]$.

This is the analytic counterpart of the geometric structure implemented in
the compactified hyperboloidal approach. QNM boundary conditions are in-built, as
regularity conditions, in the `bulk' of the operator $L$ in Eqs. (\ref{e:QNMspectralproblem}) and
(\ref{e:QNMspectralproblem_boxes}).

\subsection{Scalar product: QNMs as a non-selfadjoint spectral problem}
\label{s:scalar_product}
The outgoing boundary conditions in the present setting define a
leaky system, with a loss of energy through the boundaries --- null infinity and the black hole
horizon --- so that the system is not conservative. This suggests
that the infinitesimal generator of the evolution in Eq. (\ref{e:wave_equation_hg}),
namely the operator $L$, should be non-selfadjoint. This requires the
introduction of an appropriate scalar product in the problem. Moreover,
such identification of the appropriate Hilbert space for solutions is also key for
the regularity conditions evoked above.

Eq. (\ref{e:wave_equation_hg}) describes the evolution of each mode $\phi_{\ell m}$
in a background 1+1-Minkowski spacetime with a scattering potential $V_\ell$.
A natural scalar product in this reduced problem (cf. \cite{GasJar21} for an extended
discussion in terms of the full problem),
both from the physical and the
analytical point of view,
is given in terms of the energy associated with such scalar field mode.
In the context of the spectral problem (\ref{e:QNMspectralproblem}), we must consider generically
a complex scalar field $\phi_{\ell m}$, for which the associated stress-energy tensor writes
(dropping $(\ell, m)$ indices) is
\bea
\label{e:T_ab}
T_{ab} = \frac{1}{2}\left(\nabla_a\bar{\phi}\nabla_b\phi
-\frac{1}{2}\eta_{ab}\Big(\nabla^c\bar{\phi}\nabla_c{\phi} + V_{\ell} \bar{\phi}\phi\Big) + \mathrm{c.c}\right) \ ,
\eea
where $\eta_{ab}$ denotes the Minkowski metric in arbitrary coordinates  and ``$\mathrm{c.c}$''
indicates ``$\mathrm{complex-conjugate}$''.
In a stationary situation, the ``total energy'' contained in the spatial slice $\Sigma_\tau$
and associated with the mode $\phi$ is given \cite{Wald84} by
\bea
\label{e:Energy_Tab}
E =  \int_{\Sigma_\tau} T_{ab}(\phi,\nabla\phi) t^an^b d\Sigma_\tau \ ,
\eea
where $t^a$ is again the timelike Killing vector associated with stationarity
and  $n^a$ denotes the unit timelike normal to
the spacelike slice $\Sigma_\tau$.
Writing explicitly the energy in the compactified hyperboloidal
coordinates $(\tau,x)$  in (\ref{e:change_variables}), we get 
\bea
\label{e:energy_explicit}
&&E(\phi, \partial_\tau \phi) = \int_{\Sigma_\tau} T_{ab}(\phi, \partial_\tau \phi)  t^an^b d\Sigma_\tau\\
&&=\frac{1}{2}\int_a^b \left[(g'^2-h'^2)\partial_\tau\bar{\phi} \partial_\tau\phi
  + \partial_x\bar{\phi} \partial_x\phi + g'^2 \hat{V}_{\ell}\bar{\phi}\phi\right]\frac{1}{|g'|}dx \ ,\nn
\eea
where we identify the functions appearing in the definition of the $L_1$ operator in
(\ref{e:L_1-L_2}) and (\ref{e:functions_L1_L2}). In particular, if $g'^2-h'^2>0$ (as we have
required above)
and $\hat{V}_{\ell}>0$ (this is required for positivity of the norm)
then, identifying $\partial_\tau \phi = \psi$ as in (\ref{e:psi_u_def}),
we can write the following norm for the vector $u$ in (\ref{e:psi_u_def})
\bea
\label{e:energy_norm}
&&||u||^2_{_{E}} = \Big|\Big|\begin{pmatrix}
  \phi \\
  \psi
\end{pmatrix}\Big|\Big|^2_{_{E}} := E(\phi, \psi) \\
&&=\frac{1}{2}\int_a^b \left(w(x)|\psi|^2
  + p(x)|\partial_x\phi|^2 + q_\ell(x) |\phi|^2\right) dx \ .\nn
\eea
We refer in the following to this norm as the ``energy norm''.
We notice that $\gamma(x)$ in Eq. (\ref{e:functions_L1_L2}),
associated with $L_2$ does not enter in the norm, that is, in the energy.
This norm comes indeed from a scalar product. Rewriting, for making
its role more apparent, the $q_\ell(x)$ function as the rescaled potential $\tilde{V}_\ell$
\bea
\label{e:rescaled_V}
\tilde{V}_\ell :=  q_\ell(x) = |g'(x)|\hat{V}_\ell = \frac{\hat{V}_\ell}{p(x)}  \ ,
\eea
and under the assumption above  $\tilde{V}_\ell > 0$, we can introduce the ``energy scalar product'' for
vector functions $u$ in Eq. (\ref{e:psi_u_def}), as
\bea
\label{e:energy_scalar_product}
\langle u_1,\! u_2\rangle_{_{E}} &=& \Big\langle\begin{pmatrix}
  \phi_1 \\
  \psi_1
\end{pmatrix}, \begin{pmatrix}
  \phi_2 \\
  \psi_2
\end{pmatrix}\Big\rangle_{_{E}}  \\
&=&
\frac{1}{2} \int_a^b \!\!\!\left(\! w(x)\bar{\psi}_1 \psi_2 + p(x)  \partial_x\bar{\phi}_1\!\partial_x\phi_2 + \tilde{V}_\ell\bar{\phi}_1 \phi_2 \!\right)\!\! dx  \ , \nn
\eea
and, by construction, it holds $||u||^2_{_{E}} = \langle u, u\rangle_{_{E}}$.
This will be the relevant scalar product in our discussion.

The  full operator $L$ in (\ref{e:QNMspectralproblem}) is not selfadjoint
in the scalar product (\ref{e:energy_scalar_product}).
In fact, the first-order operator $L_2$ stands for a dissipative term encoding the energy leaking at
$\scri^+$ and the BH horizon \cite{GasJar21}. One could, at a first look, consider that this is related
to the first-order character of $L_2$, which makes it antisymmetric when
integrating by parts with a $L^2([a,b],w(x)dx)$ scalar product on $\psi$, in contrast
with the selfadjoint character of the Sturm-Liouville operator $L_1$
in $L^2([a,b],w(x)dx)$ for $\phi$ functions. However, this is misleading and actually
would suggest a wrong 'bulk' dissipation mechanism. 
When calculating the formal adjoint $L^\dagger$ of the full operator $L$ with
the scalar product (\ref{e:energy_scalar_product}), one gets
\bea
\label{e:formal_adjoint}
L^\dagger = L + L^{\partial} \ ,
\eea
where $L^{\partial}$ is an operator with support only on the boundaries of
the interval $[a,b]$, that we can formally write as
\bea
\label{e:L_dagger}
L^{\partial} =\frac{1}{i}\!
\left(
  \begin{array}{c|c}
    0 & 0 \\
    \hline 
   0 & L^\partial_2
  \end{array}
  \right) \ ,
  \eea
  with $L^\partial_2$ given by the expression
  \bea
  \label{e:L2_boundary}
  L^\partial_2 = 2\frac{\gamma(x)}{w(x)}\left(\delta(x-a)-\delta(x-b)\right) \ ,
  \eea
  where $\delta(x)$ formally denotes a Dirac-delta distribution. This
  is just a formal expression, that underlines precisely the need of a more
  careful treatment on the involved functional spaces, but it has the
  virtue of making apparent that the obstruction to selfadjointness
   lays at the boundaries, as one expects in
  our QNM problem, and not in the bulk, as one could naively conclude from the presence
  of a first-order operator $L_2$ (cf. discussion above): $L^\partial_2$ explicitly entails a boundary dissipation mechanism.
  In particular, we note that  $L$ is selfadjoint in the non-dissipative $L_2=0$ case,
  as expected, but that this has required the introduction of quite a non-trivial
  scalar product.

  As a bottomline, in this section we have cast the QNM problem as
  the eigenvalue problem of a non-selfadjoint operator.
  In the following section we discuss the implications of this.

  \section{Spectral stability and pseudospectrum}
  \label{s:spectral_stability_pseudospectrum}
 The spectrum of a non-selfadjoint operator is potentially unstable under small perturbations
of the operator. Let us consider
a linear operator $A$ on a Hilbert space with scalar product $\langle \cdot, \cdot \rangle$, 
and denote its adjoint by $A^\dagger$,  satisfying
$\langle A^\dagger u, v\rangle = \langle  u, A v\rangle$.
The operator $A$ is called normal if and only if $[A,A^\dagger]\!=\!0$.
In particular, a selfadjoint operator  $A^\dagger\!=\!A$ is normal.
In this setting, the `spectral theorem' (under the appropriate functional space assumptions)
states that a normal operator
is characterized as being unitarily diagonalizable. The eigenfunctions of $A$ form an
orthonormal basis and, crucially in the present discussion,
the eigenvalues are stable under perturbations of $A$.
The lack of such a `spectral theorem' for non-normal operators entails a severe 
loss of control on eigenfunction completeness and the potential instability
of the spectrum of the operator $A$. Here, we focus on this second aspect.

\subsection{Spectral instability: the eigenvalue condition number}
\label{s:spectral_instability}
Let us consider an operator $A$ and an eigenvalue $\lambda_i$.  Left $u_i$
and right $v_i$ eigenvectors are characterised as \footnote{In the matrix case
$u^*_i A = \lambda_i u^*_i$, with $u^*=\bar{u}^t$ , i.e. $u_i$ are indeed left-eigenvectors.}
\bea
\label{e:leftright_eigenvectors}
A^\dagger u_i= \bar{\lambda}_i u_i \ \  , \ \ A v_i = \lambda_i v_i   \ ,
\eea
with  $\bar{\lambda}_i$  the complex conjugate of $\lambda_i$.
Let us consider, for $\epsilon>0$, the perturbation of $A$ by a (bounded) operator $\delta A$
\bea
\label{e:matrix_perturbation}
A(\epsilon) = A + \epsilon \; \delta A \ \ , \ \ ||\delta A||= 1 \ .
\eea
The eigenvalues~\footnote{Specifically, we consider ``proper eigenvalues'' in the sense
  of belonging to the point spectrum $\sigma_p(A)$ of $A$, in particular not being
  part of the continuum spectrum $\sigma_c(A)$ of the operator. For simplicity, we consider
eigenvalues of multiplicity one.} in the perturbed spectral problem
\bea
\label{e:perturbed_spectral_problem}
A(\epsilon) v_i(\epsilon) = \lambda_i(\epsilon) v_i(\epsilon)  \ ,
\eea
satisfy
\bea
\label{e:eigenvalue_perturbation}
|\lambda_i(\epsilon) - \lambda_i| &=&  \epsilon  \frac{|\langle u_i, \delta A v_i(\epsilon)\rangle|}{|\langle u_i, v_i\rangle|}
= \epsilon  \frac{|\langle u_i, \delta A v_i\rangle|}{|\langle u_i, v_i\rangle|}
+ O(\epsilon^2) \\
&\leq& \epsilon \frac{||u_i|| ||\delta A v_i||}{|\langle u_i, v_i\rangle|} + O(\epsilon^2)
\leq \epsilon \frac{||u_i|| ||v_i||}{|\langle u_i, v_i\rangle|} + O(\epsilon^2)  \ ,\nn
\eea
where the first line generalizes  \cite{Kato80,trefethen2005spectra} the expression employed (for self-adjoint operators,
where $u_i=v_i$)
in quantum mechanics first-order perturbation theory, the first inequality in the second line is the Cauchy-Schwartz inequality and
in the second inequality we make explicit use of an operator norm $||\cdot||$ induced from that of the vector Hilbert space,
so that $||\delta A v||\leq ||\delta A|| ||v||$, and $||\delta A||= 1 $ in (\ref{e:matrix_perturbation}).
Then, defining the condition number $\kappa_i$ associated with the eigenvalue $\lambda_i$,
we can write the bound for the perturbation of the eigenvalue $\lambda_i$
\bea
\label{e:eigenvalue_perturbation_kappa}
|\lambda_i(\epsilon) - \lambda_i| 
\leq \epsilon \kappa_i, \quad \kappa_i = \kappa(\lambda_i) := \frac{||u_i|| \; ||v_i||}{|\langle u_i, v_i\rangle|} \ .
\eea
In the normal operator case,
$u_i$ and $v_i$ are proportional (namely, since $A$ and $A^\dagger$ commute they can
be diagonalized in the same basis).
Then, again by  Cauchy-Schwartz, $\kappa_i=1$ and we encounter spectral stability:
a small perturbation of order $\epsilon$ of the operator $A$ entails a perturbation of the same order $\epsilon$ in the spectrum. 
In contrast, in the non-normal case, $u_i$ and $v_i$ are not necessarily collinear.
In the absence of a spectral theorem nothing prevents $u_i$ and $v_i$ to become close to orthogonality
and $\kappa_i$ can become very large: small perturbations of $A$ can produce large deviations in the eigenvalues.
The relative values of $\kappa_i$ control the corresponding instability sensitivity of different $\lambda_i$'s
to an operator perturbation \footnote{Still, certain eigenvalues of a non-normal operator
(but not all) can have condition number equal to one. A 'normal eigenvalue' is defined as an eigenvalue $\lambda$ with
$\kappa(\lambda) = 1$. This notion can be helpful in the study of particular stable eigenvalues in the possibly
unstable spectrum of a non-normal operator.}.

\subsection{Pseudospectrum}
\label{s:pseudospectrum}
A complementary approach to the study of the spectral (in)stability of the
operator $A$ under perturbations consists in considering the following questions:

{\em Given the
operator $A$ and its spectrum $\sigma(A)$, which is the
set of complex numbers $\lambda \in \mathbb{C}$ that are actual eigenvalues
of ``some'' small perturbation $A + \delta A$, with $||\delta A||<\epsilon$?
Does this set extend in $\mathbb{C}$ {\em far} from the spectrum of $A$?}

In this setting, if we are dealing with an operator that is spectrally stable,
we expect that the spectrum of $A + \delta A$ will not change strongly with
respect to that of $A$, so that the set of $\lambda\in\mathbb{C}$ corresponding to the first question above will
not be far from $\sigma(A)$, staying in its vicinity at a maximum distance of order $\epsilon$.
On the contrary, if we find a tiny perturbation $\delta A$ of order $||\delta A||<\epsilon$
such that the corresponding eigenvalues of $A + \delta A$ actually reach regions in $\mathbb{C}$
at distances far apart from $\sigma(A)$, namely orders of magnitude above $\epsilon$,
we will conclude that our operator suffers of an actual spectral instability.

\medskip

\subsubsection{Pseudospectrum and operator perturbations}
\label{s:pseudospetrum_perturbations}
The previous discussion is formalized in the notion of pseudospectrum,
leading to its following (first) definition~\footnote{For the sake of
  simplicity and clarity, we dwell at the matrix level \cite{trefethen2005spectra}. For
  the discussion in general Hilbert spaces, cf. \cite{Sjostrand2019}.}.

\medskip

\noindent
     {\bf Definition 1} (Pseudospectrum: perturbative approach). Given
       $A\!\in\! M_n(\mathbb{C})$ and  $\epsilon>0$, the $\epsilon$-pseudospectrum $\sigma^\epsilon(A)$ of $A$
       is 
\bea
\label{e:pseudospectrum_def1}
 \sigma^\epsilon(A)&& \\
=&&  \{\lambda\in\mathbb{C}, \exists \; \delta A\!\in\! M_n(\mathbb{C}), ||\delta A||<\epsilon: \lambda\!\in\!\sigma(A+\delta A) \}  . \nn
\eea

\medskip

This notion of $\epsilon$-pseudospectrum $\sigma^\epsilon(A)$ is a crucial one in our study of
eigenvalue instability since it implies that points in $\sigma^\epsilon(A)$ are actual eigenvalues
of some $\epsilon$-perturbation of $A$: if $\sigma^\epsilon(A)$ extends far from the spectrum $\sigma(A)$ for
a small $\epsilon$, then a small physical perturbation $\delta A$ of $A$ can produce
large actual deviations in the perturbed physical spectrum.
The pseudospectrum becomes a systematic tool to assess spectral (in)stability, as illustrated in the hydrodynamics context~\cite{TreTreRed93}.

Although the characterization (\ref{e:pseudospectrum_def1}) of
$\sigma^\epsilon(A)$ neatly captures the notion of (in)stability of $A$,
from a pragmatic perspective it suffers from the drawback of not
providing a constructive approach to build such sets $\sigma^\epsilon(A)$
for different $\epsilon$'s (see however subsection \ref{s:pseudospectrum_random_pert}
below, for a further qualification of this question in terms of random perturbation probes).

\subsubsection{Pseudospectrum and operator resolvent}
\label{s:pseudospectrum_resolvent}
To address the construction of pseudospectra, another characterization
of the set $\sigma^\epsilon(A)$ in (\ref{e:pseudospectrum_def1}) of Definition 1 is very useful.
Such second characterization is based on the notion of the resolvent
$R_A(\lambda)=(\lambda \mathrm{Id}-A)^{-1}$ of the operator $A$.

An eigenvalue $\lambda$ of $A$ is a complex number that makes singular the operator  $(\lambda \mathrm{Id}-A)$.
More generally, the spectrum  $\sigma(A)$ of $A$ is the set $\{ \lambda\in\mathbb{C}\}$
for which the resolvent $R_A(\lambda)$ does not exist as a bounded operator
(cf. details and subtleties on this notion in e.g. \cite{Kato80,Sjostrand2019}).
This spectrum concept is a key notion for normal operators but, due to spectral instabilities discussed above,
$\sigma(A)$ is not necessarily the good object to consider for 
non-normal operators, in our context.
The notion of $\epsilon$-pseudospectrum enters then in scene.
Specifically, an equivalent characterization of the $\epsilon$-pseudospectrum set $\sigma^\epsilon(A)$
in Definition 1 is given by the following definition \cite{trefethen2005spectra,Sjostrand2019}.

 \medskip

 \noindent
 {\bf Definition 2} (Pseudospectrum: resolvent norm approach). Given
 $A\!\in\! M_n(\mathbb{C})$, its resolvent $R_A(\lambda) = (\lambda \mathrm{Id}- A)^{-1}$
 and  $\epsilon>0$, the $\epsilon$-pseudospectrum $\sigma^\epsilon(A)$ of $A$
       is characterised as
\bea
\label{e:pseudospectrum_def2}
\!\!\!\!\!\!\!\!\sigma^\epsilon(A) &=& \{\lambda\in\mathbb{C}:   ||R_A(\lambda)||
= ||(\lambda \mathrm{Id}- A)^{-1}||>1/\epsilon\}  .
\eea

 \medskip

This characterization captures that, for non-normal operators, the norm of the resolvent $R_A(\lambda)$ can be very large
far from the spectrum $\sigma(A)$. This is in contrast with the normal-operator case, where (in the $||\cdot||_2$ norm)
\bea
\label{e:resolvent_norm_bound_normal}
||R_A(\lambda)||_2\leq \frac{1}{\mathrm{dist}(\lambda, \sigma(A))} \ .
\eea
In the non-normal case, one can only guarantee (e.g. \cite{trefethen2005spectra})
\bea
\label{e:resolvent_norm_bound_nonnormal}
||R_A(\lambda)||_2\leq \frac{\kappa}{\mathrm{dist}(\lambda, \sigma(A))} \ ,
\eea
where $\kappa$ is also a condition number, different but related to the eigenvalue condition
numbers $\kappa_i$ in (\ref{e:eigenvalue_perturbation_kappa}) ($\kappa$, associated with the
matrix diagonalising $A$, provides an upper bound to the individual $\kappa_i$'s; see \cite{trefethen2005spectra}
for details). In the non-normal case, $\kappa$ can become very large and $\epsilon$-pseudospectra sets can
extend far from the spectrum of $A$ for small values of $\epsilon$.
The extension of $\sigma^\epsilon(A)$ far from
$\sigma(A)$ is therefore a signature of strong non-normality and indicates a poor analytic behavior of $R_A(\lambda)$.

The important point here is that the characterization of the $\epsilon$-pseudospectrum
in Definition 2, namely Eq. (\ref{e:pseudospectrum_def2}),
provides a practical way of calculating $\sigma^\epsilon(A)$. If we calculate the norm of the
resolvent $||R_A(\lambda)||$ as a function of $\lambda=\mathrm{Re}(\lambda)+i\mathrm{Im}(\lambda) \in\mathbb{C}$,
this provides a real function of two real variables $(\mathrm{Re}(\lambda),\mathrm{Im}(\lambda))$:
the boundaries of the $\sigma^\epsilon(A)$ sets are just the `contour lines' of the plot of this function $||R_A(\lambda)||$.
In particular, $\epsilon$-pseudospectra are nested sets in $\mathbb{C}$ around the spectrum $\sigma(A)$,
with $\epsilon$ decreasing towards the `interior' of such sets
and such that $\displaystyle\lim_{\epsilon\to0}\sigma^\epsilon(A)=\sigma(A)$.

\subsubsection{Pseudospectrum and quasimodes}
\label{s:pseudospectrum_quasimodes}
For completeness, we provide a third equivalent characterization of the
pseudospectrum in the spirit of characterising $\lambda$'s in the
$\epsilon$-pseudospectrum set $\sigma^\epsilon(A)$
as `approximate eigenvalues' of $A$, `up-to an error' $\epsilon$, with
corresponding `approximate (right) eigenvectors' $v$.
Specifically, it holds \cite{trefethen2005spectra,Sjostrand2019} that 
$\sigma^\epsilon(A)$ can be characterised also by the following (third) definition.

 \medskip
 \noindent
 {\bf Definition 3} (Pseudospectrum: quasimode approach). Given
 $A\!\in\! M_n(\mathbb{C})$ and  $\epsilon>0$, the $\epsilon$-pseudospectrum $\sigma^\epsilon(A)$ of $A$
 and its associated $\epsilon$-quasimodes $v\in\mathbb{C}^n $ are characterised by
\bea
\label{e:pseudospectrum_def3}
\sigma^\epsilon(A) 
= \{\lambda\in\mathbb{C}, \exists v\in\mathbb{C}^n: ||A v-\lambda v||<\epsilon\} \ .
\eea

\medskip

This characterisation introduces the notion of ``$\epsilon$-quasimode'' $v$ (referred to as ``pseudo-mode'' in \cite{trefethen2005spectra}),
a key notion in the  semiclassical analysis approach to the spectral study 
of $A$ \cite{Sjostrand2019}. On the other hand, this third characterization
also clearly indicates the numerical difficulty that may occur when trying to determine the actual eigenvalues of $A$,
since round-off errors are unavoidable. This signals the need of a careful treating,
whenever addressing numerically the spectral problem of a non-normal operator $A$.

\subsubsection{Pseudospectrum and choice of the norm}
\label{s:pseudospectrum_choice_norm}
In this subsection we have presented the $\epsilon$-pseudospectrum  as a notion
that may be more adapted to the analysis of non-normal operators than that
of the spectrum. We must emphasize however, that the notion of spectrum $\sigma(A)$ is intrinsic
to the operator $A$, whereas the $\epsilon$-pseudospectrum $\sigma^\epsilon(A)$ is not,
since it also depends on the choice of an operator norm. This is crucial, since
it determines what we mean by `big/small' when referring to the
perturbation $\delta A$, and therefore critically impacts the assessment of stability:
a small operator perturbation $\delta A$ in a given norm, can be a large one when considering another norm.
In the first case, from a large variation $\delta \lambda$ in the eigenvalues we would conclude instability,
whereas in the second case such variation could be consistent with stability.

In this sense, from a mathematical perspective, the study of spectral (in)stability
through pseudospectra amounts, in a good measure, to the identification of the proper
scalar product determining the norm, that is, to the identification of the
proper Hilbert space in which the operator $A$ acts. However, from a physical perspective
we might not have such a freedom to choose a mathematically conveniently rescaled norm, since
what we mean by large and small may be fixed by the physics of the problem, e.g.
by the size of involved amplitudes, intensities or the energy contained
in the perturbations. Then, the choice of an appropriate norm, both from a mathematical
and physical perspective, is a fundamental step in the analysis (cf. discussion in \cite{GasJar21}).
This is the rationale behind the choice of the energy norm $||\cdot||_{_\mathrm{E}}$ in (\ref{e:energy_norm}).
Once the norm is chosen,  the equivalent characterizations
in Definitions 1, 2 and 3, respectively Eqs. (\ref{e:pseudospectrum_def1}),
(\ref{e:pseudospectrum_def2}) and (\ref{e:pseudospectrum_def3}),
emphasize complementary aspects of the $\epsilon$-pseudospectrum notion and the $\sigma^\epsilon(A)$ sets.

\subsection{Pseudospectrum and random perturbations}
\label{s:pseudospectrum_random_pert}
When considering the construction of pseudospectra, we have presented the characterization
of $\sigma^\epsilon(A)$ in terms of the resolvent $R_A(\lambda)$ in Definition 2, Eq. (\ref{e:pseudospectrum_def2}),
as better suited than the one in terms of spectra of perturbed
operators in Definition 1, Eq. (\ref{e:pseudospectrum_def1}). 
The reason is that the former involves only the
unperturbed operator $A$, whereas the latter demands a study of the spectral problem for {\em any} perturbed operator
$A+\delta A$ with small $\delta A$: a priori, the difficulty to  explicitly control such 
space of possible $\delta A$ perturbations hinders an approach based on such characterisation
in Definition 1.

But the very nature of the obstacle suggests a possible solution, namely to consider the systematic study
of the perturbed spectral problem under random perturbations $\delta A$ as
an avenue to explore $\epsilon$-pseudospectra sets. This heuristic
expectation actually withstands a more careful analysis and constitutes the
basis of a rigorous approach to the analysis of pseudospectra \cite{Sjostrand2019}.
From a practical perspective, the systematic study of the spectral problem
of $A + \delta A$ with (bounded) random  $\delta A$ with $||\delta A||\leq \epsilon$,
has proven to be an efficient tool to explore the `migration' of eigenvalues through
the complex plane (inside the $\epsilon$-pseudospectra) \cite{trefethen2005spectra}.
This is complementary to (and 'technically' independent from) the evaluation of $\sigma^\epsilon(A)$
from the contour-lines of the norm $||R_A(\lambda)||$ of the resolvent. Such complementarity of approaches
will prove key later in our analysis  of Nollert \& Price's high-frequency perturbations
of the Schwarzschild's potential and the related QNMs.

Two important by-products of this random perturbation approach to the pseudospectrum are the following:
\begin{itemize}
\item[i)] Random perturbations help identifying instability-triggering perturbations: $\epsilon$-pseudospectra and
  condition numbers $\kappa_i$ are efficient in identifying the instability of the spectrum and/or
  a particular eigenvalue $\lambda_i$, respectively. However, they do not inform on the specific
  kind of perturbation actually triggering the instability. This can be crucial to assess the physical
  nature of the found instability. The use of families of random operators adapted to
  specific types of perturbations sheds light on this precise point. We will make critical use
  of this in our assessment of Schwarzschild's (in)stability.

\item[ii)] Random perturbations improve analyticity: a remarkable and apparently counter-intuitive effect of
  random perturbations is the improvement of the analytic behaviour
  of $R_A(\lambda)$ in $\lambda\in\mathbb{C}$~\cite{Sjostrand2019}.
  In particular, the norm $||R_A(\lambda)||$ gets reduced away from $\sigma(A)$,
  as for normal operators [cf. Eq. (\ref{e:resolvent_norm_bound_normal})], 
  so that the $\epsilon$-pseudospectra sets pattern becomes ``flattened'' (a signature of good analytic behaviour)
  below the random-perturbation scale $\epsilon$.
\end{itemize}
To complement this perspective on the relation between the two given approaches to spectral (in)stability, namely 
perturbation theory and $\epsilon$-pseudospectra, respectively subsections
\ref{s:spectral_instability} and  \ref{s:pseudospectrum},
let us connect eigenvalue condition numbers $\kappa(\lambda_i)$ with $\epsilon$-pseudospectra $\sigma^\epsilon(A)$.
The question we want to address is: how far can the $\epsilon$-pseudospectrum $\sigma^\epsilon(A)$ get away
from the spectrum $\sigma(A)$? The $\kappa_i$'s provide the answer.

Let us define the `tubular neighbourhood'  $\Delta_\epsilon(A)$ of radius $\epsilon$ around the spectrum $\sigma(A)$ as
\bea
\label{e:tubular_epsilon}
\Delta_\epsilon(A) = \{\lambda\in \mathbb{C}: \mathrm{dist}\left(\lambda,\sigma(A)\right)<\epsilon\}  \  ,
\eea
which is always contained in the $\epsilon$-pseudospectrum $\sigma^\epsilon(A)$~\cite{trefethen2005spectra}
\bea
\Delta_\epsilon(A)\subseteq \sigma^\epsilon(A) \ .
\eea
The key question is about the inclusion in the other direction. Normal operators indeed satisfy
\cite{trefethen2005spectra}
\bea
\label{e:tubular_normal}
\sigma_2^\epsilon(A) = \Delta_\epsilon(A) \ ,
\eea
where $\sigma_2^\epsilon(A)$ indicates the use of a $||\cdot||_2$ norm. That is,
a ($||\delta A||<\epsilon$) perturbed eigenvalue of a normal operator 
can move up to a distance $\epsilon$ from $\sigma(A)$.
This is what we mean by spectral stability: an operator perturbation of order $\epsilon$ induces
an eigenvalue perturbation also of order $\epsilon$.
However, in the non-normal case, where
$\kappa(\lambda_i)> 1$,  it holds (for small $\epsilon$)  \cite{trefethen2005spectra}
\bea
\label{e:tubular_error}
\sigma^\epsilon(A)\subseteq \Delta_{\epsilon\kappa}(A):=\bigcup_{\lambda_i\in\sigma(A)} \Delta_{\epsilon\kappa(\lambda_i)+O(\epsilon^2)}(\{\lambda_i\})
\ ,
\eea
so that $\sigma^\epsilon(A)$ can extend into a much larger tubular neighbourhood
of radius $\sim\epsilon\kappa(\lambda_i)$ around each eigenvalue, signaling spectral instability
if $\kappa(\lambda_i)\gg 1$. This bound is the essential content 
of the Bauer-Fike theorem relating pseudospectra and eigenvalue perturbations (cf.  \cite{trefethen2005spectra}
for a precise formulation).

\section{Numerical approach: Chebyshev's spectral methods}
\label{s:numerical_approach}
The present work is meant as a first assessment of  BH QNM (in)stability
by using pseudospectra. At this exploratory stage, we address the
construction of pseudospectra in a numerical approach.
As indicated in section \ref{s:pseudospectrum_quasimodes}, the study
of the spectral stability of non-normal operators is a challenging problem that demands
high accuracy. Spectral methods provide well-adapted tools for these calculations
\cite{trefethen2005spectra,trefethen2000spectral,canuto2007spectral}.

We discretize the differential operator $L$ in (\ref{e:wave_eq_1storder})-(\ref{e:QNMspectralproblem})
via Chebyshev differentiation matrices, built on 
Chebyshev-Lobatto $n$-point grids, producing $L^{N}$ matrix approximates
(we note systematically $n=N+1$ in spectral grids, cf. appendix \ref{a:Chebyshev_elements}).
Once the operator is discretized, the construction of the
pseudospectrum requires the evaluation of matrix norms.
A standard practical choice \cite{trefethen2005spectra,trefethen2000spectral} involves the matrix norm
induced from the Euclidean $L^2$ norm in the vector space $\mathbb{C}^n$ that,
starting from  Eq. (\ref{e:pseudospectrum_def2}) in the Definition 2 of the pseudospectrum, 
leads to the following rewriting \cite{trefethen2005spectra,trefethen2000spectral} 
\bea
\label{e:pseudospectrum_carac_L2}
\sigma^\epsilon_{_2}(A) = \{\lambda\in\mathbb{C}:   \sigma^\mathrm{min}(\lambda \mathrm{Id}- A)<\epsilon\} \ , 
\eea
where $\sigma^\mathrm{min}(M)$ denotes the smallest singular value of $M$,
that is, $\sigma^\mathrm{min}(M)=\min \{\sqrt{\lambda}:  \lambda\in \sigma(M^*M) \}$,
with $M \in M_n(\mathbb{C})$ and $M^*$ its conjugate-transpose $M^*=\overline{M}^t$.

Although Eq. (\ref{e:pseudospectrum_carac_L2}) captures the spectral instability structure
of $A$, the involved $L^2$ scalar product in $\mathbb{C}^n$ is neither faithful to the structure
of the operator $L$ in Eq.~(\ref{e:wave_eq_1storder}), nor to the physics
of the BH QNM problem (cf. discussion in section \ref{s:pseudospectrum_choice_norm}).
Instead, we rather use the natural norm in the problem, specifically the Chebyshev-discretrised
version of the `energy norm'  (\ref{e:energy_norm}), following from the
Chebyshev-discretised version of the scalar product (\ref{e:energy_scalar_product}).
Specifically, we write the discretised scalar product in an appropriate basis as (we abuse the notation, since 
we use $\langle \cdot, \cdot \rangle_{_E}$ as in (\ref{e:energy_scalar_product}),
although this is now a scalar product in a finite-dimensional space $\mathbb{C}^n$)
\bea
\label{e:matrix_scalar_product_n}
\langle u, v\rangle_{_E} = (u^*)^i G^{E}_{ij}v^j = u^*\cdot G^{E}\cdot v \ \ , \ \ u,v \in \mathbb{C}^n \ , 
\eea
where $G^E_{ij}$ is the Gram matrix corresponding to (\ref{e:energy_scalar_product})
(cf. appendix \ref{a:Chebyshev_elements} for its construction) and
we note $u^* = \bar{u}^t$.
The adjoint $A^\dagger$ of $A$ with respect to
$\langle \cdot, \cdot \rangle_{_E}$ writes then
\bea
\label{e:matrix_adjoint}
A^\dagger = \left(G^E\right)^{-1}\cdot A^*\cdot G^E \ .
\eea
The vector norm $||\cdot||_{_E}$ in $\mathbb{C}^n$ associated with $\langle \cdot, \cdot \rangle_{_E}$
in Eq. (\ref{e:matrix_scalar_product_n})
induces a matrix norm $||\cdot||_{_E}$ in $M_n(\mathbb{C})$ (again, we abuse notation by
using the same symbol for the norm in $\mathbb{C}^n$ and in $M_n(\mathbb{C})$).
Then, cf. appendix \ref{a:pseusdospectrum_energy_norm}, the
$\epsilon$-pseudospectrum $\sigma^\epsilon_{_E}(A)$ of $A\in M_n(\mathbb{C})$ 
in the norm $||\cdot||_{_E}$ writes
\bea
\label{e:pseudospectrum_carac_E}
\sigma^\epsilon_{_E} (A) = \{\lambda\in\mathbb{C}:   s_{_E}^\mathrm{min}(\lambda \mathrm{Id}- A)<\epsilon\} \ , 
\eea
where $s_{_E}$ is the smallest of the ``generalized singular values''
\bea
s_{_E}^\mathrm{min}(M) = \min \{\sqrt{\lambda}:  \lambda\in \sigma(M^\dagger M) \} \ ,
\eea
with $M \in M_n(\mathbb{C})$ and its adjoint $M^\dagger$ given by Eq. (\ref{e:matrix_adjoint}).

\section{A toy model: P\"oschl-Teller potential}
\label{s:PT}
As presented in the previous sections,
in our study of BH QNMs and their (in)stabilities,
we exploit the geometrical framework of the hyperboloidal approach to analytically impose the physical boundary conditions
at the BH horizon and at the radiation zone (future null infinity).
As discussed in section \ref{s:Hyper_approach}, a crucial feature
of such a strategy is that it allows us to cast the calculation of the
QNM spectrum explicitly as the spectral problem of a non-selfadjoint
differential operator, which is then the starting point for the
tools assessing spectral instabilities presented in section
\ref{s:spectral_stability_pseudospectrum}, namely the construction
of the pseudospectrum and the analysis of random perturbations.
Finally, spectral methods discussed in section
\ref{s:numerical_approach} are employed to study these spectral
issues through a discretisation for the derivative operators.
Prior to the study of the BH case, the goal of this section is
to illustrate this strategy in a toy model, 
namely the one given by the  P\"oschl-Teller potential.

\subsection{Hyperboloidal approach in P\"oschl-Teller}
\label{s:hyper_PT}
The P\"oschl-Teller potential~\footnote{Also known as Eckart, Rosen-Morse,
  Morse-Feshbach potential, see \cite{Boonserm:2010px} for a discussion of
the terminology.}, given by the expression
\bea
\label{e:PT}
V(\bar{x}) = \frac{V_o}{\mathrm{cosh}^2(\bar{x})} = {V_o} \, {\mathrm{sech}^2(\bar{x})}\ \ , \ \ \bar{x}\in]-\infty,\infty[ \ ,
\eea
has been widely used as a benchmark for the study
of QNMs in the context of BH perturbation
theory~(e.g. \cite{FerMas84,Beyer:1998nu,Medved:2003rg}).
Interestingly, QNMs of this potential have been very recently revisited
to illustrate, on the one hand, the hyperboloidal approach to QNMs in a
discussion much akin to the present one (cf. \cite{Bizon:2020qnd},
cast in the setting of de Sitter spacetime) or, on the other hand,
functional analysis key issues related to the selfadjointness of the
relevant operator \cite{Fabris:2020kog}.
Our interest in P\"oschl-Teller stems from
the fact that it  shares the
fundamental behavior regarding QNM (in)stability to be encountered
later in the BH context, but in a mathematically much simpler setting.
In particular, P\"oschl-Teller presents weaker singularities
than the Regge-Wheeler and Zerilli potentials in Schwarzschild,
that translates in the absence of a continuous part of the spectrum
of the relevant operator $L$
(corresponding to the ``branch cut'' in standard approaches to QNMs).

Let us consider the compactified hyperboloids given by Bizo\'n-Mach coordinates
 \cite{Bizon:2014nla,Donninger:2020sqm}
mapping $\mathbb{R}$ to $]-1,1[$
\bea
\label{e:change_variables_tanh}
\left\{
\begin{array}{rcl}
\tau &=& \bar{t} - \ln\left(\cosh \bar{x}\right) \\ 
x &=& \tanh \bar{x}  
\end{array}
\right. \ ,
\eea
or, equivalently
\bea
\left\{
\begin{array}{rcl}
\bar{t} &=& \tau - \frac{1}{2}\ln(1-x^2) \\ 
\bar{x} &=& \mathrm{arctanh}(x)  
\end{array}
\right.\ .
\eea
In the spirit of the conformal compactification along the hyperboloids described in section \ref{s:wave_eq_hyper_approach},
we add the two points at (null) infinity (no BH horizon here), namely $x=\pm 1$, so that we work with the compact
interval $[a,b]=[-1,1]$.
Under this transformation the wave equation  (\ref{e:wave_equation_tortoise})
reads
 \bea
  \label{e:wave_equation_hyperboloidal_tanh}
  &&\Big((1-x^2)\left(\partial^2_\tau + 2x \partial_\tau\partial_x + \partial_\tau + 2x \partial_x - (1-x^2)\partial^2_x\right)
     \nn \\
   &&  + V\Big)\phi = 0 \ ,
  \eea
  namely the version of Eq. (\ref{e:wave_equation_hg}) corresponding to the transformation (\ref{e:change_variables_tanh}).
 We notice that angular labels $(\ell,m)$ are not relevant in the one-dimensional P\"oschl-Teller problem.
  If $x\neq 1$  we can divide by $(1-x^2)$ and, defining
  \bea
  \label{e:potential_tilde}
  \tilde{V}(x) = \frac{V}{(1-x^2)} \ ,
  \eea
we can write    
  \bea
  \label{e:wave_equation_hyperboloidal_tanh_v2}
  \left(\partial^2_\tau + 2x \partial_\tau\partial_x + \partial_\tau + 2x \partial_x
  - (1-x^2)\partial^2_x + \tilde{V}\right)\phi = 0 \ .
  \eea
  This expression is formally valid for any given potential $V(\bar{x})$ (although analyticity issues may
  appear if the asymptotic decay is not sufficiently fast, as it is indeed the case for Schwarzschild
  potentials at $\scri^+$). If we now
  insert the P\"oschl-Teller expression (\ref{e:PT}) and notice  $\mathrm{sech}^2(\bar{x}) = 1-x^2$, 
  we get a remarkably simple effective potential $\tilde{V}$, actually a constant
  \bea
  \tilde{V}(x)=V_o \ .
  \eea
  In particular, the  P\"oschl-Teller wave equation (\ref{e:wave_equation_hyperboloidal_tanh_v2}) 
exactly corresponds to Eq. (4) in \cite{Bizon:2020qnd},
so that  the P\"oschl-Teller problem is equivalent to the Klein-Gordon equation in de Sitter spacetime
with mass $m^2=V_o$.  In the following, we choose $\lambda=1/\sqrt{V_o}$ in the rescaling (\ref{e:dimensional_rescaling}),
so that we can set
\bea
\label{e:tildeV=1}
\tilde{V}=1 \ .
\eea
  Performing now the first-order reduction in time (\ref{e:psi_u_def})-(\ref{e:wave_eq_1storder})
  we get for $w(x)$, $p(x)$, $q(x)$ and $\gamma(x)$ in Eq.~(\ref{e:functions_L1_L2}) the
  values
\bea
\label{e:functions_L1_L2_PT}
&& w(x)=1 \ \ , \ \ p(x) = (1-x^2) \ \ , \ \
q(x)=\tilde{V}= 1\nn \\
&& \gamma(x)=-x \ ,
\eea
and therefore the operators $L_1$ and $L_2$ building the operator $L$ in Eq.~(\ref{e:L_operator}) write,
in the P\"oschl-Teller case, as
\bea
\label{e:L_1-L_2_PT}
L_1 &=& \partial_x\left((1-x^2)\partial_x\right) -1 \nn \\
L_2 &=&  -\left(2x \partial_x + 1\right) \ .
\eea
As discussed in section \ref{s:regularity_BC}, the function $p(x)=1-x^2 $
vanishes at the boundaries of the interval $[a,b]=[-1,1]$,
defining a singular Sturm-Liouville operator. This is at the basis of the absence of boundary
conditions, if sufficient regularity is enforced on the eigenfunctions of the spectral problem.
Regularity therefore encodes the
outgoing boundary conditions (see below).
Finally, the scalar product (\ref{e:energy_scalar_product}) writes in this case as
\bea
\label{e:energy_scalar_product_PT}
\langle u_1,\! u_2\rangle_{_{E}} &=& \Big\langle\begin{pmatrix}
  \phi_1 \\
  \psi_1
\end{pmatrix}, \begin{pmatrix}
  \phi_2 \\
  \psi_2
\end{pmatrix}\Big\rangle_{_{E}}  \\
&=&
\frac{1}{2}\!\!\int_{-1}^1 \!\!\!\left(\! \bar{\psi}_1 \psi_2 + (1-x^2)  \partial_x\bar{\phi}_1\!\partial_x\phi_2 +\bar{\phi}_1 \phi_2 \!\right)\!\! dx \ .\nn
\eea

\subsection{P\"oschl-Teller QNM spectrum}

\subsubsection{Exact P\"oschl-Teller QNM spectrum}
\label{s:exact_PT_sectrum}
P\"oschl-Teller QNM spectrum can be obtained by solving the eigenvalue
problem (\ref{e:QNMspectralproblem})-(\ref{e:QNMspectralproblem_boxes}) with
operators $L_1$ and $L_2$ given by Eq. (\ref{e:L_1-L_2_PT}). As commented above, no boundary
conditions need to be added, if we enforce the appropriate regularity.
In this particular case, this eigenvalue problem can be solved exactly.
The resolution itself is informative, since it illustrates this regularity
issue concerning boundary conditions.

If we substitute the first component of  (\ref{e:QNMspectralproblem_boxes}) 
into the second or, simply, if we take the Fourier transform in $\tau$
in Eq. (\ref{e:wave_equation_hyperboloidal_tanh_v2}) (with $\tilde{V}=1$ from  the
chosen $\lambda$ leading to Eq.~(\ref{e:tildeV=1})),
we get
\bea
\label{e:PT-prehypergeometric}
\left[(1-x^2)\frac{d^2}{dx^2} - 2(i\omega+1)x\frac{d}{dx} - i\omega(i\omega +1) -1  \right]\phi =0 \ ,
\eea
This equation can be solved in terms of the hypergeometric function ${}_2F_1(a,b;c;z)$,
with $\displaystyle z=\frac{1-x}{2}$ (see details
in appendix \ref{a:QNM_PT}). In particular, for each value of the spectral parameter
$\omega$ we have a solution that can be written as a linear combination
of linearly independent solutions obtained from ${}_2F_1(a,b;c;z)$. Discrete QNMs are obtained
only when we enforce the appropriate regularity, that encodes the outgoing boundary conditions.
In this case, this is obtained by enforcing the solution to be analytic in $x\in[-1,1]$
(corresponding in $z$ to analyticity in the full closed interval $[0,1]$), which amounts
to truncate the hypergeometric series to a polynomial. We emphasize that such a need of truncating
the infinite series to a polynomial, a familiar requirement encountered in many different physical settings,
embodies here the enforcement of outgoing boundary conditions.
In sum, this strategy leads to the P\"oschl-Teller QNM frequencies (cf. e.g. \cite{Beyer:1998nu,Boonserm:2010px})
\bea
\label{e:PT_QNM}
\omega^\pm_n =  \pm \frac{\sqrt{3}}{2} +i\left(n +\frac{1}{2}\right) \ ,
\eea
with corresponding QNM eigenfunctions in this setting
\bea
\label{e:phi_n}
\phi^{\pm}_n(x) = P_n^{(i\omega^\pm_n, i\omega^\pm_n)}(x) \ , \ x \in [-1,1] \ ,
\eea
where $P_n^{(\alpha,\beta)}$ are the Jacobi polynomials (see appendix \ref{a:QNM_PT}).
Two comments are in order here:
\begin{itemize}
\item[i)] {\em QNMs are normalizable}: QNM eigenfunctions $\phi^{\pm}_n(x)$ are finite and regular
  when making $\bar{x}\to\pm\infty$, corresponding to $x=\pm 1$. This is in contrast with
  the exponential divergence of QNM eigenfunctions in Cauchy approaches, where the time
  slices reach spatial infinity $i^0$. This is a direct consequence of the hyperboloidal approach
  with slices reaching $\scri^+$. The resulting normalizability of the QNM eigenfunctions can be relevant in e.g.
  resonant expansions (cf. e.g. discussion in \cite{LalYanVyn17}).

\item[ii)] {\em QNM regularity and outgoing conditions}:
  In the present case, namely P\"oschl-Teller in Bizo\'n-Mach coordinates, analyticity 
  (actually polynomial structure) implements the regularity enforcing outgoing boundary conditions.
  Analyticity is too strong in the general case. But asking for smoothness is not enough
  (see e.g. \cite{Ansorg:2016ztf}). In Refs. \cite{Gajic:2019qdd,Gajic:2019oem,galkowski2020outgoing}
  this problem is approached in terms of Gevrey classes, that interpolate between
  analytic and (smooth) $C^\infty$ functions, identifying the space of  $(\sigma,2)$-Gevrey functions
  as the proper regularity notion. The elucidation of the general adequate functional space for QNMs,
  tantamount of the consistent implementation of outgoing boundary conditions, is crucial for the
  characterization of QNMs in the hyperboloidal approach.
\end{itemize}

\subsubsection{Numerical P\"oschl-Teller QNM spectrum}
\label{s:numerical_PT_sectrum}
\begin{figure}[h!]
\centering
\includegraphics[width=8.5cm]{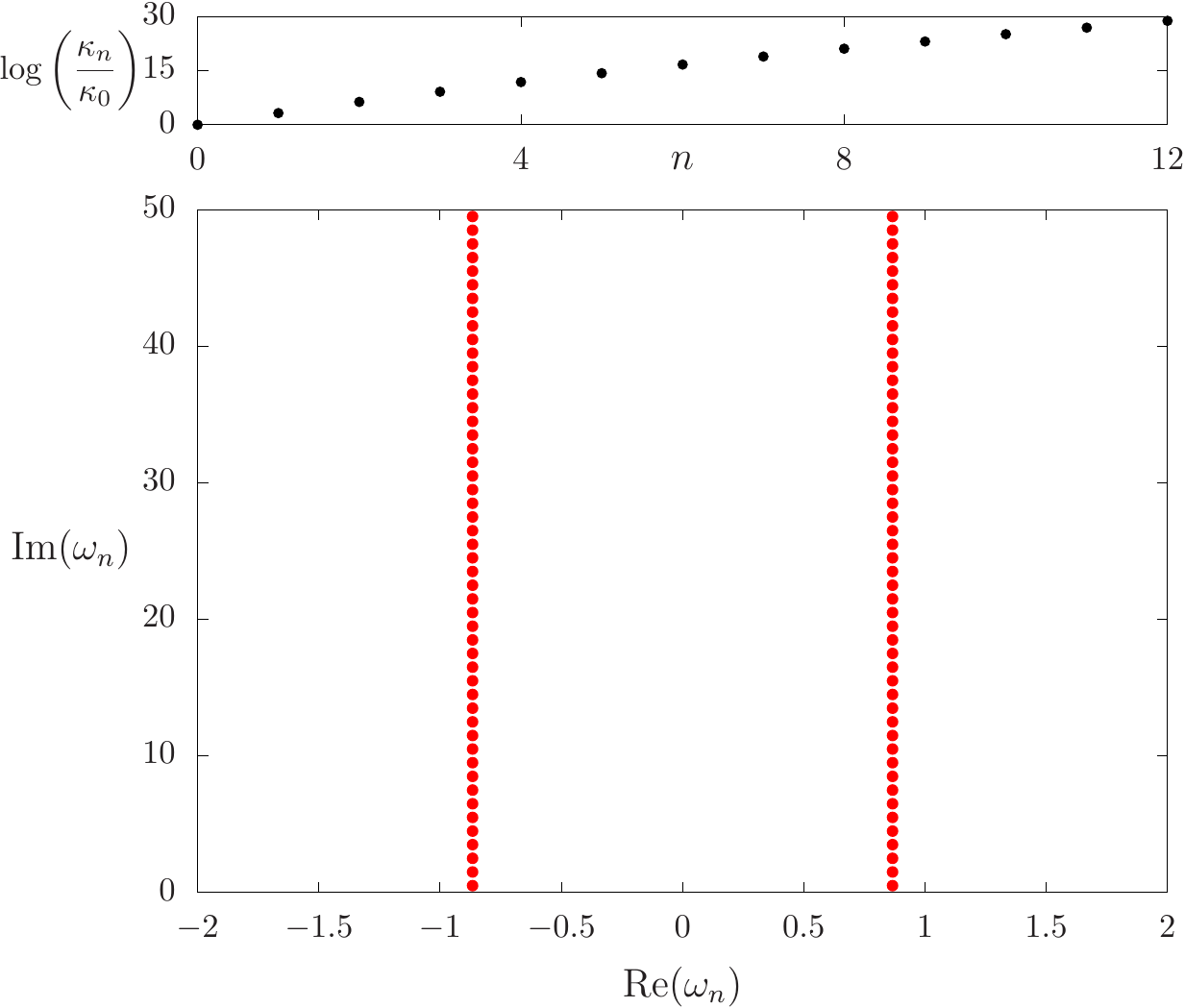}
\caption{P\"oschl-Teller QNM problem.
  {\em Bottom panel}: QNM spectrum for the P\"oschl-Teller potential, calculated in the hyperboloidal
  approach described in section \ref{s:hyper_PT}, with Chebyshev spectral methods and enhanced machine precision.
  {\em Top panel}: ratios of condition numbers $\kappa_n$ of the first QNMs
  over the condition number $\kappa_0$ of the fundamental QNM, indicating a growing spectral instability compatible
  with the need of using enhanced machine precision.}
\label{fig:Eigenvalues_PT}
\end{figure}

Fig.~\ref{fig:Eigenvalues_PT} shows the result of the numerical counterpart of the P\"oschl-Teller eigenvalue
calculation, whose exact discussion has been presented above,
by using the discretised operators $L$, $L_1$ and $L_2$
described in section \ref{s:numerical_approach} and appendix
\ref{a:Chebyshev_elements}
\bea
\label{e:L_discrete}
L^{N} v_n^{(N)}= \omega_n^{(N)}  v_n^{(N)}  \ .
\eea
This indeed recovers numerically the analytical result in Eq. (\ref{e:PT_QNM}) (we drop the ``$\pm$'' label, focusing
on one of the branches symmetric with respect the vertical axis).

We stress that the remarkable agreement between the numerical values from the bottom panel of Fig.~\ref{fig:Eigenvalues_PT}
(see also Fig.~\ref{fig:EVPert_PT} later) and the exact expression \eqref{e:PT_QNM} is far from being a trivial result,
as already illustrated in existing systematic numerical studies. This is in particular the case of Ref. \cite{BinZwo}
(where P{\"o}schl-Teller is referred to as the Eckart barrier potential), where the fundamental
mode $\omega^\pm_0$ in (\ref{e:PT_QNM}) is stable and accurately recovered, whereas {\em all} overtones
$\omega^\pm_{n\geq 1}$
suffer from a strong instability (triggered, according to the discussion in \cite{BinZwo,Zwors87},
by the $C^1$-regularity of the 
approximation modelling the P\"oschl-Teller potential) and could not be recovered.

\begin{figure}[h!]
\centering
\includegraphics[width=8.5cm]{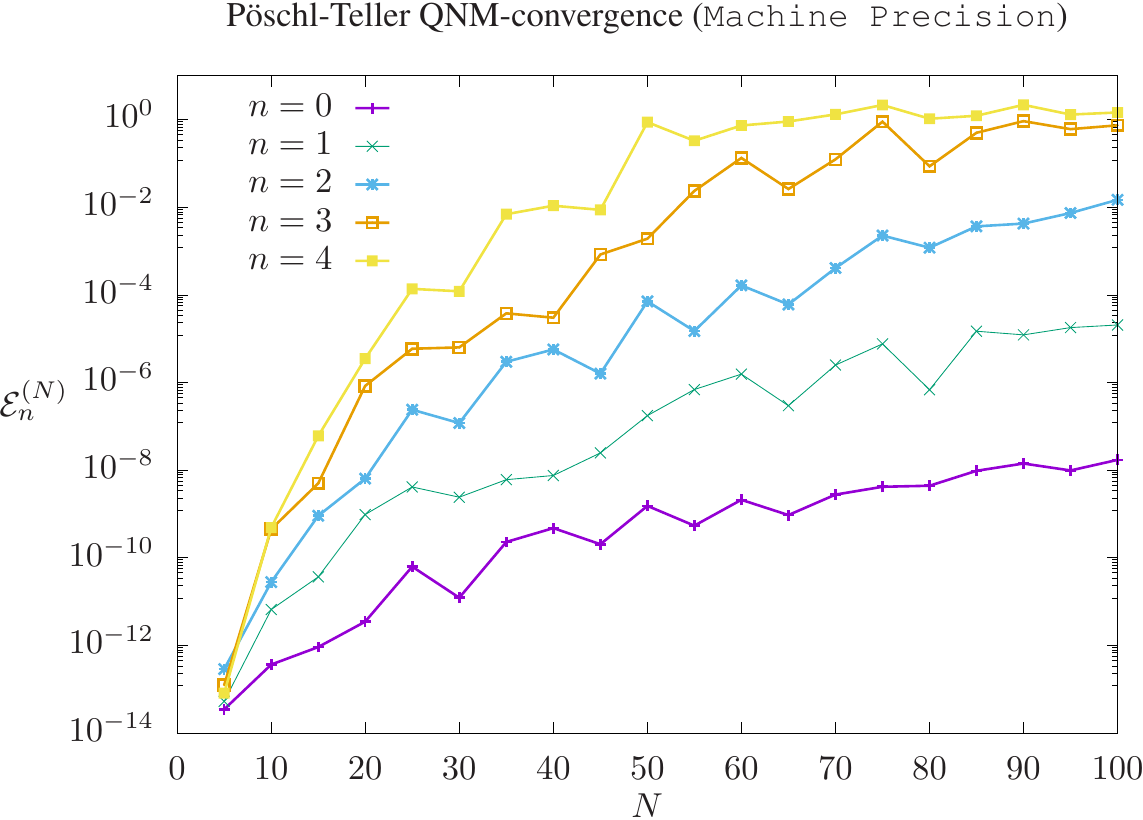}
\includegraphics[width=8.5cm]{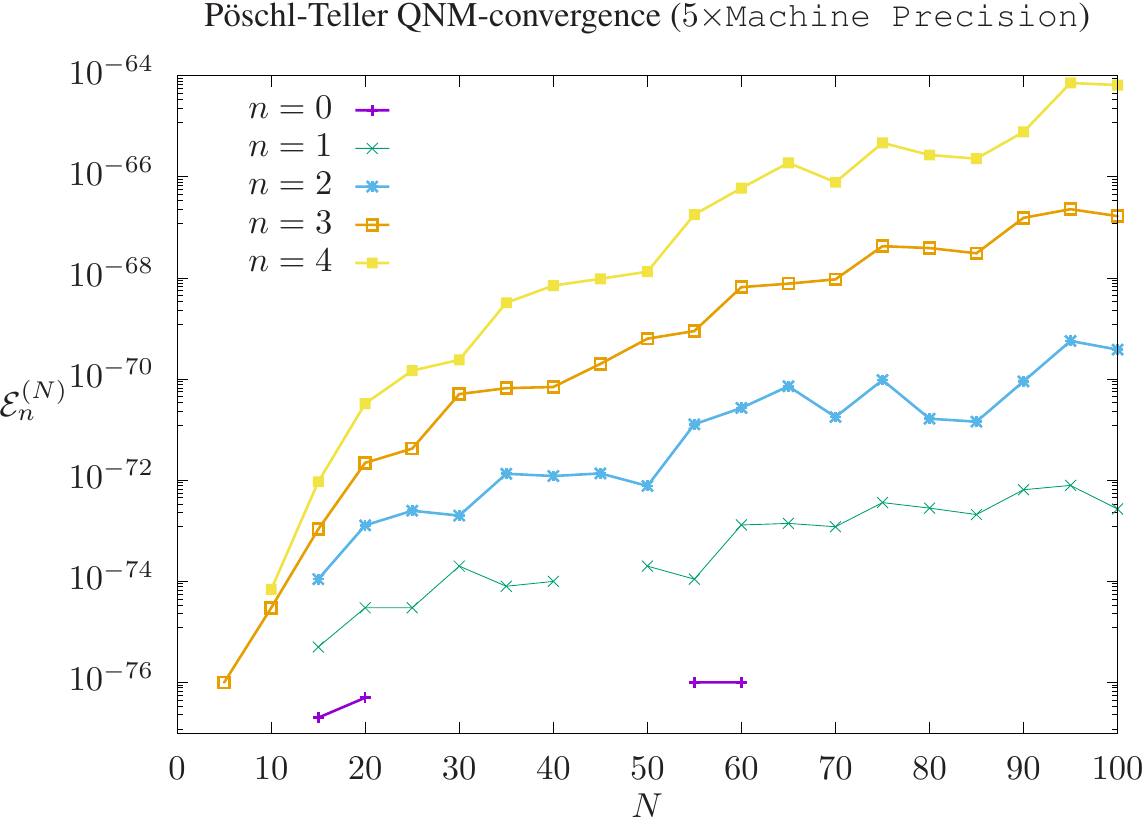}
\caption{Convergence test for the P\"oschl-Teller QNM. {\em Top panel}: double floating point operations with internal round-off error set to \texttt{Machine Precision}. {\em Bottom panel}: double floating point operations with internal round off error set to \texttt{$5\times$Machine Precision}.
  Note that missing points for $n=0$ correspond to errors that exactly vanish at the employed
    machine precision.
}
\label{fig:conv_PT}
\end{figure}

In our setting, a convergence study of the numerical values shows that the relative error 
\be
{\cal E}_{n}^{(N)} = \left| 1 - \frac{\omega^{(N)}_n}{\omega_n}\right| \ ,
\ee
between the exact QNM $\omega_n$ and the corresponding numerical approximation $\omega^{(N)}_n$
(obtained at a given truncation $N$
of the differential operator) actually {\em increases} with the resolution. This is a first hint of the instabilities
to be discussed later.
Indeed, the top panel of Fig.~\ref{fig:conv_PT} displays the error for the fundamental mode $n=0$ and the first overtones $n=1,\ldots, 4$ when the eigenvalue problem for the discretised operator is naively solved with the standard machine roundoff error for  floating point operations (typically, $\sim 10^{-16}$ for double precision).

It is astonishing how, despite the simplicity of the exact solution, the relative error grows significantly already
for the first overtones and, crucially, more strongly as the damping grows with higher overtones.
To mitigate such a drawback, one needs to modify the numerical treatment in order to allow
for a smaller roundoff error in floating point operations. The bottom panel Fig.~\ref{fig:conv_PT} shows the
error ${\cal E}_{n}^{(N)}$ when the calculations are performed with an internal roundoff error according to
\texttt{$5\times$Machine Precision}, i.e. $\sim 10^{-5\times 16}$. In this case, the fundamental QNM $n=0$ is
``exactly'' calculated at the numerical level (i.e. the difference between its exact value and the numerical
approximation vanishes in the
employed precission). The error for the overtones still grows, but in a safe range for all practical purposes.
The values displayed in the bottom panel of Fig.~\ref{fig:Eigenvalues_PT} were obtained with internal roundoff error
set to \texttt{$10\times$Machine Precision} and we can assure that the errors of all overtones are smaller than $10^{-100}$.

\subsubsection{Condition numbers of QNM frequencies}
\label{s:PT_condition_numbers}
The growth in the relative error as we move to higher overtones in Fig.~\ref{fig:conv_PT}, suggests
an increasing spectral instability in $n$ of eigenvalues $\omega^\pm_n$, triggered
by numerical errors related to machine precision, so that this instability can be reduced
(but not eliminated) by improving the internal roundoff error.

At the level of the non-perturbed spectral problem (\ref{e:L_discrete}), and in order to assess more systematically such spectral instability,
we can apply the discussion in section \ref{s:spectral_instability} to the P\"oschl-Teller approximates $L^N$.
Namely, solving the right-eigenvector problem (\ref{e:L_discrete}) together with left-eigenvector one
\bea
\label{e:L_discrete_dagger}
\left(L^{N}\right)^\dagger u_n^{(N)}= \bar{\omega}_n^{(N)}  u_n^{N}  \ ,
\eea
we can compute the condition numbers
$\kappa^{(N)}_n =\kappa(\omega^{(N)}_n)=  || v^{(N)}_n||_{_{E}} ||u_n^{(N)}||_{_E}/|\langle v^{(N)}_n, u^{(N)}_n \rangle_{_E}|$
introduced in Eq. (\ref{e:eigenvalue_perturbation_kappa}).
Notice that this is quite a non-trivial calculation, since it involves: first, the
construction of the adjoint operator $\left(L^{N}\right)^\dagger = \left(G^E\right)^{-1}\cdot (L^N)^*\cdot G^E$
and, second, the calculation of scalar products $\langle \cdot, \cdot \rangle_{_{E}}$ and (vector) energy
norms $||\cdot||_{_E}$. These calculations involve the determination of the Gram matrix $G^E$ associated
with the energy scalar product (\ref{e:energy_scalar_product_PT}) by implementing
expression (\ref{e:ScalarProd_WaveEq_2}) in appendix \ref{a:Chebyshev_elements}.
These expressions are quite non-trivial and in the
following section we provide a strong test to the associated analytical and numerical construction.

The result is shown in the top panel of Fig.~\ref{fig:Eigenvalues_PT}. The ratio of the condition numbers $\kappa_n$,
relative to the condition number of the fundamental mode $\kappa_0$, grows strongly with $n$.
This indicates a strong and increasing spectral instability consistent with the error convergence
displayed in Fig.~\ref{fig:conv_PT}. The rest of this section is devoted to address this spectral stability issue.

\subsection{P\"oschl-Teller pseudospectrum}
\label{s:PT-pseudospectrum}

\subsubsection{Motivating the pseudospectrum}
As the previous discussion makes apparent, a crucial question that arises after obtaining the QNM spectrum of the operator
$L$ in Eq. (\ref{e:L_operator}),
with $L_1$ and $L_2$ in (\ref{e:L_1-L_2_PT}) is whether such QNM eigenvalues are stable under small perturbations of $L$.
More specifically for QNM physics, and in the context of the wave equation (\ref{e:wave_equation_tortoise}),
whether the QNM spectrum is stable under small perturbations of the potential $V$.
The latter is the specific type of perturbation we are  assessing in this work.

In the numerical approach we have adopted, perturbations in the spectrum under
small pertubations in $L$ may  arise either from numerical noise resulting from the chosen
discretisation strategy, or they can originate from ``real-world sources'', namely small
physical perturbations of the considered potential $V$. Ultimately,
in the BH setting for which P\"oschl-Teller provides a toy model,
such physical perturbations could stem from a ``dirty" environment surrounding a black hole, and/or
emergent fluctuations from quantum-gravity effects. Therefore, the question of whether QNM spectrum
instability is a structural feature of the operator $L$ --- i.e. not just an artifact of a given
numerical algorithm --- is paramount for our understanding of the fundamental physics underlying the problem.

A pragmatic approach to address this question consists in explicitly introducing families of
perturbations~\footnote{In fact, as far as we are aware of the historical development,
  the path towards the interest in QNM instability 
  followed the opposite way: concerns about BH QNM spectra stability were raised only after
  modifications/approximations of the potential gave rise to unexpected results~\cite{Nollert:1996rf,Nollert:1998ys}
  (Nollert's study being itself motivated by developments in
  QNMs of leaky optical cavities \cite{LeuLiuTon94,LeuLiuYou94,ChiLeuSue95},
  namely the study of QNM completeness).},
and study their effect on the QNM spectra themselves 
\cite{Leung:1999iq,Barausse:2014tra,Cardoso:2019mqo,Hui:2019aox,Daghigh:2020jyk,Maggio:2020jml,Qian:2020cnz}.
We will make contact with this approach later in section \ref{s:PT_perturbed_spectra}, but before
that, we apply the pseudospectrum approach described in section \ref{s:pseudospectrum}  to the
P\"oschl-Teller problem.
Indeed, one of the main goals of our present work is to bring attention to and emphasise the fact that
the {\em unperturbed}
operator already contains crucial information to assess such (in)stability features.
We have already encountered this fact in the evaluation of the condition numbers $\kappa_n$
in Fig.~\ref{fig:Eigenvalues_PT}, that only depends on the unperturbed operator $L$,
but we develop this theme further with the help
of the pseudospectrum notion.
Indeed, pseudospectrum analysis provides a framework to identify the (potential) spectral instability, which is oblivious
to the particular perturbation employed. Then, in a second stage, actual perturbations of the operators with
a particular emphasis
on random perturbations along the lines in section \ref{s:pseudospectrum_random_pert},
can be used to complement and refine such pseudospectrum analysis.

\begin{figure}[t!]
\centering
\includegraphics[width=8.5cm]{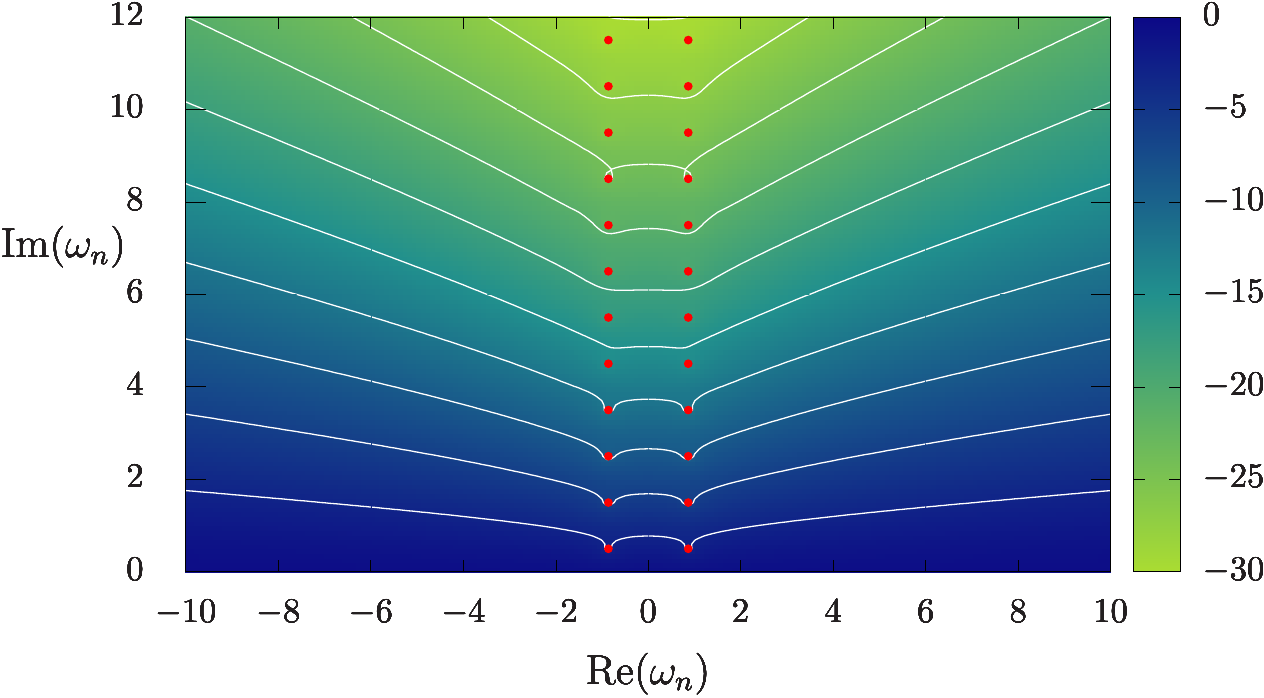}
\includegraphics[width=8.5cm]{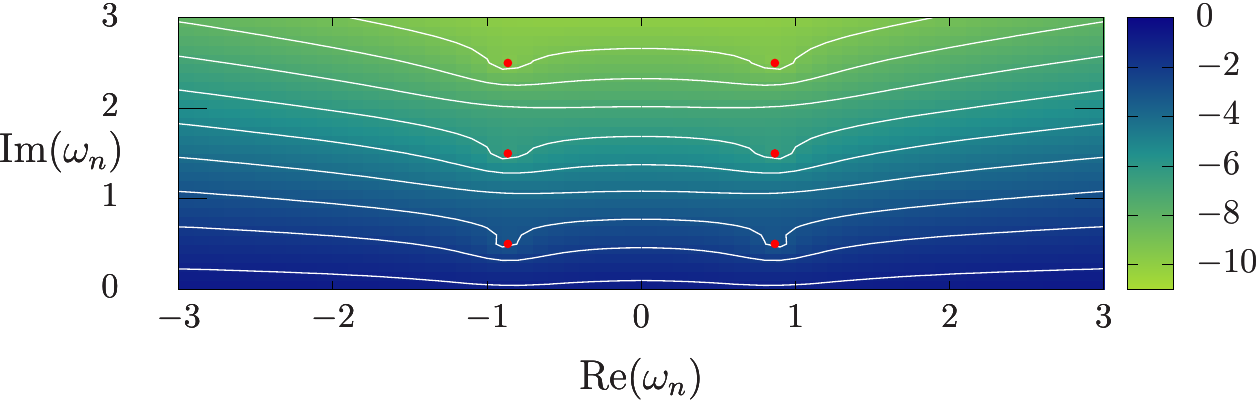}
\caption{ {\em Top panel}: Pseudospectrum for the P\"oschl-Teller potential. QNMs  (red circles)
  from Fig.~\ref{fig:Eigenvalues_PT} are superimposed
  for reference on their (in)stability.
The color log-scale corresponds to $\mathrm{log}_{10}\epsilon$, white lines indicating the
boundaries of $\epsilon$-pseudospectra sets $\sigma^\epsilon$, whose interior extends upwards in
the $\omega$-complex plane.
{\em Bottom panel}: Zoom into the region around the fundamental QNM and first overtones.}
\label{fig:Pseudospectra_PT}
\end{figure}

Fig.~\ref{fig:Pseudospectra_PT} shows the pseudospectrum for the P\"oschl-Teller potential
in the energy norm of Eq. (\ref{e:energy_norm}) associated with
the scalar product (\ref{e:energy_scalar_product_PT}). Let us explain the content of such a figure.
According to the characterization in the Definition 1, namely Eq. (\ref{e:pseudospectrum_def1}),
 of the $\epsilon$-pseudospectrum  of the operator $L$, the set $\sigma^\epsilon(L)$ is the collection of all complex
  numbers $\omega\in\mathbb{C}$ that are actual eigenvalues for some operator $L + \delta L$, where $\delta L$ is a small
  perturbation of ``size'' smaller than a given $\epsilon>0$. Consequently and crucially,
  adding a perturbation $\delta L$ with
  $||\delta L||_{_E}<\epsilon$ entails an actual (``physical'') change in the eigenvalues $\omega_n$
  that can reach up the boundary
  of  the $\sigma^\epsilon(L)$ set, marked in white lines in Fig.~\ref{fig:Pseudospectra_PT}.
  The key question is to assess if $\epsilon$-pseudospectra for small $\epsilon$
  can extend in large areas of $\mathbb{C}$ or not. This is tightly related to
  condition numbers $\kappa_n$ controlling eigenvalue  spectral (in)stabilities,
  as explicitly  estimated by the Bauer-Fike relation (\ref{e:tubular_error})
  between  $\epsilon$-pseudospectra sets and 'tubular 
  neighbourhoods' $\Delta_{\kappa\epsilon}$ of radii $\epsilon \kappa_n$ around the spectrum.
  Let us first discuss a selfadjoint test case and, in a second stage, the actual non-selfadjoint
  case~\footnote{More properly and generally~\cite{trefethen2005spectra}, one should
    distinguish the ``normal'' (indeed selfadjoint in the particular discussion in the present work)
    and the ``non-normal'' operator cases.}.
  
  \subsubsection{Pseudospectrum: selfadjoint case}
\label{s:pseudospectrum_PT_selfadjointcase}
\begin{figure}[h!]
\centering
\includegraphics[width=8.9cm]{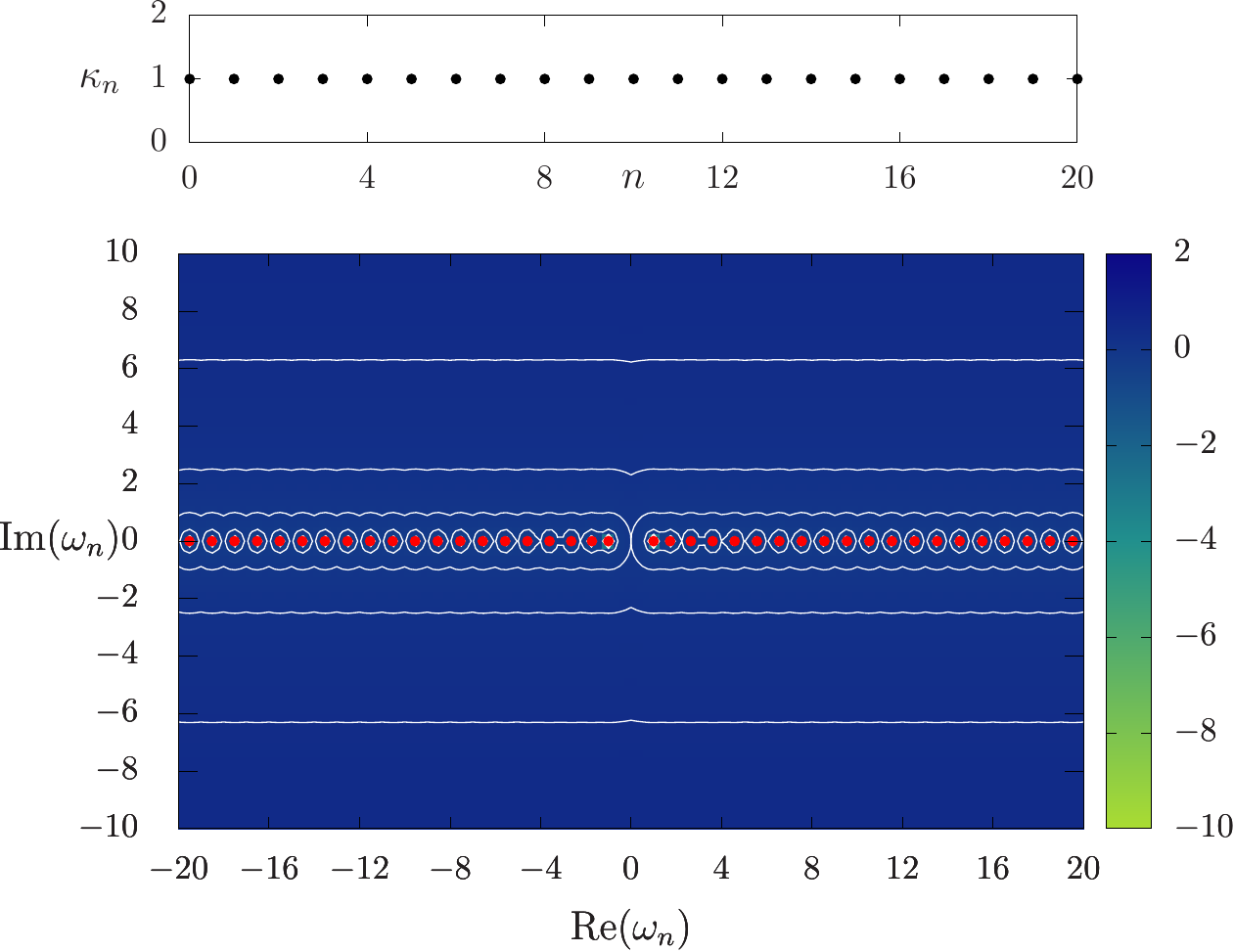}
\includegraphics[width=8.9cm]{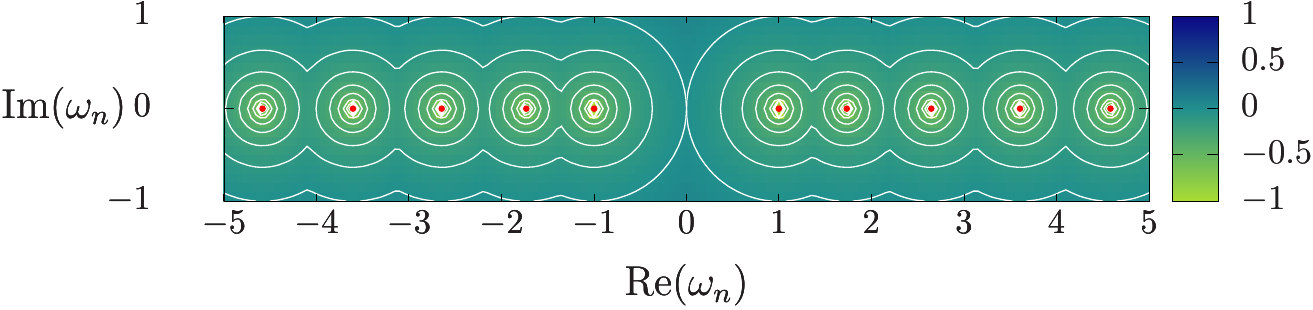}
\caption{Pseudospectrum and eigenvalue condition numbers of a self-adjoint operator (P\"oschl-Teller with $L_2=0$).
 {\em Top panel}: Condition numbers: $\kappa_n=1$ for  $\omega_n$ ($0\leq n\leq 20$).
{\em Middle panel}: Pseudospectrum: ``flat'' pattern typical of a spectrally stable (normal) operator. 
{\em Bottom panel}: Zoom near the spectrum, with concentric circles ("radius $\epsilon$"  tubular
regions around eigenvalues) characteristic of stability.}
\label{fig:Pseudospectra_Sphere}
\end{figure}

As discussed in section  \ref{s:scalar_product}, setting $L_2=0$ in Eq. (\ref{e:L_operator})
---while keeping $L_1$ as in Eq. (\ref{e:L_1-L_2_PT})---
leads to a selfadjoint operator $L$~\footnote{Such an operator is relevant by itself,
  since it corresponds actually to the azymuthal mode $m=0$
  of a wave propagating on a sphere with a constant unit potential, indeed a conservative system.
  The eigenfunctions are nothing more than
  the Legendre polynomials $\phi_n(x) = P_n(x)$, with real eigenvalues $\omega_n^\pm = \pm \sqrt{1+\ell(\ell+1)}$.
This provides a robust test case.}.
  Therefore the associated spectral problem is, cf. section  \ref{s:spectral_instability}, stable. 
    A typical pseudospectrum in the selfadjoint (more generally `normal') case
    is illustrated by Fig.~\ref{fig:Pseudospectra_Sphere}:  a ``flat'' pseudospectrum
    with large values of $\epsilon$  for $\epsilon$-pseudospectra sets,
    when moving ``slightly'' (in the $\mathbb{C}$-plane) from the  eigenvalues. Note also, in this case, the
    horizontal contour lines far from the spectrum, indicating that all eigenvalues share
    the same stability properties in the energy norm.

    Let us describe Fig.~\ref{fig:Pseudospectra_Sphere} in more detail.
    Boundaries of $\epsilon$-pseudospectra $\sigma^\epsilon(A)$ are marked in white lines,
  with $\epsilon$'s corresponding to the values in the color log-scale.
  Pseudospectra $\sigma^\epsilon(L)$ are, by construction, ``nested sets'' around the spectrum
  (red points in Fig.~\ref{fig:Pseudospectra_Sphere}),
  the latter corresponding to the ``innermost set'' $\sigma^\epsilon(L)$ when $\epsilon \to 0$.
  In this selfadjoint case, condition numbers in (\ref{e:eigenvalue_perturbation_kappa})
  must satisfy $\kappa_n=1$, as we have verified  and explicitly shown in the top panel
  of Fig.~\ref{fig:Pseudospectra_Sphere}.
  Then, and consistently  with Eq. (\ref{e:tubular_normal}), the corresponding
  nested sets $\sigma^\epsilon(L)$ are actually tubular regions $\Delta_\epsilon(L)$ of ``radius $\epsilon$''
  around the spectrum,
  so that a change $\delta L$ with a norm of order $\epsilon$ in
  the operator $L$ entails a maximum change in the eigenvalues of the same order $\epsilon$.
  Specifically,  $\epsilon$-pseudospectra sets show concentric circles around the spectra that quickly
  reach large-epsilon values, i.e.
    $\epsilon \sim O(1)$, when moving away from eigenvalues. As a consequence, one would need perturbations in the operator
  of the same order to dislodge the eigenvalues slightly away from their original values:
  we say then that $L$ is spectrally stable.
  Pseudospectra sets with small $\epsilon$ are then ``tightly packed'' in ``thin throats'' around
  the spectrum, so that light green colors are indeed so close to spectrum ``red points'' that they are not visible
  in the scale of
  Fig.~\ref{fig:Pseudospectra_Sphere}, giving rise to a typical ``flat'' pseudospectrum figure of a ``single color''.

  Horizontal boundaries of $\epsilon$-pseudospectra, when far from the spectrum, is a consequence (in this
  particular problem) of the use of the energy norm. If another norm is used, e.g. the standard one induced
  from the $L^2$ norm in $\mathbb{C}^n$, the global ``flatness'' of the pseudospectrum is still
  recovered, especially when comparing with the corresponding scales in Fig.~\ref{fig:Pseudospectra_PT},
  indicating already a much more stable situation than the general $L_2\neq 0$ case.
  But when refining the scale, one would observe that pseudospectra contour lines far from
  the spectrum are not horizontal but present a slope growing with the frequency. This indicates
  that, under perturbations of the same size in that $L^2$ norm, higher frequencies can move further that low frequencies,
  this being in tension with the equal stability of all the eigenvalues. What is going on is the  effect commented
  in section \ref{s:pseudospectrum_choice_norm} concerning the impact of the norm choice on
  the notions of ``big/small'':
  when using the $L^2$ norm, we would be marking with the same ``small'' $\epsilon$ different perturbations
  among which there exist $\delta L$ instances that actually excite strongly the high frequencies, but such a feature
  is blind to the $L^2$ norm. If using however
  a norm sensitive to high-frequency effects, as it is the case of the energy norm that has a $H^1$ character
  incorporating derivative terms, those same perturbations $\delta L$ would have a norm much larger than
  $\epsilon$, the derivative terms in the energy norm indeed weighting more as the frequency grows. What in the
  $L^2$ norm was a small perturbation $\delta L$, turns out to be a big one in the energy one, so stronger
  modifications in the eigenvalues are indeed consistent with stability. 
  In practice, in order to construct a given $\epsilon$-pseudospectrum set, such ``high-energy''
  perturbations $\delta L$ need to be renormalized to keep $\epsilon$ fixed, something that the energy norm
  does automatically. This is a neat example of how the choice of the norm affects the assessment
  of spectral stability and, in particular, of the importance of the energy norm in the present work,
  namely for high-frequency issues.  

  Fig.~\ref{fig:Pseudospectra_Sphere} may appear as a boring figure, but it is actually a tight
  and constraining test of our construction, both at the analytical and numerical level.
  First of all, panels in Fig.~\ref{fig:Pseudospectra_Sphere} correspond to different
  calculations: the top panel results from an eigenvalue calculation (actually two, one for $L$ and
  another for $L^\dagger$), whereas the ``map'' in the middle and bottom panels is
  the result of calculating the energy norm
  of the resolvent $R_L(\omega) = (\omega\mathrm{Id} - L)^{-1}$ at each point $\omega\in\mathbb{C}$.
  Both calculations depend on the construction of the Gram matrix $G^E$, but are indeed different implementations.
  The $\kappa_n=1$ values in the top panel constitute a most stringent test, since modifications in either the
  analytical structure of the scalar product (\ref{e:energy_scalar_product_PT}) or the slightest mistake
  in the discrete counterpart  (\ref{e:ScalarProd_WaveEq_2}) spoil the result.
  As discussed at the end of section \ref{s:PT_condition_numbers}, this provides a strong test both of the 
  analytical treatment and the numerical discretization of the differential operator 
  and scalar product. On the other hand, the plain flatness of the pseudospectrum
  in the middle panel is a strong test of the selfadjoint character of $L$ when $L_2=0$ that,
  given the subtleties of the spectral discretization explained in appendix \ref{a:Chebyshev_elements},
  provides a reassuring non-trivial test to the whole numerical scheme.

  \subsubsection{Non-selfadjoint case: P\"oschl-Teller pseudospectrum}
  In contrast with the selfadjoint case, when considering the actual $L_2\neq 0$
  of the  P\"oschl-Teller case, pseudospectra sets $\sigma^\epsilon(L)$ with small $\epsilon$
  extend in Fig.~\ref{fig:Pseudospectra_PT} into large regions of $\mathbb{C}$
  (with typical sizes much larger than $\epsilon$) and
    therefore the operator $L$ is spectrally unstable:
    very small (physical) perturbations $\delta L$, with $||\delta L||_{_E}<\epsilon$,
    can produce large variations in the eigenvalues
    up to the boundary of the now largely extended region $\sigma^\epsilon(L)$. Such strong variations of the
    spectrum are not a numerical artifact, related e.g. to machine precission,
    but they rather correspond to an actual structural property of the
    non-perturbed operator. Indeed, large values of
    the condition numbers $\kappa_n$ 
    in the top panel of Fig.~\ref{fig:Eigenvalues_PT} entail that  
    the tubular sets $\Delta_{\epsilon\kappa}(L)$ in Eq.~(\ref{e:tubular_epsilon}) extend
    now into large areas in $\mathbb{C}$.
  This fact on $\kappa_n$'s is consistent with the large regions in Fig.~\ref{fig:Pseudospectra_PT}
  corresponding to $\sigma^\epsilon(L)$ sets with very small $\epsilon$'s.
  Such an non-trivial pattern of $\epsilon$-pseudospectra is a strong indication
  of spectral instability, although without a neat identification of the actual nature of
  the perturbations
  triggering instabilities.

\subsubsection{Reading pseudospectra: ``topographic maps'' of the resolvent}
In practice, if one wants to read from pseudospectra ---such as those in Fig.~\ref{fig:Pseudospectra_PT}
or Fig.~\ref{fig:Pseudospectra_Sphere}---
  the possible effect on QNMs of a physical perturbation of (energy) norm of order $\epsilon$, one must first
  determine  the  ``white-line'' corresponding to that $\epsilon$ (using the log-scale). Then, eigenvalues
  can move potentially in all the region bounded by that line
  (namely, the $\epsilon$-pseudospectrum set for the non-perturbed
  operator $L$) that, in Fig.~\ref{fig:Pseudospectra_PT}, corresponds to the region ``above'' white lines.

  Pseudospectra can actually be seen as a  ``map'' of the analytical structure of the resolvent
  $R_L(\omega) = (\omega\mathrm{Id} - L)^{-1}$ of the operator $L$, taken as a function of $\omega$.
  This corresponds to the characterization in Definition 2 of the pseudospectrum,
  Eq. (\ref{e:pseudospectrum_def2}), which it is indeed the one used to effectively construct the
  pseudospectrum (specifically, its realisation (\ref{e:pseudospectrum_carac_E}) in the energy norm;
  cf. section \ref{a:charac_pseudo} for details). 
  In this view, the boundaries of the $\epsilon$-pseudospectra (white lines in Figs.~\ref{fig:Pseudospectra_PT}
  and \ref{fig:Pseudospectra_Sphere}) can be seen as ``contour lines'' of the ``height function'' $||R_L(\omega)||_{_E}$,
  namely the norm of the
  resolvent  $R_L(\omega)$. In quite a literal sense, the pseudospectrum can be read then as a topographic map,
  with stability
  characterised by very steep throats around eigenvalues fastly reaching flat zones away from the spectrum, whereas
  instability corresponds to non-trivial ``topographic patterns'' extending in large regions of the map far
  away from the eigenvalues.

  In sum, this ``topographic perspective'' makes apparent the stark contrast between 
  the flat pattern of the selfadjoint case of Fig.~\ref{fig:Pseudospectra_Sphere}, corresponding
  to stability, and the non-trivial pattern of the (non-selfadjoint) P\"oschl-Teller pseudospectrum
  in Fig.~\ref{fig:Pseudospectra_PT}, in particular indicating a (strong) QNM
  sensitivity to perturbations that increases as damping grows.

\subsection{P\"oschl-Teller perturbed QNM spectra}
\label{s:PT_perturbed_spectra}
Pseudospectra inform about the spectral stability and instability of an operator, but do not identify the
  specific type of perturbation triggering instabilities. Therefore, in a second stage, it is illuminating
  to complement the pseudospectrum information with the exploration of spectral
  instability with ``perturbative probes''
  into the operator, always under the perspective acquired with the pseudospectrum.
  A link between both pseudospectra and perturbation strategies is provided by the
  Bauer-Fike theorem~\cite{trefethen2005spectra}, as expressed in Eq. (\ref{e:tubular_error}).

  \subsubsection{Physical instabilities: perturbations in the potential $V$}
  Not all possible perturbations of the $L$ operator are physically meaningful.
  An instance of this, in the setting of our numerical approach, are machine precision error
  perturbations $\delta L^N$ to the $L^N$ matrix.
  As discussed in section \ref{s:numerical_PT_sectrum}, machine precision errors indeed trigger
  large deviations in the spectrum, consistently with the non-trivial pattern of the
  pseudospectrum in Fig.~\ref{fig:Pseudospectra_PT},
  but clearly we should not consider such effects as physical.
  They are a genuine numerical artifact, since the structure of the perturbation  $\delta L^N$ does not
  correspond to any physical or geometrical element in the problem.

  The methodology we follow to address this issue is: i) given a grid resolution $N$, we first set the
  machine precision to a value sufficiently high so as to guarantee that all non-perturbed
  eigenvalues are correctly recovered, and ii) we then add a prescribed perturbation with the specific structure
  corresponding to the physical aspect we aim at studying.

  In the present work we focus on a particular kind of perturbation, namely perturbations
  to the potential $V$ and, more specifically, perturbations $\delta \tilde{V}$
  to the rescaled potential $\tilde{V}$ in (\ref{e:rescaled_V}).
  This is in the spirit of studying the problem in \cite{Nollert:1996rf}.
  That is, we consider perturbations $\delta L$ to the $L$ operator
  of the form 
  \bea
  \label{e:delta_L_V}
\delta L =\left(
  \begin{array}{c|c}
    0 & 0 \\
    \hline 
   \delta \tilde{V} & 0
  \end{array}
  \right)  \ .
  \eea
  We note that, at the matrix level, the  $\delta \tilde{V}$ submatrix is just a diagonal matrix.
  Therefore, the structure of $\delta L$ in Eq. (\ref{e:delta_L_V}) is a very particular one.
  The pseudospectrum in Fig.~\ref{fig:Pseudospectra_PT} tells us that $L$ is spectrally unstable,
  and we know that machine precision perturbations trigger such instabilities,
  but nothing guarantees that $L$ is actually unstable under a perturbation of the particular
  form (\ref{e:delta_L_V}). It is a remarkable fact, crucial for our physical discussion,  
  that $L$ is indeed unstable under such perturbations
  and, therefore, under perturbations of the potential $V$.

  \subsubsection{Random and high-frequency perturbations in the potential $V$}
\begin{figure*}[h!]
\includegraphics[width=7.cm]{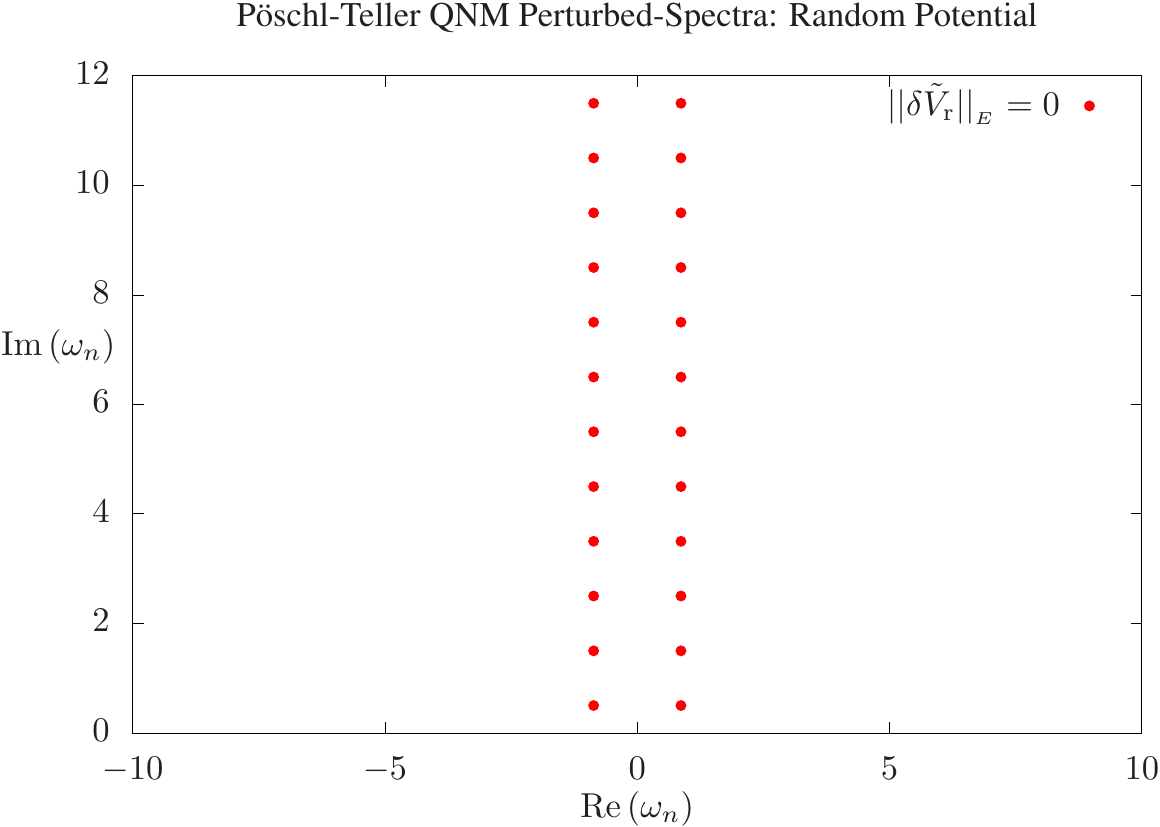}
\includegraphics[width=7.cm]{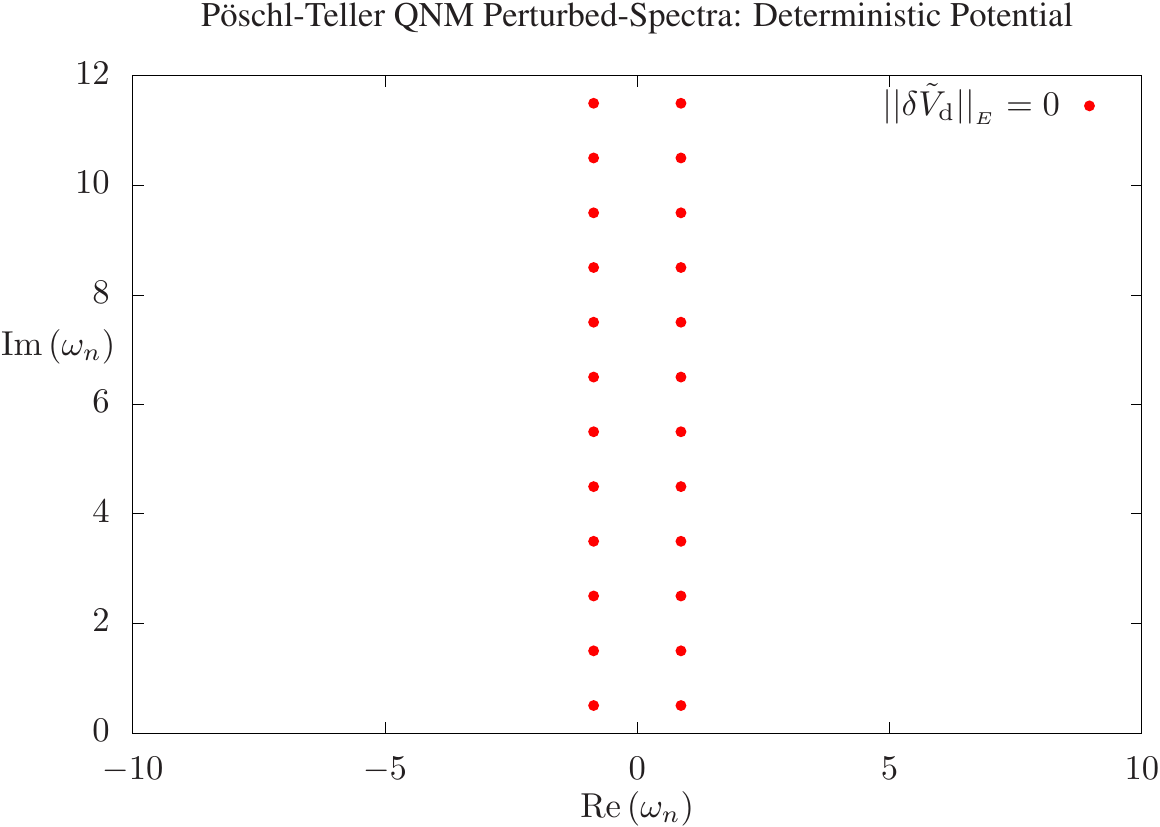} \\
\includegraphics[width=7.cm]{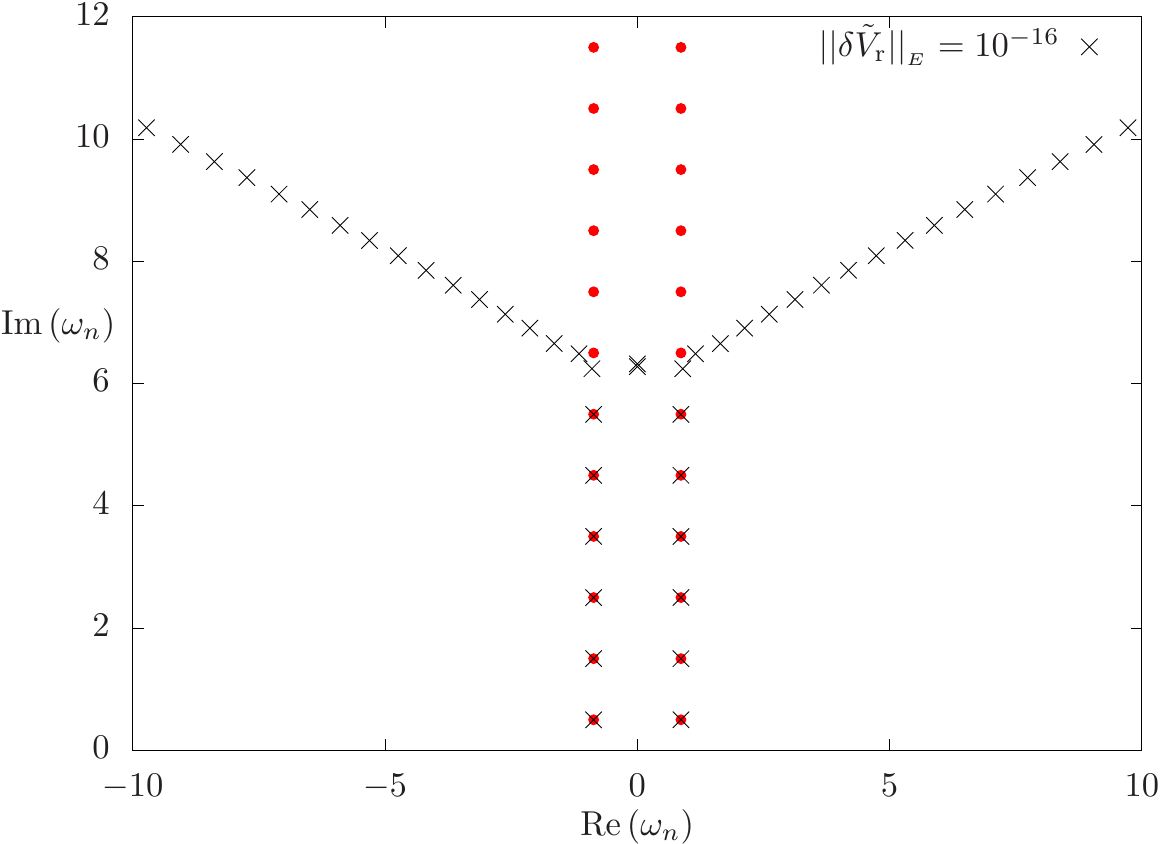}
\includegraphics[width=7.cm]{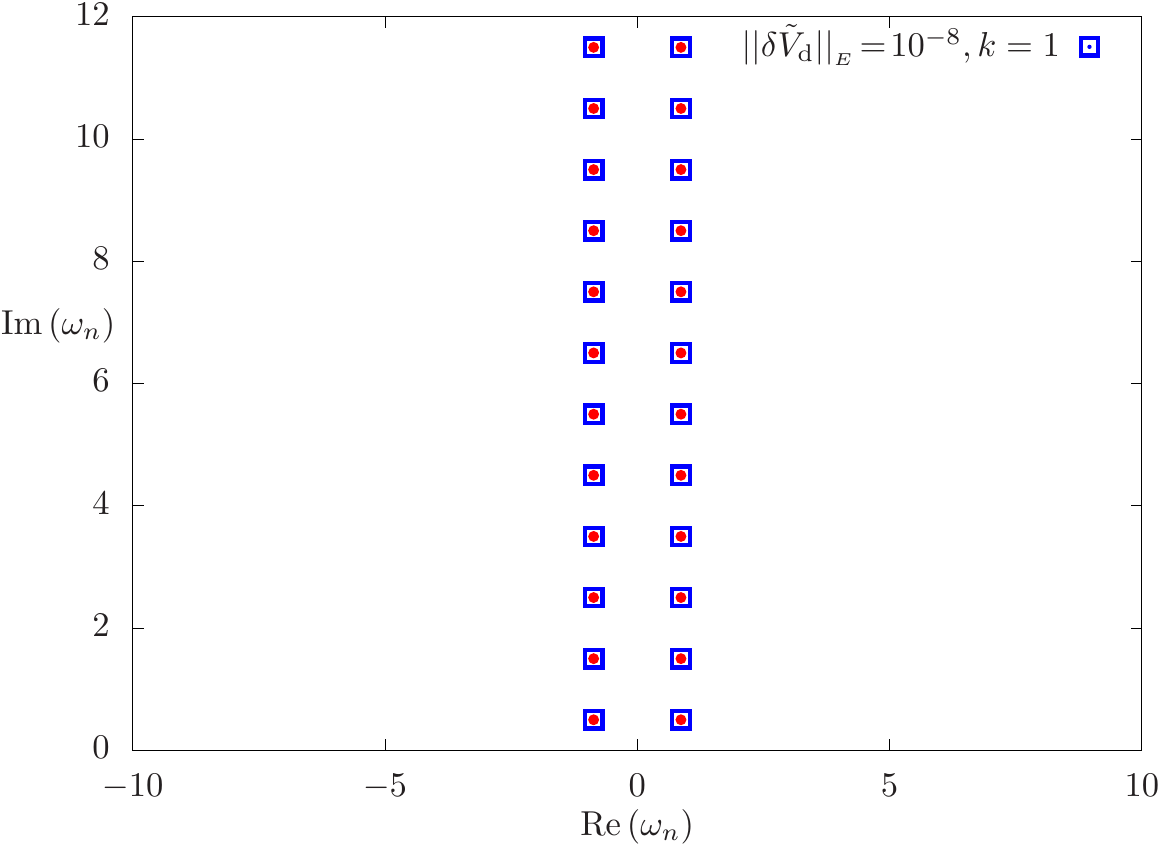} \\
\includegraphics[width=7.cm]{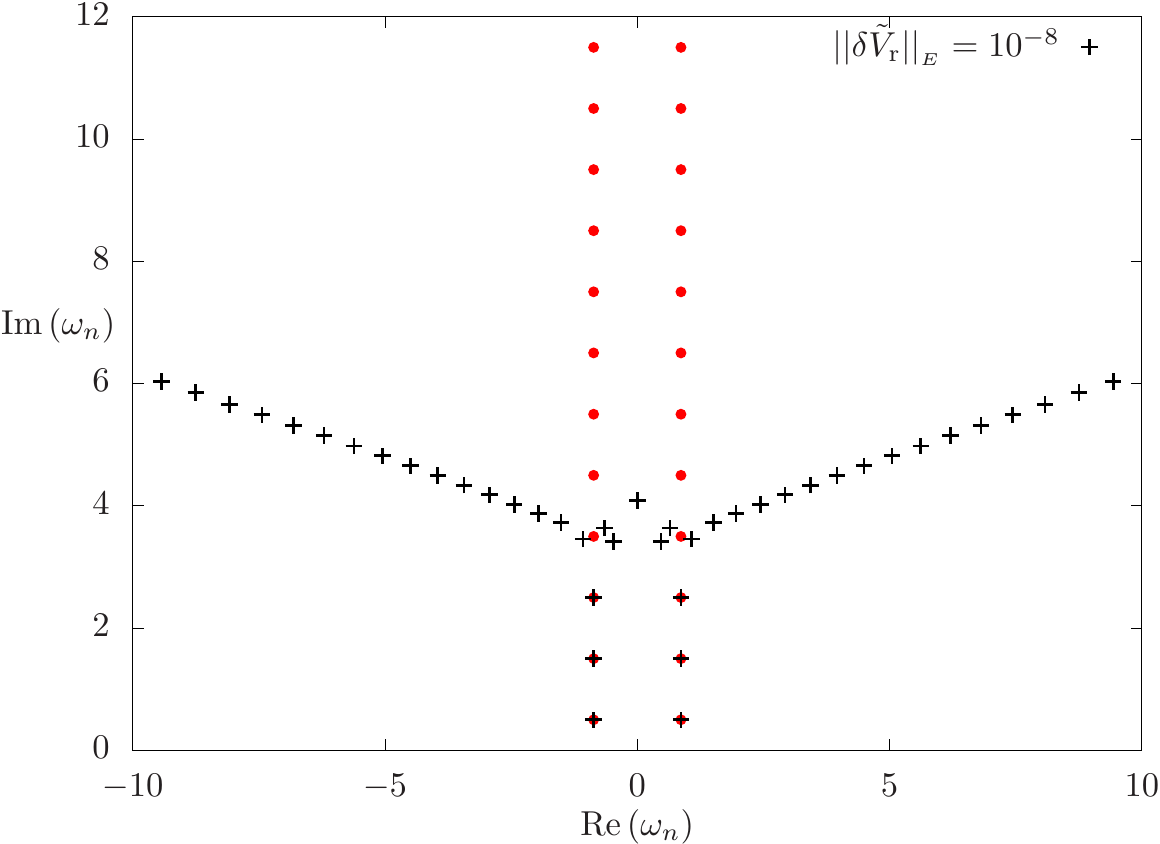}
\includegraphics[width=7.cm]{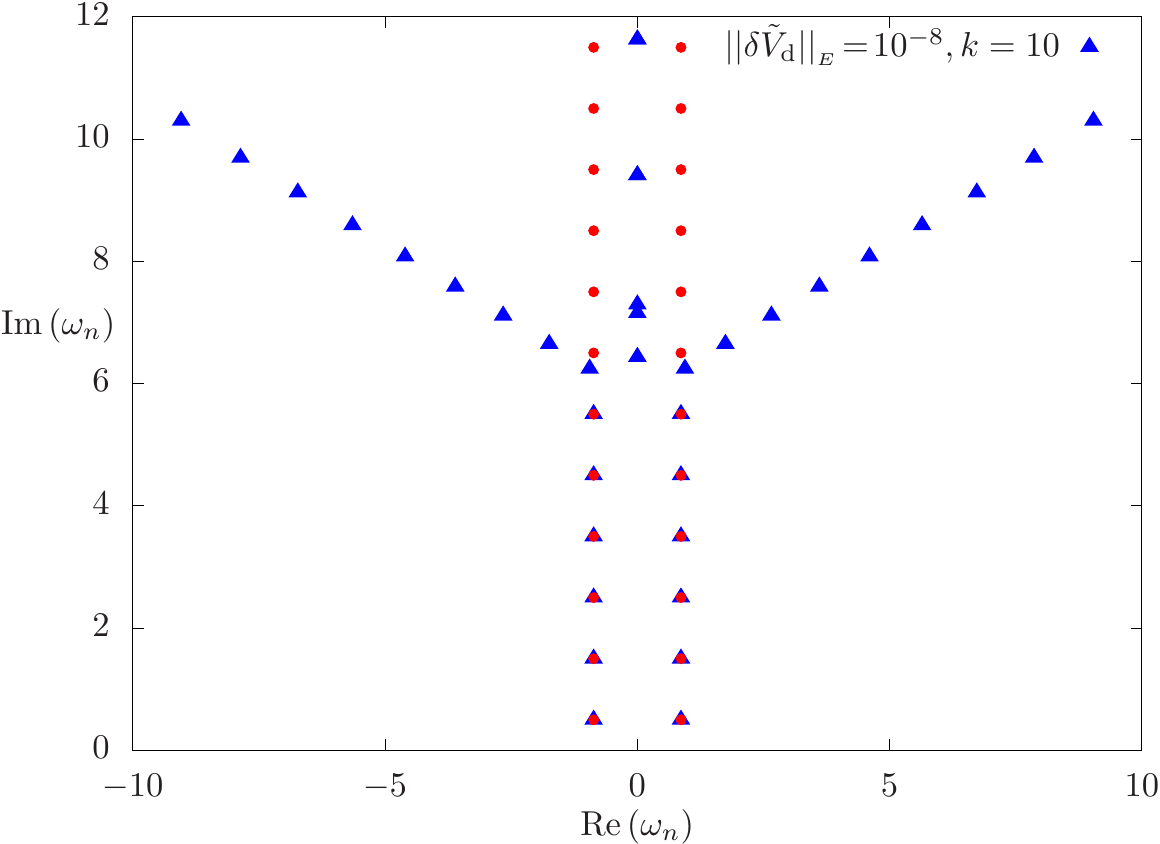} \\
\includegraphics[width=7.cm]{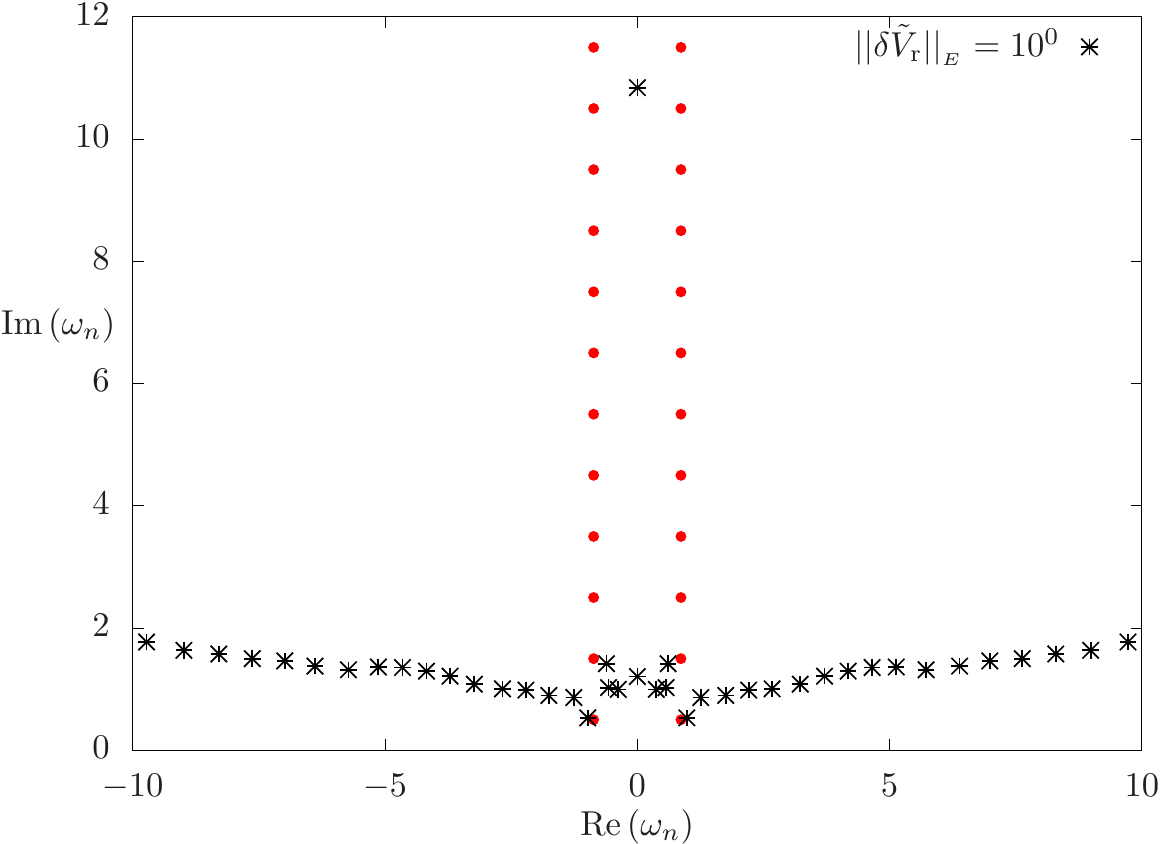}
\includegraphics[width=7.cm]{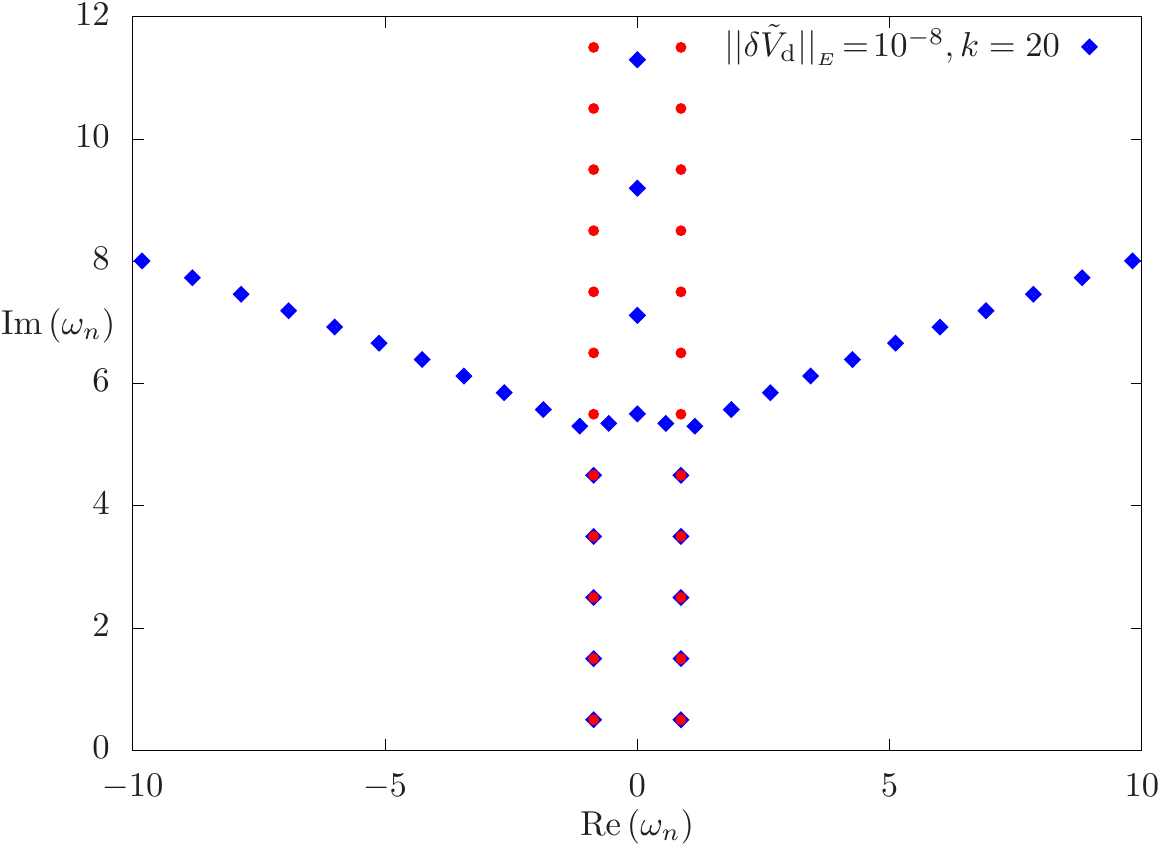}
\caption{{\em Left column}: Sequence of QNM spectra for the P\"oschl-Teller potential subject to a random perturbation
  $\delta \tilde{V}_{\rm r}$ of
  increasing ``size'' (in energy norm). The sequence shows how ``switching on'' a perturbation makes the QNMs
  migrate to a new branch (that actually follows closely a pseudospectrum contour line,
  compare with Fig.~\ref{fig:Pseudospectra_PT}),
  in such a way that the instability
  starts appearing at highly-damped QNMs and descends in the spectrum as the perturbation grows
  (unperturbed values, in red, are kept along the sequence for comparison). 
  The top panel corresponds to the non-perturbed potential shown in Fig.~\ref{fig:Eigenvalues_PT},
  the second panel shows how a random perturbation of
  with (energy) norm $||\delta \tilde{V}_{\rm r}||_{_E}=10^{-16}$ already reaches the 6th QNM overtone,
  whereas in the third panel
  a perturbation with $||\delta \tilde{V}_{\rm r}||_{_E}=10^{-8}$ already reaches the 3rd overtone. This confirms the
  instability already detected in the pseudospectrum, indicating its high-frequency nature.
  Crucially, to reach the  fundamental mode, a perturbation of the same order $O(1)$ as the variation of the
  eigenvalue is required, this demonstrating the stability of the fundamental QNM in agreement
  with the pseudospectrum in Fig.~\ref{fig:Pseudospectra_PT}. {\em Right panel}:  Sequence of QNM spectra for
  P\"oschl-Teller subject to a
  deterministic perturbation $\delta \tilde{V}_{\rm d} \sim \cos(2\pi k\, x)$. The first panel shows again
  the unperturbed potential, whereas the second one shows that a ``low frequency'' ($k=1$) perturbation
  leaves the spectrum unperturbed, in spite of the  $||\delta \tilde{V}_{\rm d}||_{_E}=10^{-8}$ norm (compare with the random
  case with the same norm): this illustrates the harmless character of ``low frequency'' perturbations.
  The third panel shows how keeping the norm of the perturbation but increasing its frequency indeed ``switches on''
  the instability, confirming the ``high frequency'' insight gained from random perturbations. The fourth panel shows how the instability increases with the
  frequency but less efficiently than with random perturbations of the same norm.
}
\label{fig:EVPert_PT}
\end{figure*}

  We have considered two types of generic, but representative, perturbations $\delta L$
  of the form given in Eq. (\ref{e:delta_L_V}): 
\begin{itemize}
\item[i)] {\em Random perturbations} $\delta\tilde{V}_{\rm r}$: we set the perturbation according to a normal Gaussian distribution on the
  collocation points of the grid. This is, by construction, a high-frequency perturbation.
  Random perturbations are a standard tool~\cite{trefethen2005spectra} to explore generic
  properties of spectral instability and there exists indeed a rich interplay between pseudospectra and random perturbations
  \cite{Sjostrand2019}.
\item[ii)] {\em Deterministic perturbations} $\delta\tilde{V}_{\rm d}$: we have chosen
  \bea
  \label{e:pert_det_cos}
   \delta \tilde {V}_{\rm d} \sim \cos(2\pi k\, x) \ ,
   \eea
    in order to address the specific impact
  of high and low frequency perturbations in QNM spectral stability, by exploring the effect of changing the
   wave number $k$.
\end{itemize}
Perturbations $\delta\tilde{V}$ are then rescaled so as to guarantee $||\delta L||_{_E}=\epsilon$.
The impact on QNM frequencies resulting from adding these perturbations is shown in  Fig.~\ref{fig:EVPert_PT}.
In both random and deterministic cases, the sequence of images in Fig.~\ref{fig:EVPert_PT} shows a high-frequency instability
of QNM overtones, that ``migrate'' towards new QNM branches. The fundamental (slowest decaying)
QNM is however stable under these perturbations.
More generally, such QNM instability is sensitive with respect to both perturbations' ``size''and frequency.

\medskip
\begin{figure}[h!]
\centering
\includegraphics[width=9.cm]{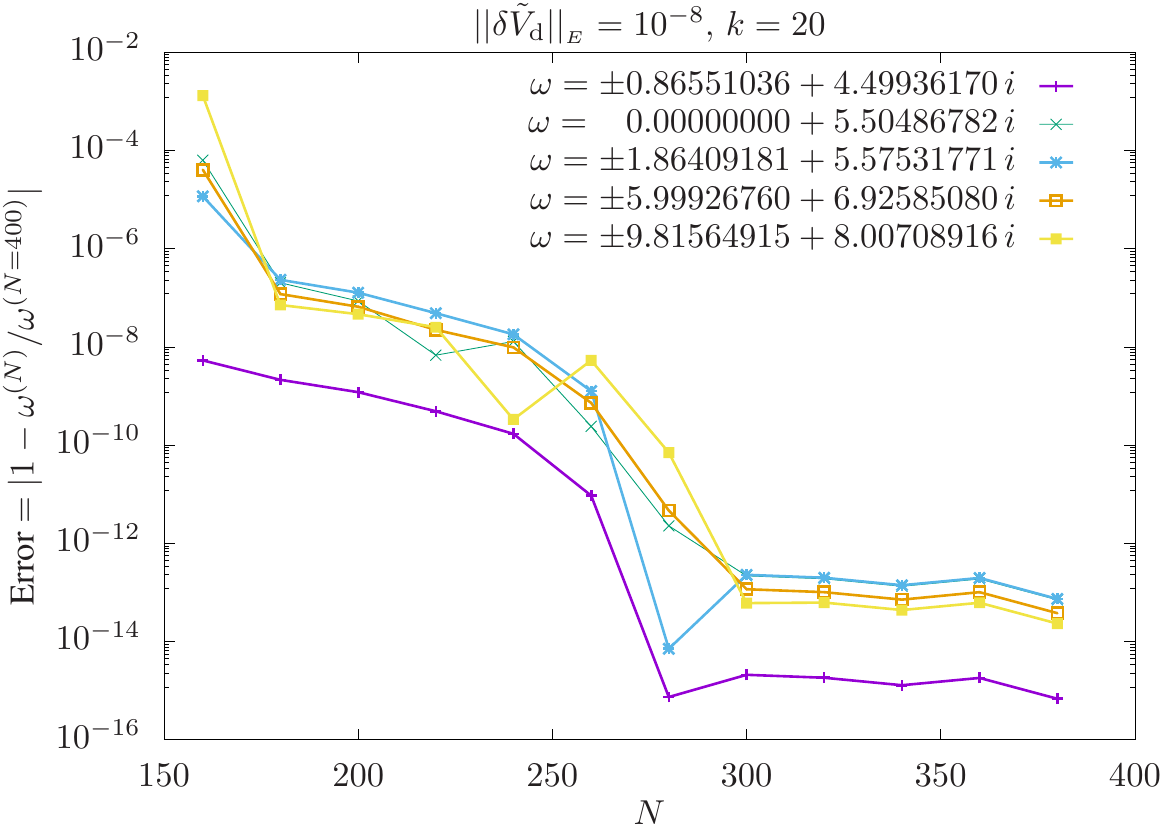}
\caption{Convergence test for five significant QNMs of P\"oschl-Teller perturbed under a deterministic
  high-frequency perturbation $\delta\tilde{V}_{\mathrm{d}}$ (cf. text). This demonstrates that the
  large QNM ``migrations'' observed in Fig.~\ref{fig:EVPert_PT} are not a numerical artifact,
  but actually very small perturbations of the potential
  can result in large variations of the QNM spectrum, consistently with the pseudospectrum in
  Fig.~\ref{fig:Pseudospectra_PT}. 
}
\label{fig:Convergence_PT_Pert}
\end{figure}

Before we further discuss the details of the QNM instability, namely the
nature of the new QNM branches, an important point must be addressed: whether the values obtained correspond to the actual eigenvalues of the new, perturbed operator $L+\delta L$, or whether
they are an artifact of some numerical noise. As in the non-perturbed case discussed
in section \ref{s:numerical_PT_sectrum}, and as explained above when introducing the employed methodology,
results are obtained with a high
internal accuracy (\texttt{$10\times$Machine Precision}), so that any numerical noise is below the range of showed values.
Proceeding systematically, Fig.~\ref{fig:Convergence_PT_Pert} presents the convergence tests for a few eigenvalues
resulting from the deterministic perturbation (random perturbations do not admit this kind
of test) with norm $||\delta\tilde{V}_{\rm d}||_{_E}=10^{-8}$ and frequency $k=20$ (bottom right panel of Fig.~\ref{fig:EVPert_PT}).
The relative error is calculated as
\be
{\cal E}_{n}^{(N)} = \left| 1 - \frac{\omega^{(N)}_n}{\omega^{(N=400)}_n}\right|, 
\ee
i.e., in the absence of exact results, we take as reference the values with a high resolution $N=400$. As representative QNMs, we have chosen:
\begin{itemize}
\item[a)] The last ``unperturbed'' overtone, whose value is actually very close to the (truly) unperturbed QNM $\omega_4$.
\item[b)] The first new QNM on the imaginary axis.
\item[c)] Three QNMs along the new branch with values spread in $1\lesssim {\rm Re}(\omega_n) \lesssim 10$ and $5\lesssim {\rm Im}(\omega_n) \lesssim 8$.
\end{itemize}

One observes a systematic convergence, with the relative error dropping circa $10$ orders of magnitudes when the numerical resolution
increases~\footnote{Compare this decrease of the error as numerical resolution increases (the ``expected'' behaviour)
  with the anomalous growth in  Fig.~\ref{fig:conv_PT}. This reflects that the ``perturbed operator'' has indeed
  improved spectral
stability properties, as compared with the spectrally unstable ``unperturbed'' Poeschl-Teller operator.} from $N=150$ to $N=400$. This result confirms that the spectrum corresponds indeed to the new, perturbed operator, and is not a numerical artifact. This neatly shows the unstable nature of the QNM spectrum of the unperturbed P\"oschl-Teller operator: eigenvalues indeed migrate to new branches under very small perturbations.

\subsubsection{Perturbed QNM branches and pseudospectrum}
\begin{figure}[t!]
\centering
\includegraphics[width=8.5cm]{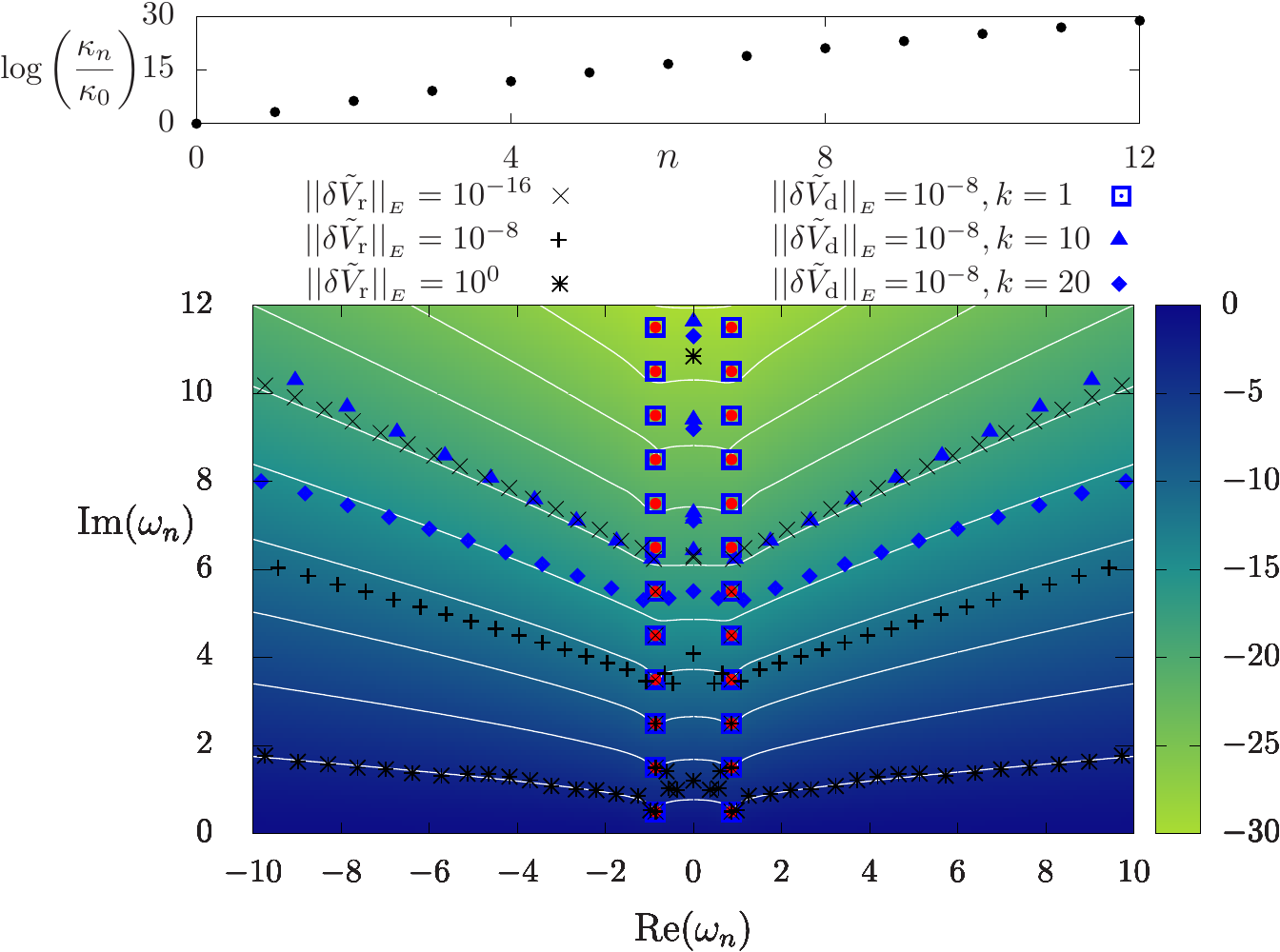}
\caption{QNM spectral instability of  P\"oschl-Teller potential.
  Combination of Figs.~\ref{fig:Eigenvalues_PT},~\ref{fig:Pseudospectra_PT}
  and~\ref{fig:EVPert_PT}, corresponding to three independent calculations,
  respectively: condition numbers  ratios $\kappa_n/\kappa_0$ ({\em top panel}),
  pseudospectrum and perturbed QNM spectra ({\em bottom panel}).
  The bottom pannel demonstrates the high-frequency nature of the spectral instability,
  as well as the migration of P\"oschl-Teller QNMs towards pseudospectrum contour lines
  under high-frequency perturbations.}
\label{fig:Pseudospectrum_perturbations}
\end{figure}

High-frequency perturbations trigger the migration of QNM overtone frequencies to  new perturbed QNM branches.
Fig.~\ref{fig:Pseudospectrum_perturbations} displays
the perturbed QNM spectra on the top of the pseudospectra for the unperturbed operator. The remarkable
``predictive power'' of the pseudospectrum becomes apparent: perturbed QNMs ``follow'' the boundaries of pseudospectrum sets.
That is, QNM overtones ``migrate'' to new branches closely tracking  the $\epsilon$-pseudospectra contour lines.
This happens for both random and deterministic high-frequency perturbations.
Crucially, no such instability is observed for low-frequency deterministic 
perturbations, with small wave number $k$.
Consequently, we shall refer in the following to this effect as an ultraviolet instability of QNM overtones.

Remarkably, such high-frequency QNM instability 
is not limited to highly damped QNMs but indeed reaches the lowest overtones,
the random perturbations being more effective in reaching the slowest decaying 
overtones for a given norm $||\delta{V}||_{_E}=\epsilon$.
This result is qualitatively consistent with  analyses in  \cite{Nollert:1998ys,Hui:2019aox} for
Dirac-delta potentials (compare e.g.,~perturbed QNM branches in Fig.~\ref{fig:Pseudospectrum_perturbations}
here 
with Fig.~1 in Ref. \cite{Hui:2019aox}). 
These findings advocate the use of pseudospectra to probe QNM instability,
demonstrating its capability to capture it already at the level of the non-perturbed
operator. At the same time, pseudospectra are oblivious to the nature of the
perturbation triggering instabilities. A complementary perturbation analysis, in particular
through random perturbations, has been then necessary to identify the high-frequency
nature of the instability, confirming its physicality in the sense of being associated with
actual perturbations of the potential $V$.

\subsubsection{High-frequency stability of the slowest decaying QNM}
\label{s:Slowest_QNM_PT}
The high-frequency instability observed for QNM overtones is absent in the fundamental QNM.
The slowest decaying QNM is therefore ultraviolet stable.
Such stability is already apparent in the pseudospectrum in Fig.~\ref{fig:Pseudospectra_PT},
where the order of the $\epsilon$'s corresponding to $\epsilon$-pseudospectra sets around the fundamental QNM
reaches the values in the stable self-adjoint case in Fig.~\ref{fig:Pseudospectra_Sphere}.
This high-frequency stability is then confirmed in the perturbation analysis.
Indeed,  Fig.~\ref{fig:Pseudospectrum_perturbations} demonstrates
the need of large perturbations in the operator in order to reach the
fundamental QNM, namely (random) perturbations
with a `size' $||\delta\tilde{V}||_{_E}$ of the same order
as the induced variation in $\omega^\pm_0$. This behaviour is a tantamount of spectral stability.

\begin{figure}[t!]
\centering
\includegraphics[width=8.5cm]{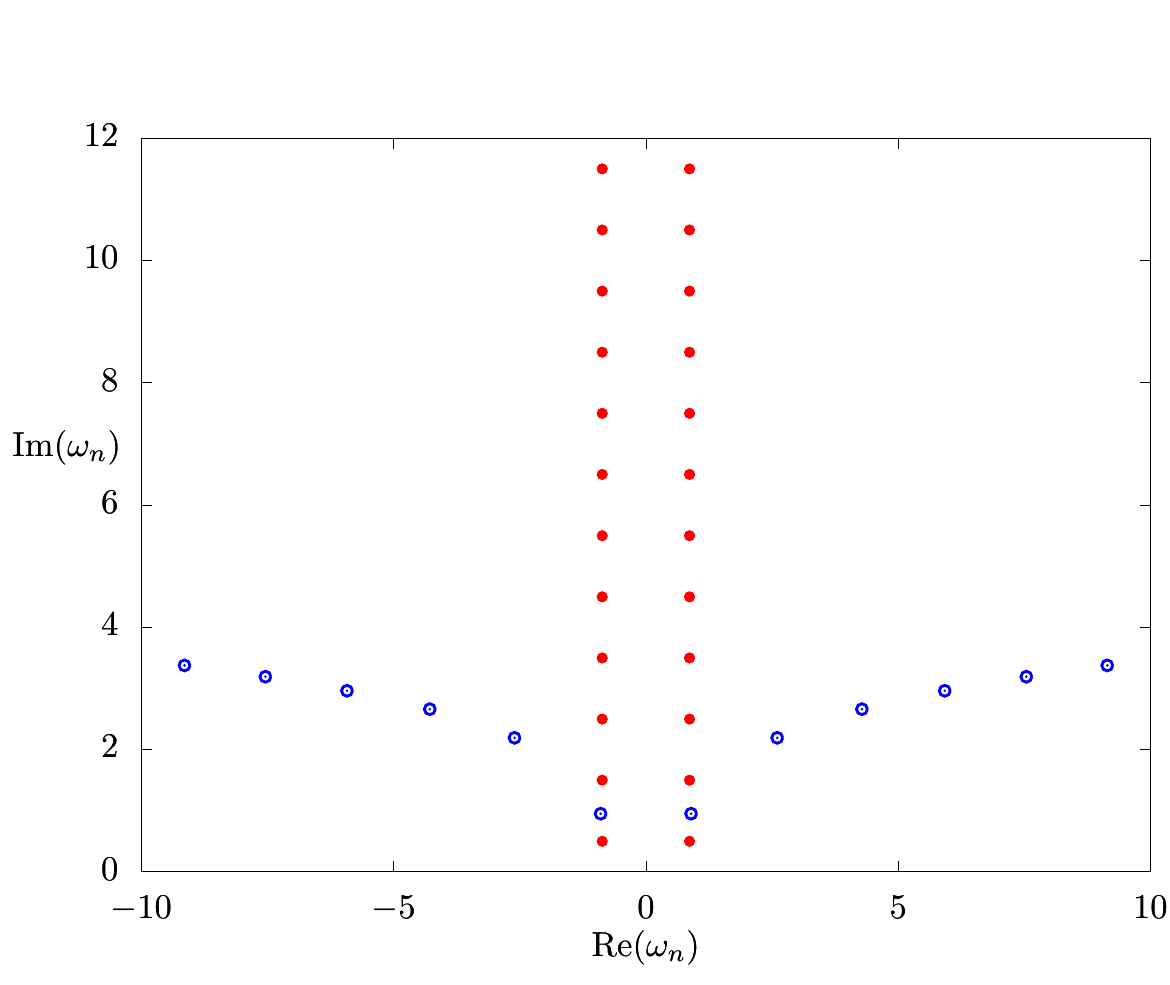}
\caption{QNMs of P\"oschl-Teller ``cut'' potential.  Setting P\"oschl-Teller potential to
  zero outside an interval $[x_{\mathrm{min}},x_{\mathrm{max}}]$ introduces
  high-frequency perturbations that make QNM overtones migrate towards pseudospectrum
  contour lines, as well as an ``infrared'' modification that alters the fundamental QNM frequency.
  Whereas the latter tends to the non-perturbed P\"oschl-Teller value as
  $x_{\mathrm{min}}\to -\infty$ and $x_{\mathrm{max}}\to \infty$, QNM
  overtones remain always strongly perturbed.}
\label{fig:Infrared_effect_PT}
\end{figure}

The contrast between the high stability of $\omega^\pm_0$ and the instability of overtone resonances
$\omega^\pm_{n\geq 1}$
has already been evoked in \ref{s:numerical_PT_sectrum}, when referring to the large condition number
ratios $\kappa_n/\kappa_0$, in particular referring to 
Bindel \& Zworski's discussion in \cite{BinZwo,Zwors87}. This high-frequency stability of the fundamental
mode is in tension with the instability found by Nollert in \cite{Nollert:1996rf}
for the slowest decaying mode for Schwarzschild. We will revisit this point
in section \ref{s:Schwarzschild_infrared}. For the time being, we simply emphasize that the observed stability
relies critically on the faithful treatment of the asymptotic structure of the
potential, that is in-built in the adopted hyperboloidal approach 
permitting to capture the long-range structure of the potential up to null infinity $\scri^+$.
It is only when we enforce a modification of the potential at ``large distances'' that the ``low frequency''
fundamental QNM is affected. This is illustrated in  Fig. \ref{fig:Infrared_effect_PT} (see
also \cite{Qian:2020cnz,SheJar20}),
corresponding to a P\"oschl-Teller potential set to zero beyond a compact interval
$[x_{\mathrm{min}},x_{\mathrm{max}}]$: such ``cut'' introduces high frequencies that make migrate the overtones
to the new branches and, crucially, alters the asymptotic structure so that
the fundamental QNM is also modified. Such ``infrared'' effect  is however compatible with the spectral
stability of the fundamental QNM, since such ``cut'' of the potential does not
correspond to a small perturbation in $\delta L$.

\subsubsection{Regularization effect of random perturbations}
\begin{figure}[t!]
\centering
\includegraphics[width=8.5cm]{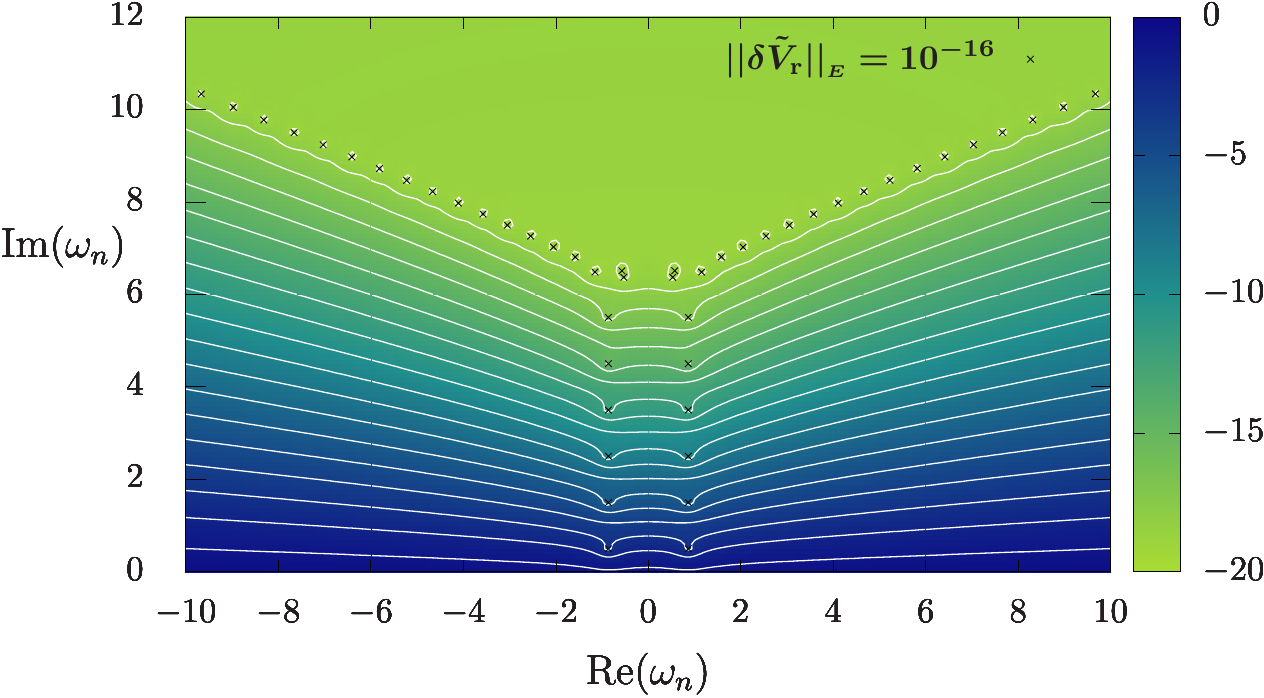}
\includegraphics[width=8.5cm]{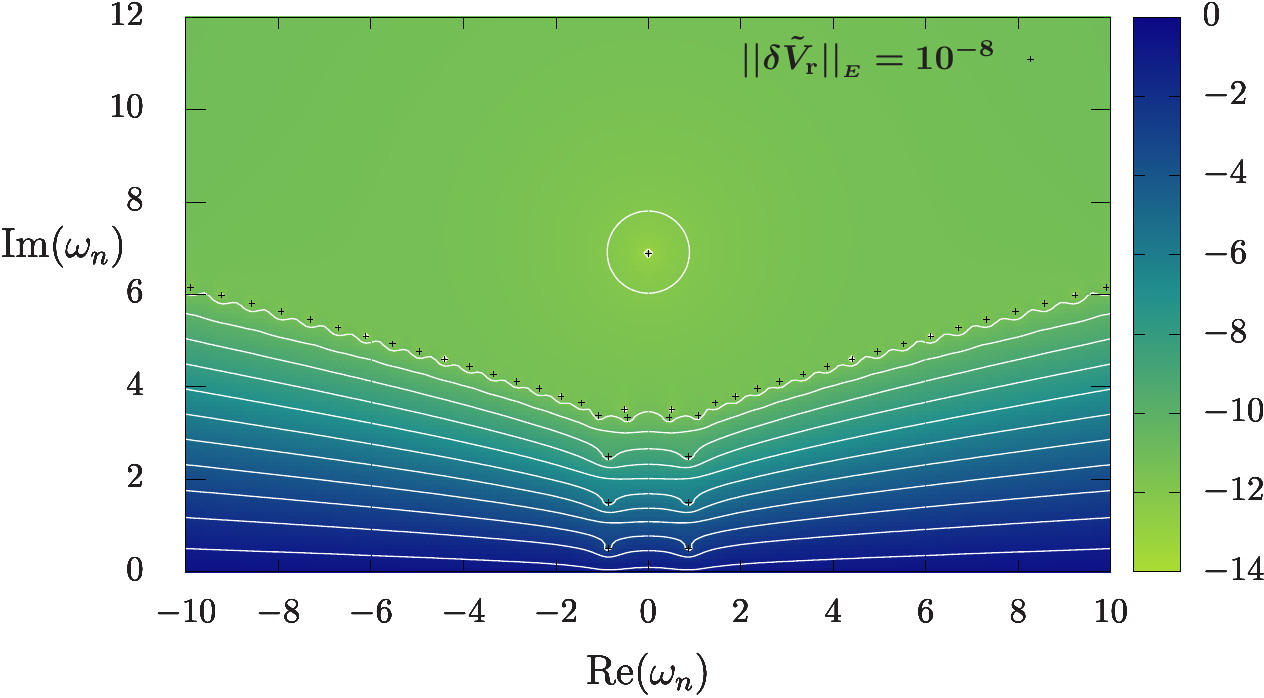}
\includegraphics[width=8.5cm]{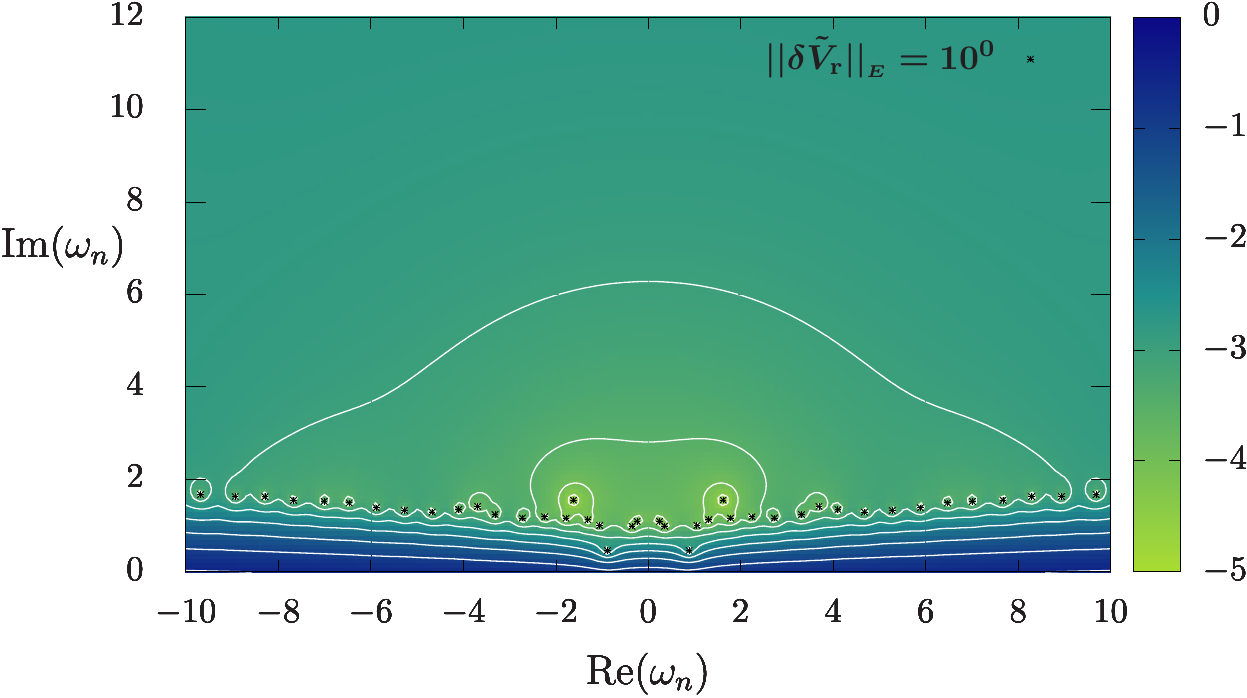}
\caption{Pseudospectra of P\"oschl-Teller under random perturbations $\delta \tilde{V}_{\mathrm{r}}$
  of increasing norm, demonstrating the ``regularizing'' effect of random perturbations:
  pseudospectra sets $\sigma^\epsilon$ bounded by that ``contour line''
  reached by perturbed QNMs become ``flat'', a signature of improved analytic behaviour of
  the resolvent, as illustrated in Fig.~\ref{fig:Pseudospectra_Sphere}. 
  Pseudospectra sets not attained by the perturbation remain unchanged.
Regularization
  of $R_{L+\delta L}(\omega)$ increases as $||\delta \tilde{V}_{\mathrm{r}}||_{_{E}}$ grows.}
\label{fig:Random_regularization}
\end{figure}

Before proceeding to discuss the BH case, let us briefly comment on an
apparently paradoxical phenomenon resulting from the interplay between
random perturbations and the pseudospectrum. In contrast with what
one might expect, the addition of a random perturbation
to a spectrally unstable operator $L$ does not worsen
the regularity properties of $L$ but, on the contrary, it improves
the analytical behaviour of its resolvent $R_L(\omega)$
\cite{hager05,Hager06a,Hager06b,HagSjo06,Borde08,BorSjo10,Borde11,Borde13,Vogel16,NonVog18,Sjostrand2019}.
This is illustrated in Fig.~\ref{fig:Random_regularization},
that shows a series of pseudospectra corresponding to random
perturbations of the P\"oschl-Teller potential with increasing  $||\delta\tilde{V}_{\rm r}||_{_E}$.
In addition to the commented migration of QNM overtones towards
pseudospectra contour lines, we observe two phenomena: i) $\epsilon$-pseudospectra
sets with $\epsilon>||\delta\tilde{V}_{\rm r}||_{_E}$ are not affected by the perturbation, whereas
ii) the pseudospectrum structure for $\epsilon<||\delta\tilde{V}_{\rm r}||_{_E}$ is smoothed into a ``flat pattern''.
As we have discussed in Fig.~\ref{fig:Pseudospectra_Sphere}, such flat pseudospectra
patterns are the signature of spectral stability, a tantamount of
regularity of the resolvent $R_L(\omega)$. The resulting improvement in the
spectral stability of $L+\delta L$, as compared to $L$, is indeed consistent
with the convergence properties of the respective QNM spectra,
as illutrated by the contrast between the corresponding convergence tests in Figs.~\ref{fig:Convergence_PT_Pert}
and~\ref{fig:conv_PT}. In sum, random perturbations improve regularity,
an intriguing effect seemingly related  intimately to a Weyl law occurring in the large-$n$ asymptotics
of QNMs \cite{Zwors87,Sjost14}, with suggestive physical implications in the QNM setting, e.g. in (semi)classical
limits to smooth spacetimes from (random) structures at Planck scales.

\section{Schwarzschild QNM (in)stability }
\label{s:QNM_Schwarzschild}
We address now the physical BH case, namely the stability of QNMs in
Schwarzschild spacetime. Whereas the previous section has been 
devoted, to a large extent, to discuss some of the technical issues in
QNM stability, the spirit in this section is to focus more
on the physical implications, in particular in the perspective
of assessing the pioneering work in \cite{Nollert:1996rf,Nollert:1998ys}.

\subsection{Hyperboloidal approach in Schwarzschild}
\label{s:hyper_Schw}
The attempt to implement the QNM stability analysis in the coordinate
system employed for P\"oschl-Teller, namely the Bizo\'n-Mach chart (\ref{e:change_variables_tanh}),
is unsuccessful. The reason is the bad analytic behaviour at null infinity of Schwarzschild
potential(s) in the corresponding coordinate $x$. Instead of this, we resort to
the `minimal gauge' slicing~\cite{Ansorg:2016ztf,PanossoMacedo:2018hab,PanossoMacedo:2018gvw},
devised to improve regularity in the Schwarzschild(-like) case.

We start by considering standard Schwarzschild $(t,r)$ coordinates in the line element
(\ref{e:spherically_symmetric_metric}),
with $f(r) = \left(1-2M/r\right)$ and BH horizon at $r=2M$. ``Axial'' and ``polar'' Schwarzschild gravitational
parities are described by the wave equation (\ref{e:wave_equation_tortoise}) with, respectively, Regger-Wheeler
$V^{\mathrm{RW},s}_{\ell}(r)$ and
Zerilli $V^{\mathrm{Z}}_{\ell}(r)$ potentials \cite{Regge57,Zeril70,Chandrasekhar:579245,Kokkotas:1999bd,maggiore2018gravitational}.
Specifically, we have
\bea
\label{e:Schwarzschild_potential_RG}
V^{\mathrm{RW},s}_{\ell}(r) =  \left(1-\frac{2M}{r}\right)  \left(\frac{\ell(\ell + 1)}{r^2} +(1-s^2) \frac{2M}{r^3} \right)\ , 
\eea
for the axial case, where $s=0,1,2$ correspond to the scalar, electromagnetic
and (linearized) gravitational cases, and
\bea
\label{e:Schwarzschild_potential_Z}
&&V^{\mathrm{Z}}_{\ell}(r) =  \left(1-\frac{2M}{r}\right) \nn \\
&&\left(\frac{2n^2(n+1)r^3 + 6n^2Mr^2+18nM^2r+18M^3}{r^3(nr+3M)^2} \right) \ , 
\eea
with
\bea
n = \dfrac{(\ell-1)(\ell+2)}{2} \ .
\eea
for the polar case.

To construct
horizon-penetrating coordinates reaching null infinity, one defines a height function $h$
in (\ref{e:change_variables}) by first considering an advanced time coordinate built
on the rescaled tortoise coordinate $\bar{x}=r^*/\lambda$, with $r_* = r + 2M\ln(r/2M-1)$,
so that the BH horizon is at $\bar{x}\to -\infty$,
and then enforcing a deformation of the Cauchy slicing into a hyperboloidal one
through the choice of a 'minimal gauge', prescribed under the guideline
of preserving a good analytic behavior at $\scri^+$. In a second stage, the function $g$
in (\ref{e:change_variables}) implementing the compactification
along hyperboloidal slices is implicitly determined by
(note that instead of $x$ in (\ref{e:change_variables}), we rather use $\sigma$ for the spatial coordinate,
so as to keep the standard usage in \cite{Ansorg:2016ztf,PanossoMacedo:2018hab,PanossoMacedo:2018gvw})
\bea
\label{e:r_sigma}
r = \frac{2M}{\sigma} \ .
\eea
Choosing $\lambda = 4M$ in the rescaling $\bar{x}=r^*/\lambda$
of Eq. (\ref{e:dimensional_rescaling}), the steps above
result (see details in  \cite{Ansorg:2016ztf,PanossoMacedo:2018hab,PanossoMacedo:2018gvw}) in
the 'minimal gauge' hyperboloidal coordinates for the transformation (\ref{e:change_variables})
\bea
\label{e:minimal_gauge}
\left\{
\begin{array}{rcl}
  \displaystyle
\bar{t} &=& \tau - \frac{1}{2}\left(\ln\sigma + \ln(1-\sigma) -\frac{1}{\sigma}\right) \\ 
\bar{x} &=& \frac{1}{2}\left(\frac{1}{\sigma}+\ln(1-\sigma)-\ln\sigma \right)
\end{array}
\right. \ ,
\eea
that, upon addition of the BH horizon and $\scri^+$ points,  maps $\bar{x}\in[-\infty,\infty]$ to the
compact interval $\sigma\in[a,b]=[0,1]$, with the BH horizon at $\sigma=1$ and future null infinity at $\sigma=0$.

Implementing transformation (\ref{e:minimal_gauge}) in 
 the first-order reduction in time in Eqs. (\ref{e:psi_u_def})-(\ref{e:wave_eq_1storder}),
 we get for $w(\sigma)$, $p(\sigma)$, $q_{\ell}(\sigma)$ (now explicitly depending on $\ell$)
 and $\gamma(\sigma)$ in Eq.~(\ref{e:functions_L1_L2}) 
\bea
\label{e:functions_L1_L2_Schw}
\begin{array}{rllcrcl}
w(\sigma)&=&2 \left( 1+\sigma \right) \ &,& \ p(\sigma)&=&2 \sigma^2(1-\sigma) \ ,  \\
\displaystyle q_{\ell}(\sigma)&=&\displaystyle \frac{ (4M)^2V_{\ell}}{2\sigma^2(1-\sigma)}\ &,& \   \gamma(\sigma)&=& 1 -2\sigma^2 \ , 
\end{array}
\eea
leading to the $L_1$ and $L_2$ operators building $L$ in Eq.~(\ref{e:L_operator})
\bea
\label{e:L_1_L_2_Schwarzschild}
L_1 &=& \frac{1}{2(1+\sigma)}\left[\partial_\sigma\left(2\sigma^2(1-\sigma)\partial_\sigma\right)
 - \tilde{V}_{\ell} \right] \nn \\
L_2 &=& \frac{1}{2(1+\sigma)}\left(2(1-2\sigma^2)\partial_\sigma -4\sigma\right) \ , 
\eea 
where the rescaled potential $\tilde{V}_{\ell}(\sigma):=q_{\ell}(\sigma)$ results, in the respective axial and polar cases,
in the explicit expressions
\bea
\label{e:Vtilde_Schw}
\tilde{V}_\ell^{\mathrm{RW},s} &=& 2\Big (\ell(\ell + 1) + (1-s^2)\sigma \Big) \nn \\
\tilde{V}_\ell^{\mathrm{Z}} &=& 2\Bigg(\sigma + \dfrac{2n}{3}\left(1 + 4n\dfrac{3+2n}{(2n+3\sigma)^2}\right)\Bigg) \ .
\eea
Finally, from Eqs. (\ref{e:functions_L1_L2_Schw}) and (\ref{e:energy_scalar_product}),
the energy scalar product is
\bea
\label{e:energy_scalar_product_Schw}
&&\langle u_1,\! u_2\rangle_{_{E}} = \Big\langle\begin{pmatrix}
  \phi_1 \\
  \psi_1
\end{pmatrix}, \begin{pmatrix}
  \phi_2 \\
  \psi_2
\end{pmatrix}\Big\rangle_{_{E}}  \\
&&= 
\int_{0}^1 \!\!\!\left(\! (1+\sigma) \bar{\psi}_1 \psi_2 + \sigma^2(1-\sigma) \partial_x\bar{\phi}_1\!\partial_x\phi_2 +\frac{\tilde{V}_\ell}{2}\bar{\phi}_1 \phi_2 \!\right)\!\! d\sigma \ , \nn
\eea
where the weight $\tilde{V}_\ell$ is fixed by Eq. (\ref{e:Vtilde_Schw}) for each polarization.

\subsection{Schwarzschild QNM spectrum}
\label{s:QNM_spectrum_Schw}
As discussed in section \ref{s:regularity_BC}, outgoing boundary conditions
have been translated into regularity conditions on eigenfunctions.
Specifically, as we have seen in the P\"oschl-Teller case,
the operator $L_1$ in (\ref{e:L_1_L_2_Schwarzschild}) is a singular Sturm-Lioville operator,
namely the function $p(\sigma)= \sigma^2(1-\sigma)$ vanishes 
at the boundaries of the interval $[a,b]=[0,1]$ consistently with  Eq.  (\ref{e:vanishing_p_borders}).
This translates into the fact that no boundary conditions can be imposed if  enough regularity is required.

But there is a key difference between the P\"oschl-Teller and the BH case:
whereas in  P\"oschl-Teller the function $p(x)=(1-x)(1+x)$ vanishes
linearly at the boundaries, and therefore $x=\pm 1$ are regular singular points, in
Schwarzschild this is true for $\sigma=1$ (BH horizon) but not
for $\sigma=0$ ($\scri^+$), due to the quadratic $\sigma^2$ term.
Null infinity is then an irregular singular point. This is the counterpart,
in our compactified hyperboloidal formulation,
of the power-law decay of Schwarzschild potentials responsible for the
branch cut in the Green function of Eq.~(\ref{e:wave_equation_tortoise}),
with its associated ``tails'' in late decays of scattered
fields. In the context of our spectral problem for the operator $L$,
this translates into the appearance of a (``branch cut'') continuous part in the spectrum.
This has an important impact on the numerical approach, since the
continuous branch cut is realized in terms of actual eigenvalues
of the discretised approximates $L^N$. 
Such eigenvalues are not QNMs and can indeed be unambiguously identified,
but their presence has to be taken into account when performing the
spectral stability analysis, that becomes a more delicate problem than in P\"oschl-Teller.
In this context, the latter becomes a crucial benchmark to guide the analysis in the BH case.

\begin{figure}[t!]
\centering
\includegraphics[width=8.5cm]{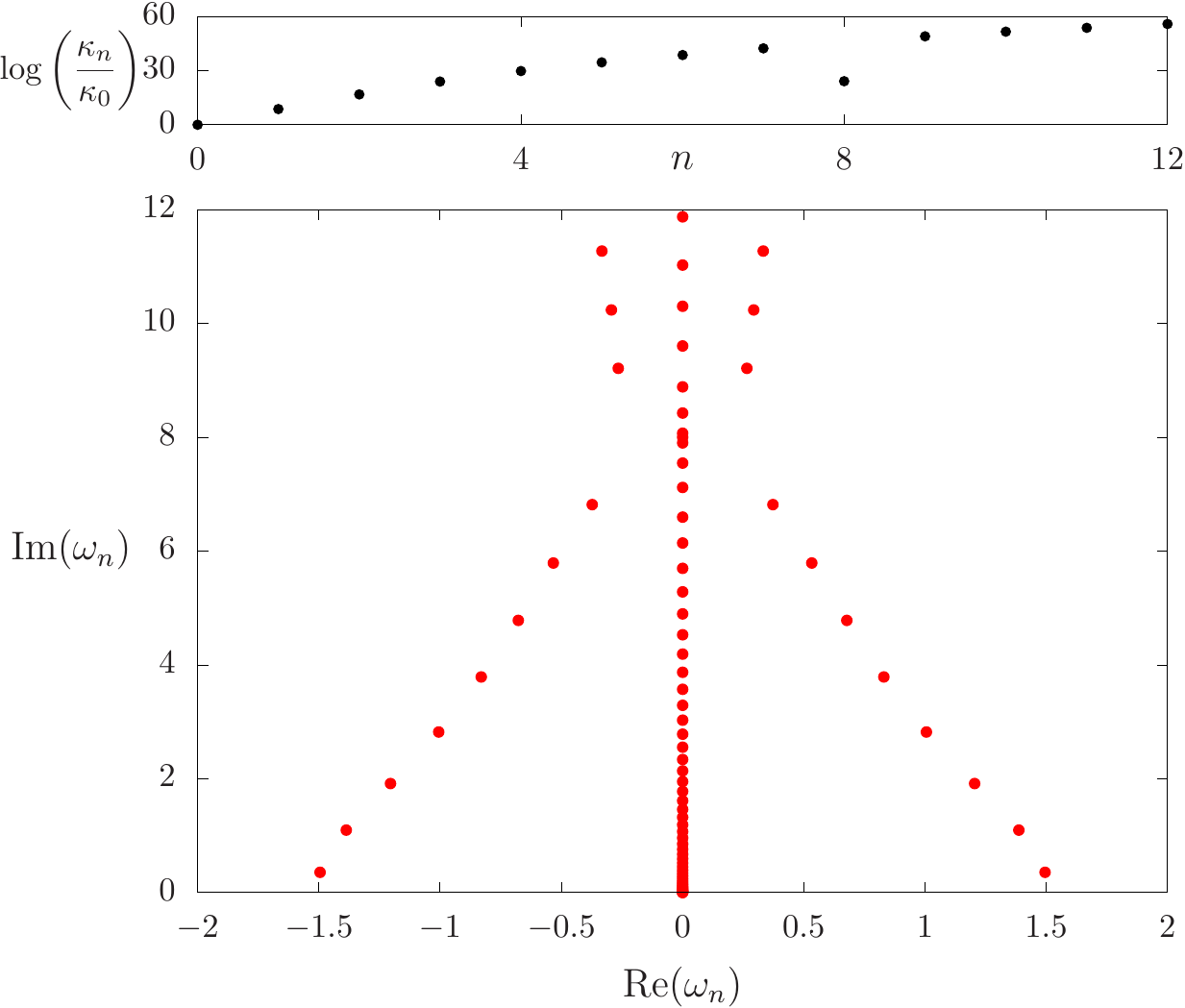}
\caption{Schwarzschild QNM problem. {\em Bottom panel}:
  QNMs for the $\ell=2$ axial and polar
  gravitational modes of Schwarzschild spacetime, corresponding respectively to the (isospectral)
  Regge-Wheeler and Zerilli potentials (eigenvalues
  along the imaginary upper half-line are the numerical counterpart of the  Schwarzschild branch cut, but also
  the algebraically special QNM $\omega_{n=8}$; see \cite{Ansorg:2016ztf} for a discussion of this).
  Note the normalization $4M\omega_n$, consistent with $\lambda=4M$ after Eq. (\ref{e:r_sigma}).
  {\em Top panel}: condition numbers $\kappa_n$ normalized to the condition number
  $\kappa_0$ of the fundamental QNM. Note the relative enhanced stability of the
  algebraically special QNM.}
\label{fig:Eigenvalues_Schwarzschild}
\end{figure}

The Schwarzschild (gravitational) QNM spectrum (for $\ell=2$)
is shown in Fig.~\ref{fig:Eigenvalues_Schwarzschild}, that presents the result
of the numerical calculation of the spectrum of the $L$ operator defined by
(\ref{e:functions_L1_L2_Schw}). This is obtained either for the Regge-Wheeler
or the Zerilli rescaled potentials in (\ref{e:Vtilde_Schw}), corresponding respectively
to potentials (\ref{e:Schwarzschild_potential_RG}) and (\ref{e:Schwarzschild_potential_Z}).
This provides a crucial internal consistency check for the analytical and numerical construction,
since both potentials are known to be QNM-isospectral (see below in section \ref{s:axial_polar}).
The branch cut structure is apparent in the eigenvalues along the upper imaginary axis.
Such ``branch cut'' points can be easily distinguished from the special QNM
corresponding to $\omega_{n=8}$, also in the imaginary axis, simply by changing the resolution: branch points move
``randomly'' along the vertical axis, whereas $\omega_{n=8}$ stays at the same frequency
(see later \ref{s:high_freq_pert} for a more systematic approach to establish the ``non-branch''
nature of $\omega_{n=8}$, when we will  consider high-frequency perturbations to QNMs). Moreover, eigenfunctions associated with
algebraically special modes are polynomials, as shown in the detailed studies
of these modes for Schwarzschild and Kerr in ~\cite{Ansorg:2016ztf,Cook16}.

%see~\cite{Cook16} for a detailed study of algebraically special modes in Kerr, showing that eigenfunctions associated with such modes are polynomial, in accordance with
%the results for Schwarzschilld obtained in the hyperboloidal setting~\cite{Ansorg:2016ztf}.

Due to the lack of an exact expression for the Schwarzschild QNMs,
one must compare the obtained values against those available in the literature via alternative approaches
--- see, for instance~\cite{Kokkotas:1999bd,Nollert:1999ji,Berti:2009kk,Konoplya:2011qq,BertiWebSite,CardosoWebSite,BHPToolkit,Stein2019}.
An estimative for the errors when QNMs are
calculated with the methods from this work is found in Ref.~\cite{PanossoMacedo:2018gvw}.
From the practical perspective, and regardless of the numerical methods, it is well known that the
difficulty to accurately calculate
numerically a given QNM overtone $\omega^\pm_n$ increases significantly with $n$. For instance, convergence
and machine precision
issues similar to the ones commented above are reported in 
Refs.~\cite{Lin:2016sch,Jansen:2017oag,Fortuna:2020obg}, a control of the internal roundoff accuracy being required.
Alternatively, iterative algorithms such as Leaver's continued fraction method~\cite{Leaver85} require an initial seed relatively near a given QNM, which must be carefully adapted when dealing with the overtones~\cite{BHPToolkitPaper}.
The bottomline is that the calculation of BH QNM overtones is a challenging and very delicate issue.

In our understanding, the latter challenge is not a numerical hindrance
but the consequence of a structural feature of the underlying analytical problem,
namely the spectral instability of the Schwarzschild QNM problem. This is manifested already at the
present stage of analysis, namely the calculation of QNM frequencies of non-perturbed Schwarzschild, in the
eigenvalue condition numbers $\kappa_n$'s shown in the top panel of Fig.~\ref{fig:Eigenvalues_Schwarzschild}:
we encounter again the pattern found in the P\"oschl-Teller case, cf. Fig.~\ref{fig:Eigenvalues_PT},   
with a  growth of the spectral instability as the damping increases, with the notably anomaly 
of an enhanced stability for the algebraically special QNM frequency, with $n=8$.
We devote the rest of the section to explore this spectral instability with the
tools employed for P\"oschl-Teller.

\subsection{Schwarzschild pseudospectrum}
\label{s:Schwarzschild pseudospectrum}
\begin{figure}[h!]
\centering
\includegraphics[width=8.5cm]{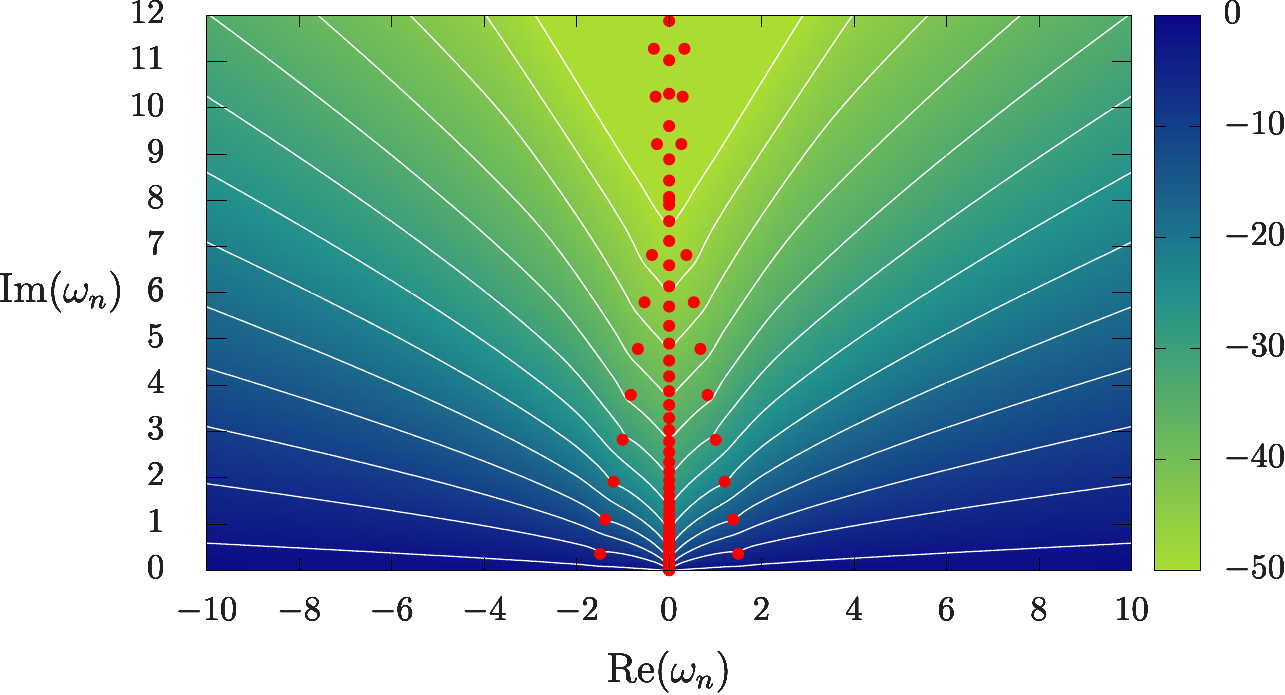}
\includegraphics[width=8.5cm]{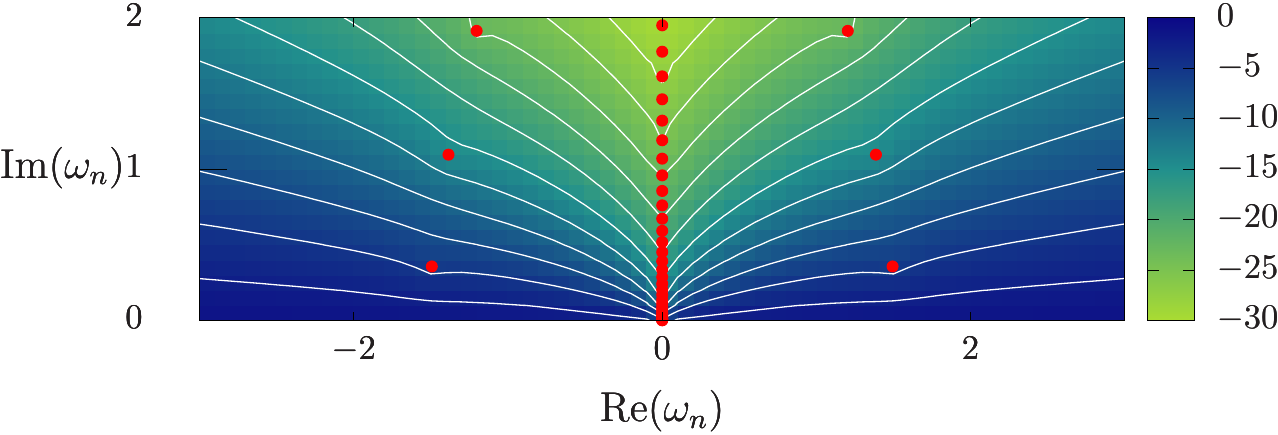}
\caption{ {\em Top panel}: Pseudospectrum of Schwarzschild spacetime ($\ell=2$  gravitational modes,
  from Regger-Wheeler potential, similar   for Zerilli). Again, QNM frequencies  (red circles)
  from Fig.~\ref{fig:Eigenvalues_Schwarzschild} are superimposed
  for reference on their (in)stability.
  The pattern of  $\epsilon$-pseudospectra sets $\sigma^\epsilon$ is qualitative similar to the
  P\"oschl-Teller one (cf. Fig.~\ref{fig:Pseudospectra_PT}),
  though presenting an enhanced spectral instability indicated by the 
  smaller $\epsilon$ values of $\epsilon$-pseudospectra contour lines (cf. range in color log-scale for
  $\mathrm{log}_{10}\epsilon$ in  Fig.~\ref{fig:Pseudospectra_PT}).
  {\em Bottom panel}: Zoom into the region around the fundamental QNM and first overtones.}
\label{fig:Pseudospectra_Schw}
\end{figure}

The pseudospectrum of Schwarzschild is presented in Fig.\ref{fig:Pseudospectra_Schw}.
As illustrated in P\"oschl-Teller, the pseudospectrum provides a systematic
and global tool to address QNM spectral instability, already at the level of the
unperturbed potential. A ``topographic map'' of the analytic structure of the
resolvent,  where regions associated with small $\epsilon$-pseudospectra (light green)
correspond to strong spectral instability, whereas regions
with large $\epsilon$ (namely $O(\epsilon)\sim1$, dark blue) indicate spectral stability.
The superposition of the QNM spectrum shows the respective spectral stability
of QNM frequencies.

 We can draw the following conclusions
 from Fig.~\ref{fig:Pseudospectra_Schw}:
\begin{itemize}
\item[i)] The Schwarzschild  pseudospectrum indicates a strong
  instability of QNM overtones, an instability that grows fast with the damping.
  White-line boundaries corresponding to $\epsilon$-pseudospectra with very
  small  $\epsilon$'s  extend in large regions of the complex plane.
  This is compatible with the results in \cite{Nollert:1996rf}, providing a
  rationale ---already at the level of the unperturbed potential---
  for the QNM overtone instability discovered by Nollert.

\item[ii)] The slowest decaying  QNM is spectrally stable. Fig.~\ref{fig:Pseudospectra_Schw}
  tells us that changing the fundamental QNM frequency requires perturbations
  in the operator of order $||\delta L||_{_E}\sim 1$. This corresponds
  to spectral stability and is in tension with the results in \cite{Nollert:1996rf},
  where the  fundamental  QNM is found to be unstable. We will address this point below.

\item[iii)] Schwarzschild and P\"oschl-Teller potentials show qualitatively the same pseudospectrum pattern,
 with large ``green regions'' producing patterns in
 stark contrast with the flat selfadjoint case. On the one hand, this
 reinforces the usage of P\"oschl-Teller as a convenient guideline for understanding the stability
 structure of BH QNMs and, on the other hand, it points towards an instability mechanism
 independent, at least in a certain measure, on some of the details of the potential.

\end{itemize}
We can conclude that Fig.~\ref{fig:Pseudospectra_Schw} demonstrates ---at the level of the unperturbed operator---
the main features of the stability structure of the BH QNM spectrum, namely the QNM overtone instability
and the stability of the fundamental QNM.
However, the pseudospectrum does not inform us about the particular type of the perturbations that
trigger the instabilities. This is addressed in the following subsection.

\subsection{Perturbations of Schwarzschild potential}
\label{s:QNMpert_QNM}
Once the Schwarzschild pseudospectrum, together with the condition numbers $\kappa_n$, have
presented evidence of QNM spectral instability at the level of the unperturbed operator,
in this section we address the question about the actual physical character of perturbations
triggering such instabilities.

\subsubsection{Ultraviolet instability of BH QNM overtones}
\label{s:high_freq_pert}
The qualitative agreement between  P\"oschl-Teller and Schwarzschild pseudospectra, cf.  Figs.~\ref{fig:Pseudospectra_PT}
and \ref{fig:Pseudospectra_Schw},
together with the experience gained in the study of  P\"oschl-Teller perturbations regarding the high-frequency
instability of all QNM overtones and the stability of the fundamental QNM, guide our steps in the analysis of the BH setting.

\paragraph {Random perturbations: spoils from the ``branch cut''.}
The presence of a ``branch cut'' in the Schwarzschild spectrum, discussed
in section \ref{s:QNM_spectrum_Schw}, translates into a methodological subtlety
when considering random perturbations in the BH case, as
compared with the  P\"oschl-Teller one. The difficulty stems from the fact that
not only the QNM eigenvalues, but also the eigenvalues associated with the discretized
version of the branch cut, are sensitive to random perturbations $\delta\tilde{V}_{\rm r}$ of the potential.
As a consequence, the possible contamination from eigenvalues from the branch cut
complicates the analysis of the impact of random perturbations on QNM frequencies.
This is an artifact of our particular numerical approach, and not a problem of the differential
operator itself, but it limits our capability to assess the  triggering by random perturbation
of the QNM migration to new branches, that was observed in the
P\"oschl-Teller case (cf. left column of Fig.~\ref{fig:EVPert_PT}).
Other tools, either numerical refinements and/or analytical methodologies, are required to address this specific
issue in Schwarzschild.

This does not mean that random perturbations have no use in our
BH discussion. An illustrative example is the study of the stability of the
algebraically special Schwarzschild QNM $\omega_{n=8}$. Whereas random perturbations
move ``branch cut'' eigenvalues away from the imaginary axis, the
algebraically special QNM stays stable. This methodology provides a powerful and
efficient tool to probe the ``physicality'' of specific eigenvalues
in very general settings  (cf. e.g. Fig. 4 in \cite{bizo2019dynamics}).

\paragraph{Deterministic perturbations.}
Given the limitations for random $\delta\tilde{V}_{\rm r}$'s, in the present study we have focused on the class of deterministic
perturbations to the potential $\delta\tilde{V}_{\rm d}$ provided by Eq. (\ref{e:pert_det_cos}). Crucially,
such perturbations do not perturb the ``branch eigenvalues'' as (much as) random $\delta\tilde{V}_{\rm r}$ do, by-passing then
the associated spectral instability contamination.
Despite their simplicity, they provide a good toy-model to explore the effects of
astrophysically motivated perturbations (assessment of ``long range/low frequency'' versus ``small scale/high frequency''
perturbations), as well as those arising from generic approaches to quantum gravity (``small scale/high frequency''
effective fluctuations). They are, therefore, conveniently suited to address these instability issues.

\begin{figure*}[t!]
\centering
\includegraphics[width=8.5cm]{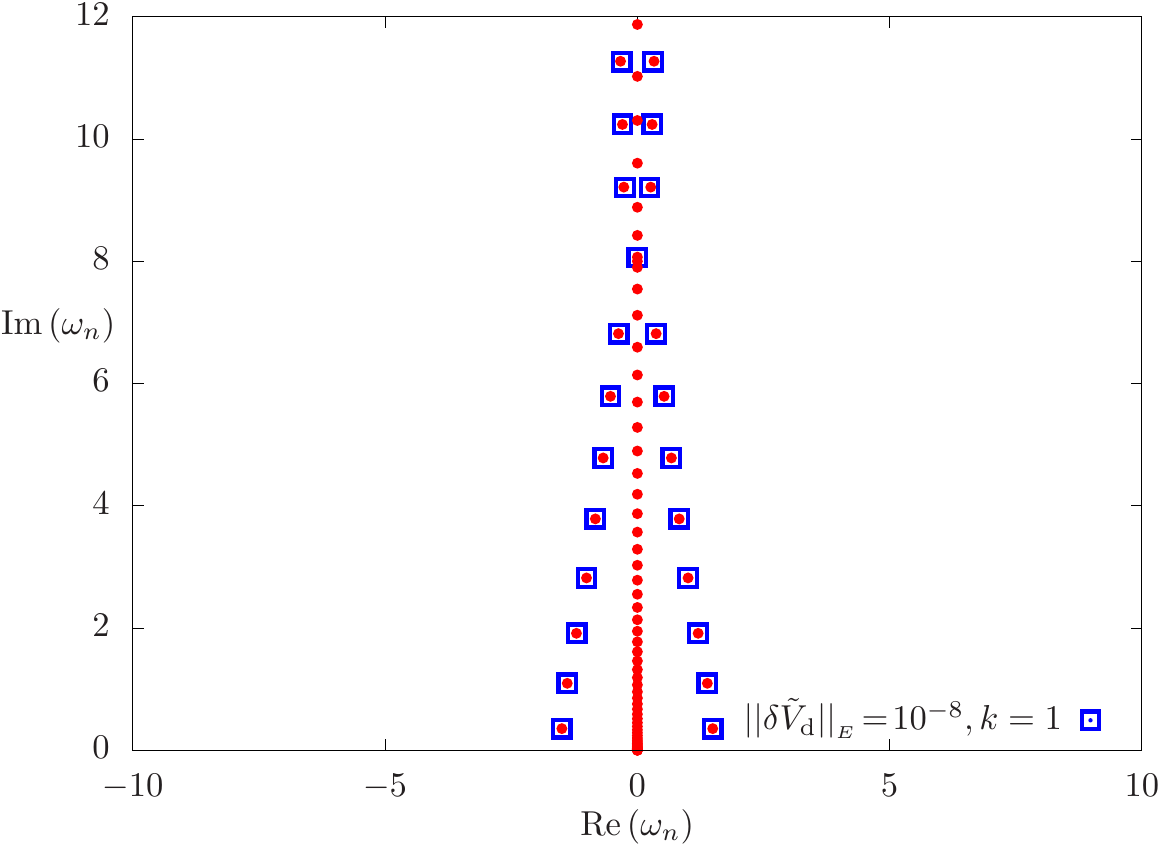}
\includegraphics[width=8.5cm]{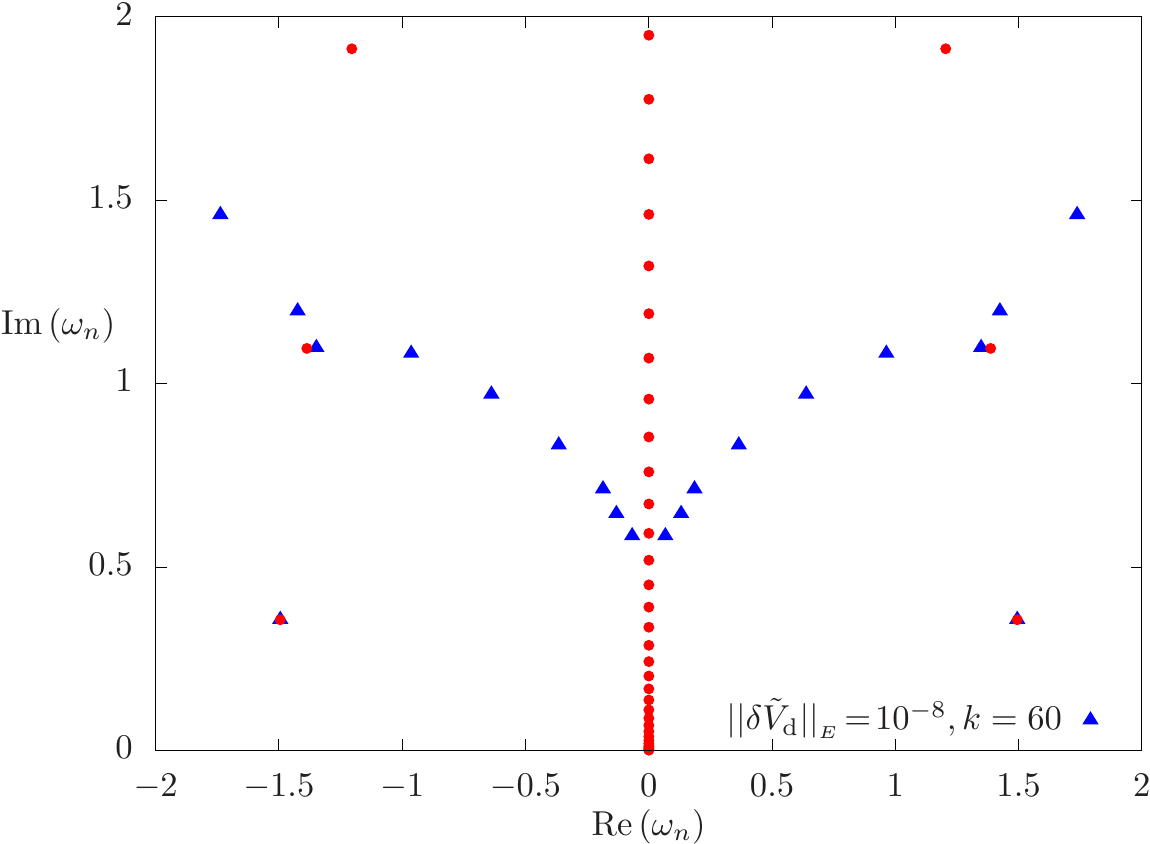}
\includegraphics[width=8.5cm]{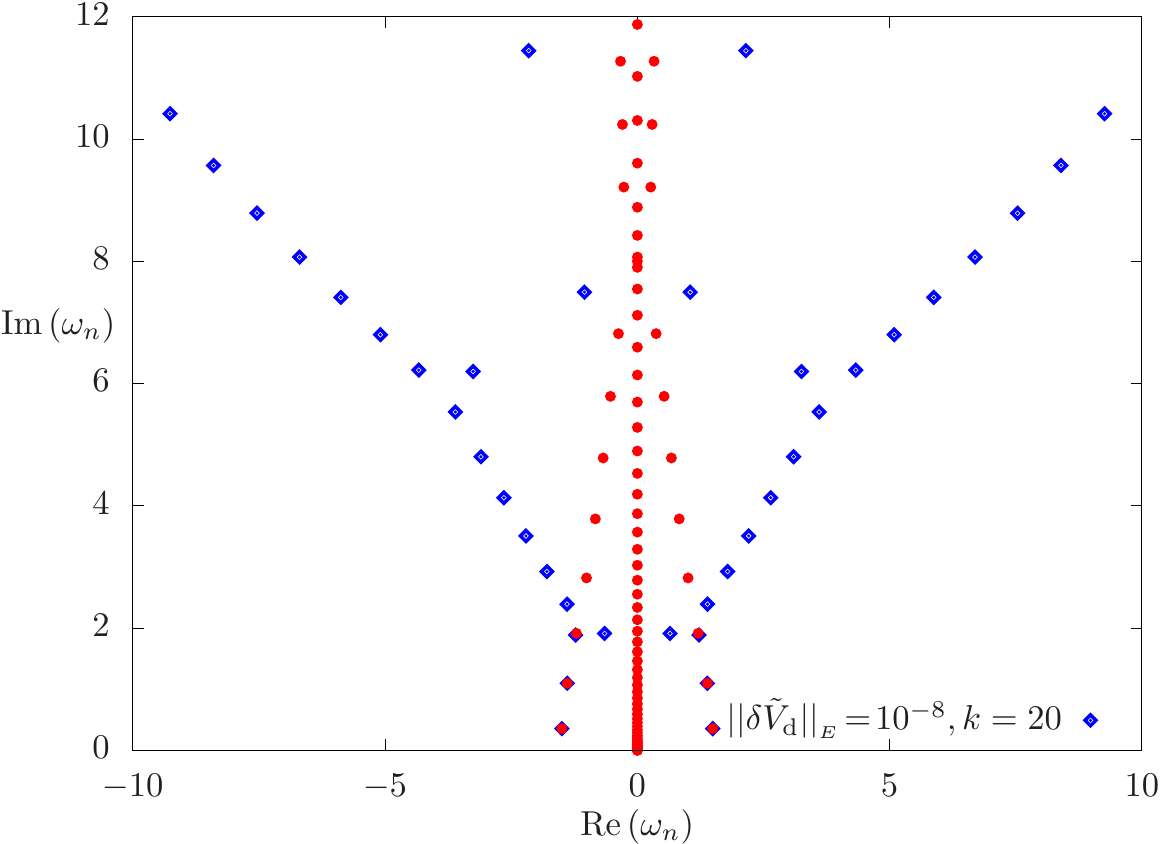}
\includegraphics[width=8.5cm]{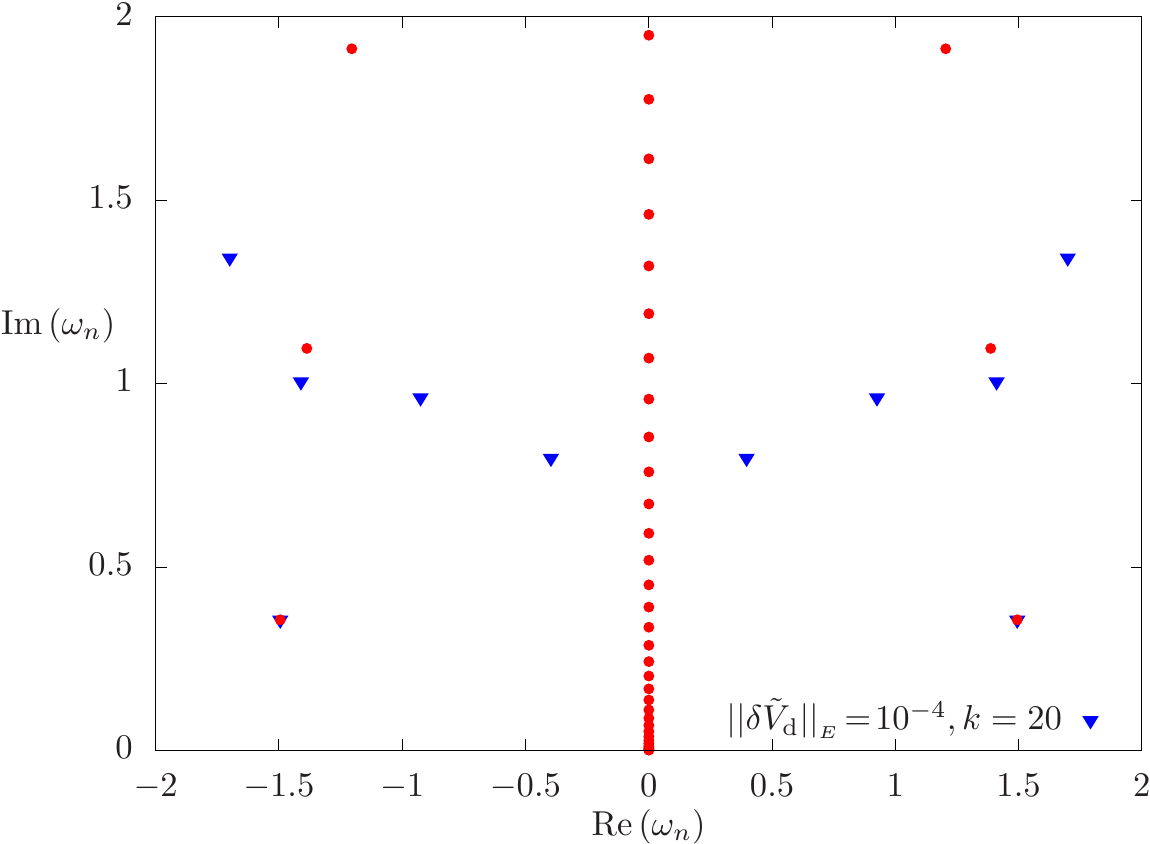}
\caption{QNM spectra for deterministic perturbations $\delta\tilde{V}_{\mathrm{d}}$
  of Schwarzschild $\ell=2$  gravitational modes (here Regge-Wheeler, similar behaviour for
  Zerilli, cf. Fig.~\ref{fig:PolarAxial_QNM}), superimposed over the unperturbed values (red).
  {\em Left column}: stability under low frequency perturbation (top panel) versus high-frequency instability of
  QNM overtones (bottom panel).
  {\em Right column}: zoom into the first QNM overtones, showing the
  instability of the first overtone by increasing i) the frequency of the  perturbation (top panel),
  and ii) the (energy) norm of the perturbation (bottom panel).}
\label{fig:EVPert_Schwarzschild}
\end{figure*}

The left column in Fig.~\ref{fig:EVPert_Schwarzschild} depicts (with $||\delta\tilde{V}_{\rm d}||_{_E}\sim 10^{-8}$)
the stability of the first overtones against
low frequency perturbations ($k=1$, top-left panel) in contrast with the instability resulting from high-frequency
perturbations ($k=20$, bottom-left panel). Pushing along this line, the right column in Fig.~\ref{fig:EVPert_Schwarzschild}
zooms in to study the very first overtones, which are paramount for the incipient field of black-hole spectroscopy.
Assessing the (in)stability of the very first overtones is therefore crucial for current research programs in gravitational astronomy.
It becomes apparent that the first overtones, this including the very first overtone, are indeed affected without
any extraordinary or fine tuned perturbations $\delta\tilde{V}_{\mathrm{d}}$. In particular, and taking the left column as a reference,
the first overtone is reached: i) either by considering
a ``slightly'' more intense perturbation
($||\delta\tilde{V}_{\rm d}||_{_E}\sim 10^{-4}$, $k=20$), or ii) perturbations with sufficiently  high frequency
($||\delta\tilde{V}_{\rm d}||_{_E}\sim 10^{-8}$, $k=60$).

From this perturbation analysis of the BH potential we conclude: i) all QNM overtones are ultraviolet unstable, as in P\"oschl-Teller,
  the instability reaching the first overtone for sufficiently high frequency;
ii) QNMs are stable under low frequency perturbations, this illustrating
that spectral instability does not mean instability under ``any'' perturbation, in
particular long-wave perturbations not affecting the QNM spectrum; 
iii) the slowest decaying QNM is ultraviolet stable, a result 
in tension with the instability of the fundamental QNM found in \cite{Nollert:1996rf}.
  We revisit this point in section \ref{s:Schwarzschild_infrared} below.

\subsubsection{Isospectrality loss: axial versus polar spectral instability}
\label{s:axial_polar}
Regge-Wheeler and Zerilli potentials for axial and polar perturbations
are known to be isospectral in the QNM spectrum
(cf. \cite{ChaDet75,Chandrasekhar:579245,Anderson:1991kx,Glampedakis:2017rar}; see also
\cite{maggiore2018gravitational}).
In particular, Chandrasekhar identified (cf. point 28 in \cite{Chandrasekhar:579245})
a necessary condition for two (one-dimensional) potentials $V_1(\bar{x})$ and $V_2(\bar{x})$,
with $\bar{x}\in]-\infty,\infty[$ as the rescaled tortoise coordinate, 
to have the same transmission amplitude and present the same QNM spectrum.
Specifically, both potentials must render the same values when evaluating
an infinite hierarchy of integrals
\bea
\label{e:KdV_conservedquantities}
C_n = \int_{-\infty}^\infty v_n(\bar{x}) d\bar{x}\ ,
\eea
with
\bea
v_1 &=& V \ \ , \ \ v_3 = 2V^3 + V'^2 \nn \\
v_5 &=& 5V^4 + 10 V V'^2 + V''^2 \ \ , \ \ v_{2n+1} = \ldots
\eea
These quantities turn out to be the conserved quantities of the
Korteweg-de Vries equation and connect the Schwarzschild QNM isospectrality
problem to integrability theory through the
inverse scattering transform of Gelfand-Levitan-Marchenko (GLM) theory (cf. \cite{Dunajski:2010zz};
see \cite{Glampedakis:2017rar} for an alternative approach in terms of Darboux transformations).

\begin{figure}[t!]
\centering
\includegraphics[width=8.cm]{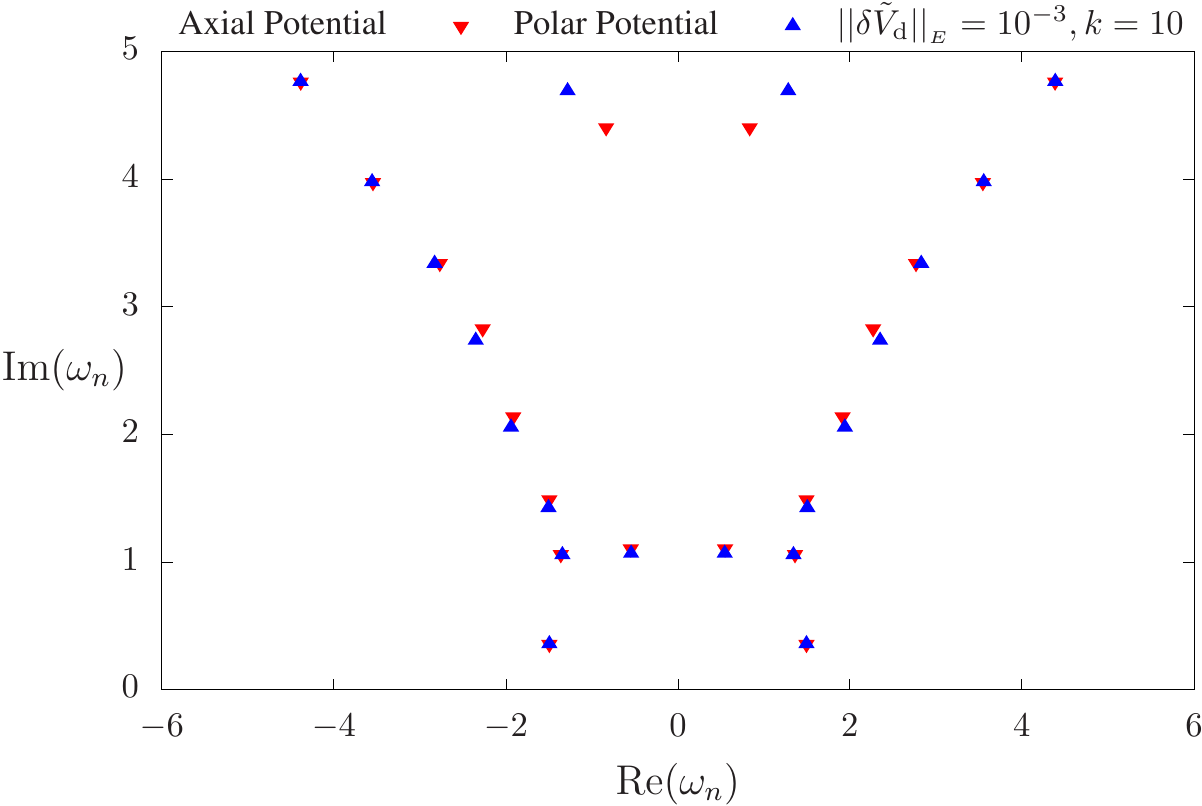}
\includegraphics[width=8.5cm]{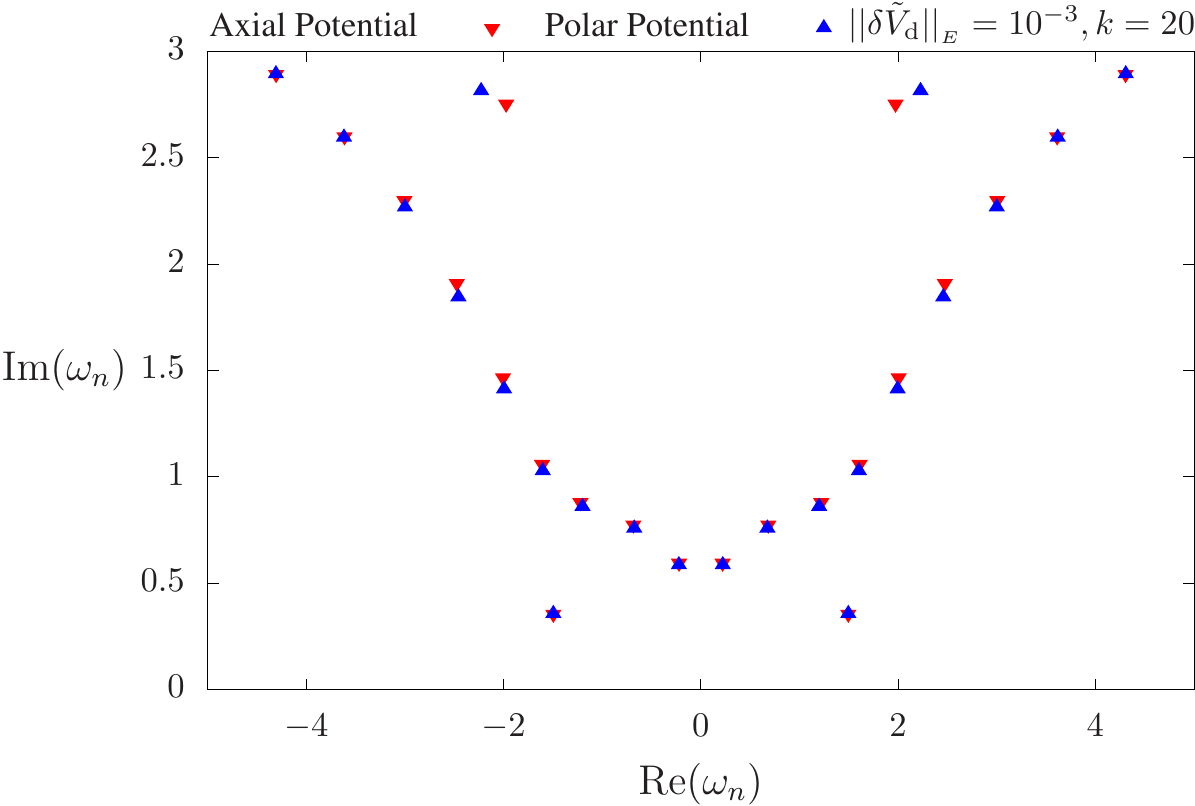}
\includegraphics[width=8.5cm]{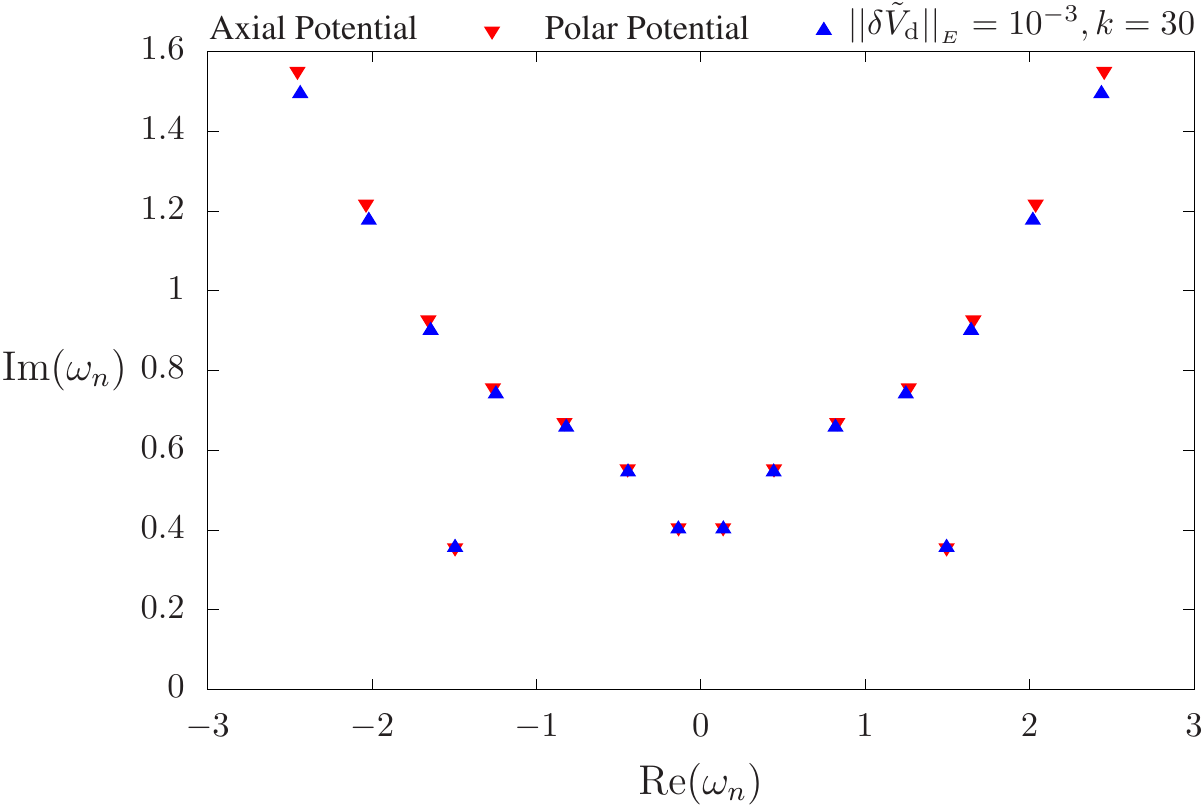}
\caption{Loss of isospectrality in Schwarzschild, under
  high-frequency perturbations. The sequence of figures shows a zoom
  into the perturbation of lowest  $\ell=2$ axial and polar QNM overtones (the branch cut has been removed),
  with $\delta \tilde{V}_{\mathrm{d}}$ fixed to a value reaching the
  first overtone, and then increasing the frequency. The breaking of axial and polar
  isospectrality is demonstrated, with perturbed axial overtones slightly
  more damped than polar perturbed counterparts, though both laying over the same
  perturbed QNM branches (actually tracking the  pseudospectra contour lines,
  cf. Fig.~\ref{fig:Fig1} below). The fundamental QNM remains unchanged,
  consistently with its stability, so the dominating ringdown frequency
  remains ``isospectral''.
   }
\label{fig:PolarAxial_QNM}
\end{figure}

The key point for our spectral stability analysis of $L$ is that axial and polar QNM isospectrality is the
consequence of a subtle and ``delicate'' integrability property of stationary
BH solutions, so we do not expect it to be robust under generic perturbations of $V$.
In particular, given the non-linear dependence in $V$ of the conserved quantities
$C_n$ in (\ref{e:KdV_conservedquantities}), 
we would expect either random $\delta\tilde{V}_{\rm r}$ or deterministic $\delta\tilde{V}_{\rm d}$
perturbations to render different values of $C_n$, therefore resulting in a loss of QNM isospectrality.
Fig.~\ref{fig:PolarAxial_QNM} confirms this expectation: whereas the fundamental QNM mode
remains stable under high-frequency perturbations, 
isospectrality is broken for the overtones with a slight, but systematic, enhanced damping
in the axial case. Other mechanisms for BH isospectrality loss have been envisaged, e.g.
in the study of the imprints of modified gravity theories \cite{Cardoso:2019mqo}, ultraviolet
QNM overtone instability
providing a possible mechanism inside general relativity. In sum, isospectrality loss
provides an interesting probe into QNM instability, with potential observable consequences
and will be the subject
of a specifically devoted study elsewhere.

\subsubsection{``Infrared instability'' of the fundamental QNM}
\label{s:Schwarzschild_infrared}
Both the pseudospectrum and the explicit perturbations of the potential indicate
a strong spectral stability of the slowest decaying Schwarzschild QNM.
This is tension with the results in \cite{Nollert:1996rf,Nollert:1998ys}, where the instability affects the
whole QNM spectrum, this including the slowest decaying QNM.
This is a fundamental point to establish, since it directly
impacts the dominating frequency in the late BH ringdown signal.

In our understanding, and as it was the case of the P\"oschl-Teller potential discussed in
section \ref{s:Slowest_QNM_PT}, the instability of the fundamental QNM
frequency found by \cite{Nollert:1996rf}  is an artifact of the implemented
perturbations, namely step-like approximations 
to the Schwarzschild potential (in particular
Regge-Wheeler, but the same applies for Zerilli) that modify
the potential at large distances.
Specifically, $V_{\ell}$ is set to zero  beyond $[x_{\mathrm{min}}, x_{\mathrm{max}}]$,
fundamentally altering the long-range nature of Schwarzschild potential
that becomes of compact support.
What we observe in Fig.~\ref{fig:EVPert_Schwarzschild} is that keeping  a
faithful treatment of the asymptotic structure at infinity
through the compactified hyperboloidal approach keeps spectral stability.

\begin{figure}[t!]
\centering
\includegraphics[width=8.5cm]{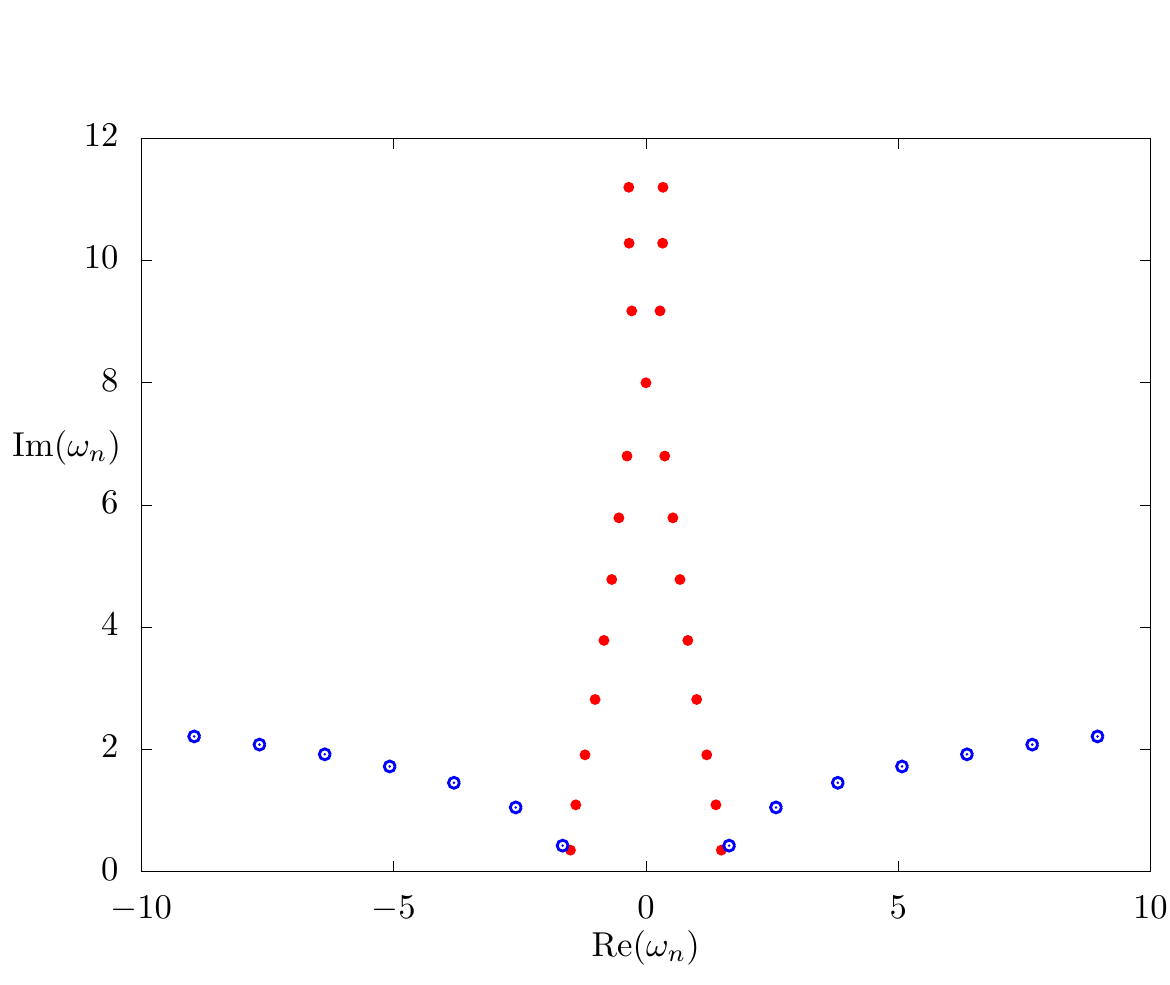}
\caption{``Infrared'' modification of the Schwarzchild fundamental QNM.
  As in the P\"oschl-Teller case, cutting the Schwarzschild potential
  ($\ell=2$, either Regge-Wheeler or Zerilli) outside a compact interval
  $[x_{\mathrm{min}}, x_{\mathrm{max}}]$ modifies the fundamental QNM,
    this accounting for its ``instability'' found in \cite{Nollert:1996rf}.
    All QNM overtones are strongly perturbed due to the high-frequencies in the
Heaviside cut, whereas (only) the fundamental QNM is recovered as $x_{\mathrm{min}}, x_{\mathrm{max}}\to \mp \infty$.}
\label{fig:Infrared_effect_Schwarzschild}
\end{figure}

To test this idea (cf. also the recent
\cite{Qian:2020cnz}, as well as \cite{SheJar20}), and as we did in P\"oschl-Teller,
we have implemented a ``cut Schwarzschild'' potential in
our hyperboloidal approach, setting the potential to zero from a given distance
(both towards null infinity and the BH horizon). The result is shown
in Fig.~\ref{fig:Infrared_effect_Schwarzschild}, showing a similar qualitative
behaviour to P\"oschl-Teller in Fig.~\ref{fig:Infrared_effect_PT}.
Overtones are strongly perturbed into the  QNM branches  already observed
in Fig.~\ref{fig:EVPert_Schwarzschild}, consistently with the  high-frequencies
introduced by the Heaviside cut. 
But, crucially, now the fundamental QNM is indeed also modified,
in contrast with its stable behaviour under high-frequency perturbations.
This reinforces the
understanding of this effect as a consequence of the ``suppression'' of the
 large-scale asymptotics of the potential~\footnote{Such suppression must
  be stronger than exponential, since Poeschl-Teller shows stability of the fundamental QNM.}.
However, the observed
modification of the fundamental QNM frequency is not as dramatic as the one in 
\cite{Nollert:1996rf}. We do not have a good explanation for this, but
it may relate to the fact that the analysis in \cite{Nollert:1996rf,Nollert:1998ys}
deals directly with Eq. (\ref{e:wave_equation_tortoise}), in particular in the
setting of a Cauchy slicing getting to spatial infinity $i^0$. Such asymptotic framework
may be more sensitive to the modification of the potential that
the hyperboloidal one, related to null infinity $\scri^+$.
In this setting, and lacking a better expression, we refer to this effect as an
``infrared instability'' of the fundamental QNM.

Enforcing the compact support nature of $V$ is naturally motivated in physical contexts such
as optical cavities, and will be studied systematically in such 
settings \cite{SheJar20}.
In gravitation the physicality of such an effect is more difficult to assess, 
since gravity is a long-range interaction that, in contrast to the electromagnetic one,
is not screened. In any case, insofar as a pertinent gravitational scenario may be envisaged
for a such ``cut potential'', then the ``infrared instability'' shown for the first
time in \cite{Nollert:1996rf} would constitute a physical effect.

\subsection{Nollert-Price BH QNM branches: instability and universality }
\label{s:Nollert_universality}
We revisit the results in \cite{Nollert:1996rf,Nollert:1998ys} (see also \cite{Daghigh:2020jyk,Qian:2020cnz}),
under the light of the elements introduced for the study of QNM spectral stability.
Fig.~2 in \cite{Nollert:1996rf} presents the migration of Schwarzschild QNMs
to new branches, as the result of perturbing the (Regge-Wheeler) Schwarzschild potential
with a step-like approximation with an increasing number ``$N_{\mathrm{st}}$'' of steps  (cf. Fig. 1 in \cite{Nollert:1996rf}).
A salient feature of Nollert's Fig. 2, further analysed with Price in \cite{Nollert:1998ys}, is that the new
QNM branches distribute in a perfectly structured family of lines in the complex plane, unbounded in the real
part of the frequency, that ``move down'' in the complex plane as $N_{\mathrm{st}}$
(i.e. the frequency in the perturbation) increases~\footnote{The Nollert case $N_{\mathrm{st}}=1$
  in his method ``iii)''  seems special. It corresponds precisely to
  the ``cut potential'' in section \ref{s:Schwarzschild_infrared} and may require a separate discussion.
  It connects also with section \ref{s:Slowest_QNM_PT}, since  method ``iii)'' in \cite{Nollert:1996rf}
  ``regularizes''  Schwarzschild with a Poeschl-Teller factor, cf. Eq.~(7) in \cite{Nollert:1996rf}.}.
A comparison with Schwarzschild's pseudospectrum in our Fig.~\ref{fig:Pseudospectra_Schw}
shows two remarkable features: i) the pattern of the new branches found and studied by Nollert and Price
is qualitatively similar to the contour lines of $\epsilon$-pseudospectra, ii) the effect of increasing the frequency
perturbation indeed corresponds to an increment in the $\epsilon$ of the corresponding
contour line (namely the ``energy  size'' of the pertubation that, as a $H^1$ norm,
includes the frequency). In other words, Nollert and Price's BH QNM branches seem indeed to be closely
related to $\epsilon$-pseudospectrum contour lines.

\begin{figure*}[t!]
\centering
\includegraphics[width=16cm]{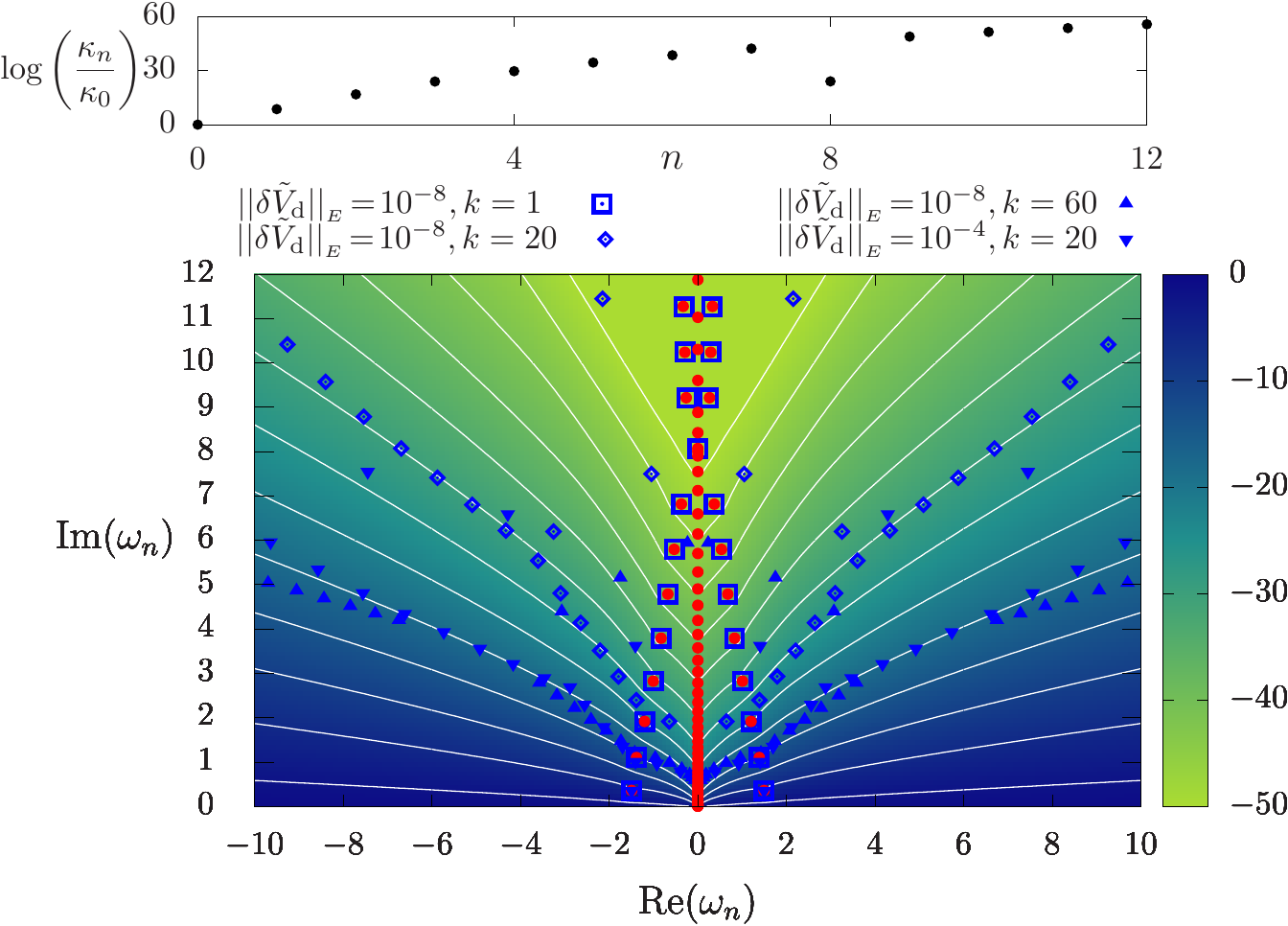}
\caption{Gravitational QNM spectral (in)stability in Schwarzschild spacetime (here, $\ell=2$ axial case corresponding to
  the Regge-Wheeler potential, same behaviour for polar modes with Zerilli potential).
  The figure shows the superposition of the pseudospectrum of Fig.~\ref{fig:Pseudospectra_Schw}, perturbed QNM spectra
  in Fig.~\ref{fig:EVPert_Schwarzschild}, together with exact QNMs and condition numbers $\kappa_n$ from
  Fig.~\ref{fig:Eigenvalues_Schwarzschild}. Employed norms follow from the energy scalar product
  in Eq. (\ref{e:energy_scalar_product_Schw}), i.e. energy
  defines ``big'' and ``small''.   
  The pseudospectrum pattern, with  $\epsilon$-pseudospectra sets with small $\epsilon$ extending into large regions
  of the complex plane, indicates spectral instability of QNM overtones, consistently with the fastly growing
  $\kappa_n$'s.
  Perturbations in the potential demonstrate the  high-frequency (ultraviolet) instability of all overtones
  and their stability under low-frequency perturbations. Both pseudospectrum and perturbations in the potential
  show the ultraviolet stability of the fundamental QNM.
  Ultraviolet instability induce QNM overtones to migrate towards $\epsilon$-pseudospectra contour lines,
  a pattern consistent with ``Nollert-Price QNM branches'' \cite{Nollert:1996rf,Nollert:1998ys}
  here illustrated up to the lowest overtone.
  Universality of this pattern is further supported by comparison with P\"oschl-Teller in
  Fig.~\ref{fig:Pseudospectrum_perturbations}.
 }
\label{fig:Fig1}
\end{figure*}

In order to test this picture, we bring our perturbation analysis in  section \ref{s:QNMpert_QNM} into
scene. Fig.~\ref{fig:Fig1} presents the 
superposition of perturbed QNM spectra in Fig.~\ref{fig:EVPert_Schwarzschild}
onto the Schwarzschild pseudospectrum in Fig.~\ref{fig:Pseudospectra_Schw}.
As in the P\"oschl-Teller case, perturbed QNMs closely track $\epsilon$-pseudospectra
lines,  demonstrating the insight gained above on Nollert's QNM instability by using
the pseudospectrum:  Nollert-Price QNM branches are identified as actual probes
into the analytical structure of the non-perturbed wave operator.
Moreover, the correlation of $\epsilon$-contour lines with the ``size/frequency''
of the perturbations, endows the pseudospectrum not only with an explicative but also with a predictive power, as
a tool to calibrate the relation between spacetime perturbations and QNM frequency changes.
The conceptual frame encoded in Fig.~\ref{fig:Fig1} is, in our understanding, the main contribution in this work.

\subsubsection{QNM structural stability,  universality and asymptotic analysis}
\label{s:universality}
Building on Nollert and Price's work,
our analysis 
strongly suggests that BH QNM overtones are indeed structurally unstable
under high-frequency perturbations:
BH QNM branches migrate to a qualitatively different class of 
QNM branches.
Noticeably and in contrast with this,  the
pseudospectrum analysis combined with the perturbation tools 
also suggests that the new class of ``Nollert-Price BH QNM branches'' presents
structural stability features pointing to a kind of 'universality' in
the QNM overtone migration pattern.

\paragraph{``Universality'' in the high-frequency perturbations.} The QNM migration pattern seems independent
of the detailed nature of the  high-frequency perturbation in the Schwarzschild potential.
First, such universality is manifested by the similar
QNM perturbation pattern produced by very different perturbations: 
step-like perturbations in \cite{Nollert:1996rf}, the sinusoidal deterministic ones showed in Fig.~\ref{fig:Fig1}
and also random perturbations (not presented here due to ``blurring'' issues, consequence of the ``branch cut'' contamination).
Second, the new branches follow closely the pseudospectra contour lines, a key point in this
 universality discussion, since it is completely prior to  and independent of perturbations.

\paragraph{``Universality'' in the potential.} Perhaps more importantly, universality seems to go beyond the insensitivity to the nature of the perturbation:
it seems to be shared by a whole class of potentials.
First, the same pattern  of perturbed branches 
is found in P\"oschl-Teller, cf. Fig.~\ref{fig:Pseudospectrum_perturbations}.
More dramatically,  Nollert and Price's  analysis in \cite{Nollert:1998ys} is particularly
illuminating in this respect. 
They considered a toy model capturing the effect of a
(Dirac-delta) high-frequency perturbation on a BH-like potential,
referred to as ``truncated dipole potential'' (TDP), that 
contains only two QNMs. Adding  the singular (high-frequency) ``spike'' creates an infinite number of QNMs, again
following a QNM branch pattern compatible
with our pseudospectra contour lines (cf. Fig. 5 in \cite{Nollert:1998ys} and see below).

But more noteworthy, and again noticed by Nollert \cite{Nollert:1996rf},
beyond the BH setting the new BH QNM branches are strikingly similar to (curvature) $w$-modes in neutron-star QNMs
(cf. e.g. Fig. 3 in \cite{Kokkotas:1999bd} and the systematic study in Ref. \cite{ZhaWuLeu11}). This is remarkable, suggesting
that exact but unstable BH QNMs migrate to perturbed but stable QNMs 
branches whose qualitative pattern may be shared by generic  compact
objects~\footnote{\label{footnote:Regge_poles}Beyond $w$-modes of compact objects, such perturbed
  BH 'universal' branches share also features with QNMs of convex obstacles, where the asymptotic form
  of QNM branches (under a 'pinched curvature assumption') can be established \cite{SjoZwo99,Zworski99}
  as $\mathrm{Im}(\omega_n) \sim K |\mathrm{Re}(\omega_n)|^{\frac{1}{3}} + C$, for $n\gg 1$.
  Focusing on the spherical obstacle case \cite{STEFANOV2006111} (see
  also \cite{zworski2017mathematical,dyatlov2019mathematical}),
  if considering  all angular $\ell$'s modes and taking
  $\ell$ as the spectral parameter (while keeping
  $n$ fixed), the similar qualitative pattern between the corresponding branches 
  and the perturbed BH QNM branches raises an intriguing
  question about a possible duality between QNM and Regge poles
%  in a complex angular momentum setting 
  (cf. e.g. \cite{Decanini:2010fz,Decanini:2011xw,Raffaelli:2014ola,Dolan:2009nk}
  in a complex angular momentum setting). In particular,
  the asymptotic logarithm pattern of perturbed-BH \cite{Nollert:1998ys}
  and compact object \cite{ZhaWuLeu11} QNMs
  is exactly recovered for Regge poles of compact objects in \cite{OuldElHadj:2019kji}
  (cf. \cite{Daud__2015} for related asymptotics).}.

\paragraph{Asymptotic analysis and universality.} How to address systematically
a possible universality in the qualitative pattern of the perturbed QNM branches?
Asymptotic analysis provides a sound approach. 
The study of the spiked TDP QNMs by Nollert and Price \cite{Nollert:1998ys} provides an excellent
illustration, with
the identification of the large-$n$ asymptotic form of perturbed QNM branches,
according to the logarithm dependence
\bea
\label{e:log_branches}
\mathrm{Im}(\omega_n) \sim C_1 + C_2 \ln \big(\mathrm{Re}(\omega_n) + C_3\big) \ \ , \ \ n\gg 1 \ ,
\eea
with $C_1$, $C_2$ and $C_3$ appropriate constants (note that $C_3$ can be put to zero
for sufficiently high $n$,  as in \cite{Nollert:1998ys}, since $\mathrm{Re}(\omega_n)\to \infty$ as $n\to\infty$; we prefer to
keep it to account for intermediate asymptotics \cite{JarMacShe21}).
It is suggestive that this makes direct contact with the possible universality of perturbed BH QNMs
and (non-perturbed) QNMs of compact objects evoked above. Indeed, as shown in Ref. \cite{ZhaWuLeu11},
$w$-modes of (a class of) neutron stars present exactly this logarithm pattern \footnote{We thank
B. Raffaelli for signaling this and also Ref. \cite{OuldElHadj:2019kji}.}.
Even more, this makes (an unexpected) contact with P\"oschl-Teller, where
the spectral instability discussed in section \ref{s:numerical_PT_sectrum}
is explained \cite{BinZwo,Zwors87,zworski2017mathematical,dyatlov2019mathematical}
in terms of so-called broad ``Regge resonances'' (not to confuse with ``Regge poles''),
precisely described by such a logarithmic dependence~\cite{Regge58} and explained in terms
of the loss of continuity at a $p$-th order derivative, i.e. in terms of an underlying reduced $C^p$ regularity
(with $p<\infty$). Along these lines of $C^p$ regularity, and in a WKB semiclassical analysis,
such logarithmic branches have been also recovered in \cite{Qian:2020cnz}
in their recent discussion of Nollert's original work \cite{Nollert:1996rf}.
It would be therefore tempting to refer to the perturbed
BH QNM branches as Nollert-Price-(Regge) QNMs, but this requires an elucidation of the
role of the reduced $C^p$ regularity in the generic perturbations we have studied here, that in particular
include $C^\infty$ regular (high-frequency) sinusoidal deterministic perturbations (\ref{e:pert_det_cos}).
In sum, the asymptotic pattern (\ref{e:log_branches}) provides a starting point to
probe, in gravitational wave signals, the physical properties (e.g. energy, frequency) of small scale perturbations \cite{JarMacShe21}.

Beyond specific models,
this kind of universal behaviour, independent
of the high-frequency perturbation detailed nature and for a large class of
potentials, invites for systematic semi-classical analyses of highly-damped scattering resonances,
in terms of the wave operator principal part \footnote{We thank N. Besset for signaling this point.},
including boundary behaviors. 
  In the spirit adopted in this work, we expect asymptotic tools in the semiclassical analysis of the pseudospectrum
  to provide a systematic approach to assess the universality of perturbed BH
  QNM branches~\footnote{Such an approach is very
    much in the spirit of the ``asymptotic reasoning'' advocated in \cite{Batte01}, where
    asymptotic analysis is understood as an efficient and systematic tool to unveil
    structurally stable patterns underlying universality behaviour.}.

\subsubsection{Overall perspective on Schwarzschild QNM instability}
\label{s:BH_QNM_summary}
The main result of this article is summarized in Fig.~\ref{fig:Fig1}.
Specifically, it combines Figs.~\ref{fig:Eigenvalues_Schwarzschild}, \ref{fig:Pseudospectra_Schw} and \ref{fig:EVPert_Schwarzschild}
to demonstrate QNM spectral (in)stability
through their respective three distinct calculations:  i) the calculation of the eigenfunctions of the exact
spectral problem to calculate condition numbers $\kappa_n$'s,
ii) the evaluation of  operator matrix norms to generate the pseudospectrum, and iii) the calculation of
eigenvalues of the perturbed spectral problem. Calculations i) and ii)
work at the level of the unperturbed problem, whereas iii) deals with the perturbed problem.
The three calculations fit consistently through  the Bauer-Fike theorem
that constrains through  Eq. (\ref{e:tubular_error})
the relation between the pseudospectrum and the tubular regions
around the spectrum. They lead to these main results:

\medskip

\noindent i) QNM overtones:
\begin{itemize}
\item[i.1)]  {\em QNM overtones are ultraviolet unstable, including the lowest overtones}.
  The pseudospectrum provides a systematic explanatory and predictive framework
  for QNM spectral instability, confirming the result by Nollert and Price
\cite{Nollert:1996rf,Nollert:1998ys}.
Such instability is indeed realised by physical  high-frequency perturbations
in the effective potential $V$, reaching the first overtone for
sufficiently high frequencies and/or amplitudes in the perturbation.

\item[i.2)] {\em QNM overtones are stable under low frequency perturbations.}
No instability appears for low/intermediate frequency perturbations of $V$, 
consistently with studies \cite{FerMas84,Leung:1999iq,Barausse:2014tra,Cardoso:2019mqo,Hui:2019aox}
on astrophysical BH environments.

\end{itemize}

\noindent ii)  Slowest decaying (fundamental) QNM:

\begin{itemize}
\item[ii.1)] {\em The slowest decaying QNM is ultraviolet stable}.
  This feature critically relies on keeping a faithful description of the asymptotic structure at infinity
  through the compactified hyperboloidal approach.
  This result is in contrast with conclusions in \cite{Nollert:1996rf,Nollert:1998ys},
  but no contradiction appears since the latter implement a step-potential approximation
  fundamentally modifying $V$ at large distances, resulting rather
  in an ``infrared probe'' into QNMs.

\item[ii.2)] {\em The slowest decaying QNM is stable under low and intermediate frequency perturbations
in the potential}.
  This property is shared by the whole QNM spectrum.

\item[ii.3)] {\em The slowest decaying QNM is ``infrared unstable''}.
  The instability of the fundamental QNM observed in 
  \cite{Nollert:1996rf,Nollert:1998ys} is physical inasmuch as fundamental
  modifications of the large-distance structure of the potential are allowed.

\end{itemize}

\noindent iii) Structural stability and QNM isospectrality.

\begin{itemize}
\item[iii.1)] {\em `Nollert-Price BH QNM branches' track pseudospectrum contour lines}.
  The QNM BH spectrum is ultraviolet structurally unstable, migrating to
  perturbed branches tracking  $\epsilon$-contour lines
  of pseudospectra. Such migration pattern is largely independent of the detailed nature
  of high-frequency perturbations and potential. Once on such
  `Nollert-Price branches', QNMs are spectrally stable.  
  These structural stability properties result in the universality
  of perturbed QNM branches.

\item[iii.1)] {\em QNM isospectrality ultraviolet  loss}. High-frequency perturbations
  spoil the integrability of Regge-Wheeler and Zerilli potentials, resulting
  in a slightly enhanced damping of axial modes with respect to polar ones. 
  
\end{itemize}

\section{Conclusions and perspectives}
\label{s:conclusions_perspectives}

\subsection{Conclusions}
\label{s:conclusions}
We have demonstrated:
i) the fundamental BH QNM is stable under high-frequency (ultraviolet) perturbations, while unstable
under (infrared) modifications of the asymptotics, the latter consistent with
\cite{Nollert:1996rf}; ii) (all) BH QNM
overtones are unstable under high-frequency (ultraviolet) perturbations, quantifiable in terms of the energy
content (norm) of the perturbation, extending results in
\cite{Nollert:1996rf,Nollert:1998ys} to show isospectrality loss; and iii) pseudospectrum contour lines provide the
rationale underlying the structurally stable pattern of perturbed `Nollert-Price QNM BH branches'.
Pseudospectra, together with tools from the
analysis of non-selfadjoint operators, have revealed
the analytic structure underlying such (in)stability properties of BH QNMs,
offering an integrating and systematic approach to encompass a priori disparate phenomena.
The soundness of the results relies on the use of a  compactified hyperboloidal approach to QNMs,
with the key identification of the relevant scalar product in the problem as associated with the physical
  energy, combined with accurate spectral numerical methods.

\subsubsection{Caveats in the current approach to QNM (in)stability}
\label{s:numerical_caveats}
Beyond the soundness of the results, key questions remain:
\begin{itemize}
\item[i)] How much does the instability depend on the hyperboloidal approach?
  In other words, is the instability a property of the equation or rather
  of the employed scheme to cast it? This is a legitimate and crucial question,
  requiring specific investigation. In spite of this, we are confident in the soundness of
  our conclusions: as discussed in detail, the same qualitative behaviour
  is found systematically by other studies not relying on the hyperboloidal
  approach, in particular Nollert and Price's pioneer work.
  Details may change from scheme to scheme, but the (in)stability properties
  seem robust.
  
\item[ii)] A numerical demonstration is not a proof. Moreover, numerical discretizations
  introduce their own difficulties and limitations. In particular, spectral issues
  in the passage from matrix approximations to the actual differential operator
  is a most delicate question. Again, we are confident in our results, as a consequence
  of mutual consistency of existing results and non-trivial tests like the ones described in the text. 
  Definitely, proofs will require the use of other methods and techniques.
  
\item[iii)] Could the observed QNM spectral instability be an effect of
  regularity loss, namely a $C^p$ effect? It may be, but it
  is difficult to conclude at this stage.  $C^p$ regularity provides indeed  a sufficient condition
  for logarithmic branches (\ref{e:log_branches}) that can be traced to works by
  Regge \cite{Regge58}, Berry \cite{Berry82,BerMou72} or Zworski \cite{Zwors87} and manifests in our setting in  
  Nollert \& Price's analysis of BH QNM instability \cite{Nollert:1998ys} (complemented in \cite{Qian:2020cnz}),
  broad ``Regge resonances'' in P\"oschl-Teller QNM instability
  \cite{BinZwo,zworski2017mathematical,dyatlov2019mathematical}, or in neutron star $w$-modes \cite{ZhaWuLeu11}
  (cf. also \cite{OuldElHadj:2019kji} in related Regge poles).
  But we also attest the same instability phenomenon for regular sinusoidal perturbations of sufficiently high-frequency.
  Moreover, the pseudospectrum already informs of the instability (cf. contour lines)
  at the unperturbed ``regular'' stage.
  If high-frequency is actually the basic mechanism, then
  $C^p$ would provide a sufficient, but not necessary condition for QNM instability.
  This point must be addressed.

\end{itemize}

\subsection{Perspectives}
\label{s:perspectives}
While the pseudospectrum framework is already employed in physics
(cf. e.g.~\cite{TreTreRed93,trefethen2005spectra,KreSieTat15,Sjostrand2019,ColRomHan19}), there seems to be
(up to our knowledge) no systematic application in the gravitational context.
The introduction of pseudospectra in gravitational physics opens an
avenue to interbreed the study of (in)stability and transients with  other domains in physics (and beyond),
by using pseudospectrum
analysis as a common methodological frame. In the following we mention some possible lines of exploration in
different gravitational settings,
from astrophysics and fundamental gravity physics to mathematical relativity, closing the discussion
with a perspective beyond gravity.

\label{s:perpectives}
\subsubsection{Astrophysics and cosmology}
The astrophysical status of the ultraviolet QNM overtone instability, that reaches the lowest overtones
for generic perturbations of sufficiently high frequency and energy, requires to assess whether actual astrophysical 
(and/or fundamental spacetime) perturbations are capable of triggering it.
Some problems in which this question is relevant are the following:
\begin{itemize}
\item[a)] {\em BH spectroscopy}.
  If such instability is actually present, this should be taken into account in current approaches to BH spectroscopy.
  The stability of the slowest decaying QNM guarantees that the dominating ringdown frequency is unaltered.
  But regarding QNM overtones, note that in we have not referred at all to late time ringdown frequencies,
  but to QNM frequencies: since such two sets of frequencies can actually
  decouple~\cite{Nollert:1996rf,Nollert:1998ys,Khanna:2016yow,Cardoso:2016rao,Konoplya:2016hmd,Daghigh:2020jyk,Konoplya:2020fwg,Qian:2020cnz}
  and, as already noticed by Nollert \cite{Nollert:1996rf}, the propagating (scattered) field itself is not much
  affected by high-frequency perturbations, finding the signature of perturbed QNMs in the gravitational
  wave signal may pose a very challenging problem \cite{JarMacShe21}. Awareness of this potential effect
  in the GW signal may however lead to specifically tailored data analysis tools. 
  
\item[b)] {\em BH environment}.
The arrangement of perturbed QNM branches along (a priori known) $\epsilon$-contour lines of pseudospectra
opens the possibility of probing, in an `inverse scattering' spirit, environmental BH
perturbations. One can envisage to read the ``size'' of the physical perturbations
by comparing observational QNM data with the ``a priori'' calibrated pseudospectrum.
This may help to assess ``dry'' versus ``wet'' BH mergers, a point of cosmological relevance in LISA science.

\item[c)] {\em Universality of compact object QNMs}.
  The combination of the ``universality'' of the perturbed ``Nollert-Price QNM BH branches'' with Nollert's
  remark on their similarity to neutron star ``$w$-modes'', together with
  the demonstrated loss of BH QNM axial/polar isospectrality, poses a natural question:
  do QNM spectra of all generic compact objects share a same pattern?
    
  Schemes such as \cite{Maggio:2020jml} may provide a systematic frame
  for the analysis of the astrophysical implications.

\item[d)] {\em BH QNM (in)stability in generic BHs}. A natural and necessary extension of the present
  work is the study of QNM (in)stability in the full BH Kerr-Newman family, in particular
  understanding how it intertwines with superradiance instability and
  the approach to extremality.
  
\end{itemize}

\subsubsection{Fundamental gravitational physics}
We note some possible prospects at the fundamental level:

\begin{itemize}
\item[a)] {\em (Sub)Planckian-scale physics}. Planck scale spacetime fluctuations
  seem a robust prediction of different models of quantum gravity. They represent ``irreducible''
  ultraviolet perturbations potentially providing a probe into Planck
  scale physics that, given the universality of BH QNM overtone instability,
  may be  `agnostic' to an underlying theory of quantum gravity. Such 
a search of quantum gravity signatures in BH gravitational wave physics is akin to
\cite{Agullo:2020hxe}.
Actually, it would suffice that a Planck scale ``cut-off'' induces an effective
$C^p$ regularity in the otherwise smooth low-energy description, to trigger the instability phenomenon.
BH QNM instability might then provide a particular probe into
'discreteness' of spacetime (e.g. \cite{Perez:2017krv} are references therein).

\item[b)] {\em QNMs and (strong) cosmic censorship}. In the setting of cosmological
  BHs, the assessment of the extendibility through the Cauchy horizon in
  Reissner-Nordstr\"om de Sitter is controlled by the parameter $\beta=\alpha/\kappa_-$,
  where $\alpha$ is the spectral gap (the imaginary part of the fundamental QNM in our setting)
  and $\kappa_-$ is the surface gravity of the Cauchy horizon~\cite{HinVas17,CarCosDes18}.
  Therefore, a good understanding  in this setting of the (in)stability properties of the
  slowest decaying QNM, and more generally of the  QNM spectrum,
   may be enlightening in the assessment of the thresholds for Cauchy horizon stability.

\item[c)]  {\em Random perturbations and spacetime semiclassical limit}. 
  The ``regularization effect'' of random perturbations
  \cite{hager05,Hager06a,Hager06b,HagSjo06,Borde08,BorSjo10,Borde11,Borde13,Vogel16,NonVog18,Sjostrand2019}
  in the scattering Green's function
  is an intriguing  phenomenon that may play a role in the transition to a semiclassical
  smooth effective description of fundamental gravitational degrees of freedom described
  in  a more basic (quantum) theory, possibly including an irreducible randomness ingredient. 
  Again, the universality of the phenomenon may play a key role.

\end{itemize}

\subsubsection{Mathematical relativity}
The presented numerical evidences need to be transformed into actual proofs. Some mathematical issues
to address are:

\begin{itemize}

  \item[a)]  {\em Regularity conditions and QNM characterization}. The mathematical
    study of QNMs entails subtle functional analysis issues.
    In the present hyperboloidal approach this involves, in particular, the choice of
    appropriate regularity conditions and the associated functional space.
    This connects our pseudospectrum study with the identification in \cite{Ansorg:2016ztf} of the full
    upper-complex plane as the actual QNM spectrum, if general $C^\infty$ eigenfunctions are allowed.
    More regularity must therefore be enforced. An analysis along the lines in
\cite{Gajic:2019qdd,Gajic:2019oem,galkowski2020outgoing}, where Gevrey classes are identified as
the proper functional spaces to define QNMs,
is therefore required. Likewise, a systematic comparison with QNM stability in the framework of 
\cite{HinVas17,Hafner:2019kov} is needed 
(cf. also \cite{zworski2017mathematical,dyatlov2019mathematical}).

\item[b)] {\em Semiclassical analysis and QNM (in)stability}. The interest of 
  asymptotic tools, in the study of QNM stability, is twofold. On the
  one hand, an ``asymptotic reasoning'' \cite{Batte01} built on the
  semiclassical analysis of QNMs (a subject taken to full maturity in Sj\"ostrand's works
  \cite{HelSjo86,Sjost02,dimassi1999spectral,DenSjoZwo03,davies2005semi})
  with a small parameter defined in terms of highly-damped QNM frequencies,
  can help to assess universality patterns of perturbed Nollert-Price BH QNM branches.
  On the other hand, asymptotic analysis provides powerful tools to prove
  rigorously spectral instability and non-trivial pseudospectra (cf. e.g. \cite{Davies99}). 
  In particular, the recent work \cite{BonFujRam20} provides an explicit example of scattering resonance
  (or QNM) instability, sharing much of the spirit of the discussion in this work.

\end{itemize}

\subsubsection{Beyond gravitation:``gravity as a crossroad in physics'' }
The disclosure of BH QNM instability \cite{Nollert:1996rf} resulted from the fluent interchange 
  between gravitational and optical physics \cite{LeuLiuTon94,LeuLiuYou94,ChiLeuSue95,ChiLeuMaa98,Leung:1999iq},
  again a key 'flow channel' in our work,  e.g. to understand the 'infrared' instability of the fundamental QNM 
  \cite{SheJar20}. In this  spirit, the present work can offer some hints for further boosting
  such kind of transversal research in physics.

The hyperboloidal approach, with its  explicit formulation of the dynamics in terms of a
  non-selfadjoint operator, provides a scheme of interest whenever
  dealing with an open physical system with losses at a radiation zone, a recurrent situation
  throughout  physics (e.g. in optics, acoustics, physical oceanography, to cite some settings).
  A specific lesson of the present work, to be exported to other physical contexts,
  is the identification of the relevant scalar product in terms of the system's energy,
  thus casting an a priori technical issue into neat physical terms. Moreover,
  when studying QNMs, the normalizability of the QNM eigenfunctions in the hyperboloidal
  approach may open an alternative avenue to the characterization of the so-called 'mode volume' $V_n$
  of a QNM. This is relevant e.g. in the setting of photonic/plasmonic resonances \cite{LalYanVyn17}:
  together with the notion of 'quality factor' $Q_n$, given in terms of the
 ratio between the real and imaginary parts of a QNM (see e.g. \cite{Pook-Kolb:2020jlr} for its
connection with BH gravity physics), it characterizes the Purcell factor $F_n\sim Q_n/V_n$
controlling the enhancement of spontaneous emission of a quantum system, a key notion in 'cavity quantum electrodynamics' \cite{Walther_2006}.

Regarding the pseudospectrum, this notion is relevant whenever a non-Hermitian
  (or more generally non-selfadjoint operator) enters into scene, as it is typically
  the case in open systems \cite{Ashida:2020dkc}. In the context of non-Hermitian
  quantum mechanics, it has been proposed \cite{KreSieTat15} to endow the  pseudospectrum
  with a guiding central role in the theory, in a setting in which spectral instability
  makes insufficient the standard notion of spectrum  to fully characterize the relevant operators.
  Apart from spectral instability, the pseudospectrum underlies purely dynamical phenomena
  \cite{TreTreRed93,trefethen2005spectra}, in particular 
  accounting for so-called nonmodal instability~\cite{Schmi07} in the
  setting of hydrodynamic stability theory and turbulence. Beyond hydrodynamics, the latter feature turns the pseudospectrum
  into a powerful tool for studying both spectral and dynamical stability issues 
  in (open) physical systems that ``trace'' over a part of the total degrees of freedom
  and, as a result, are governed by non-selfajoint operators. Such systems occur all over
  physics (e.g. condensed matter, optics, plasmonics, acoustics, nanophysics...~\cite{Ashida:2020dkc}),
  offering a natural arena for extending the already large range of applications of pseudospectra
  \cite{EmbTre_webpage}.

Gravitational physics is remarkable in its capacity to ``provide a framework that calls for the
  interchange of ideas, concepts and methodologies from very different communities''~\cite{AldBarJar11} in physics.
The hyperboloidal approach and the pseudospectrum here discussed realize an instance of this
understanding of ``gravity as a crossroad in physics''~\cite{AldBarJar11}.

%%%%%%%%%%%%%%%%%%%%%%%%%%%
%%%   ACKNOWLEDGMENTS   %%%
%%%%%%%%%%%%%%%%%%%%%%%%%%%

\bigskip

\noindent{\em Acknowledgments.} We thank M. Ansorg, P. Bizo\'n, O. Reula and J. Sj\"ostrand for key insights.
We also thank
J. Olmedo, C. Barcel\'o, L. Garay (and the rest of Carramplas-2019 participants), 
L. Andersson, A. Ashtekar, E. Berti, N. Besset, I. Booth, Y. Boucher, V. Cardoso,  G. Colas des Francs,
M. Colbrook, A. Coutant,  G. Cox, T. Daud\'e, K. Destounis, G. Dito,
J. Frauendiener, H. Friedrich, D. Gajic, S. Gu\'erin,
D. H\"afner, M. Hitrik, A. Iantchenko,  H.R. Jauslin, J. Jezierski, B. Krishnan, J. Lampart, J. Lewandowski,
M. Maliborski,   M. Mokdad, J.-P. Nicolas, I. Racz, B. Raffaelli, A. Rostworowski, B. Sah, O. Sarbach,  B.S. Sathyaprakash,
J. Slipantschuk, A. Soumaila,
J.A. Valiente-Kroon and A. Zenginoglu. This work was supported by the French ``Investissements d'Avenir'' program through
project ISITE-BFC (ANR-15-IDEX-03), the ANR ``Quantum Fields interacting with Geometry'' (QFG) project (ANR-20-CE40-0018-02), 
the EIPHI Graduate School (ANR-17-EURE-0002), 
  the Spanish FIS2017-86497-C2-1 project (with FEDER contribution),  
  the European Research Council Grant 
  ERC-2014-StG 639022-NewNGR ``New frontiers in numerical general relativity" and the European Commission
  Marie Sklodowska-Curie grant No 843152 (Horizon 2020 programme). The project used
  Queen Mary's Apocrita HPC facility, supported by QMUL Research-IT, and CCuB computational resources
  (universit\'e de Bourgogne).

\appendix

\section{Energy scalar product and adjoint operator $L^\dagger$}
\label{a:energy_scalar_adj}

\subsection{Energy scalar product }
\label{a:energy_scalar_product}
We start by considering the energy  contained
in the hyperboloidal slice $\Sigma_\tau$, defined by $\tau=\mathrm{const}$
in Eq. (\ref{e:change_variables}), and associated with a mode $\phi_{\ell m}$
satisfying the effective Eq.~(\ref{e:wave_equation_tortoise}), namely
propagation in Minkowski with a potential $V_{\ell}$ (see also \cite{GasJar21}). In this stationary situation
this energy is given \cite{Wald84} by Eq. (\ref{e:Energy_Tab})
\bea
\label{e:Energy_Tab_app}
E =  \int_{\Sigma_\tau} T_{ab} t^an^b d\Sigma_\tau \ .
\eea
The stress-energy tensor $T_{ab}=T_{ab}(\phi_{\ell m},\nabla\phi_{\ell m})$ of a (generally complex) scalar field is given
by Eq. (\ref{e:T_ab}),
with $\eta_{ab}$ the Minkowski metric in arbitrary coordinates (dropping $(\ell,m)$)
\bea
\label{e:T_ab_app}
T_{ab} = \frac{1}{2}\left(\nabla_a\bar{\phi}\nabla_b\phi
-\frac{1}{2}\eta_{ab}\left(\nabla^c\bar{\phi}\nabla_c{\phi} + V \bar{\phi}\phi \right) + \mathrm{c.c}\right) \ ,
\eea
with ``$\mathrm{c.c}$'' denoting ``$\mathrm{complex-conjugate}$''.
Coming back to (\ref{e:Energy_Tab_app}),
and using coordinates $(\tau, x)$ adapted to
$\Sigma_t$ and defined in Eq. (\ref{e:change_variables}), 
the timelike Killing is $t^a=\partial_t =\dfrac{1}{\lambda} \partial_\tau $, and we have
\bea
\label{e:timelike_normal}
n^a &=& \frac{1}{\sqrt{g'^2-h'^2}}\left( \frac{g'^2-h'^2}{|g'|}\partial_\tau -  \frac{h'}{|g'|}\partial_x\right) \\
&=& \frac{1}{\sqrt{g'^2-h'^2}}\left( w(x)\partial_\tau - \gamma(x)\partial_x\right) ,
\eea
for the timelike normal $n^a$, with $w(x)$ and $ \gamma(x)$ defined in Eq.~(\ref{e:functions_L1_L2}).
Finally, the radial part of the  metric integration measure $d\Sigma_\tau$ induced
in the hyperboloidal slice $\Sigma_\tau$ (see details in \cite{GasJar21} for the
handling of the angular terms) is given by
\bea
\label{e:Sigma_tau_integration_measure}
d\Sigma_\tau = \lambda \sqrt{g'^2-h'^2} \; dx \ .
\eea
Inserting these elements in (\ref{e:Energy_Tab_app}), a straightforward calculation
leads to Eq.~(\ref{e:energy_explicit}), that we can rewrite as
\bea
\label{e:energy_explicit_app}
E &=&\frac{1}{2}\int_a^b \left(\frac{g'^2-h'^2}{|g'|}\partial_\tau\bar{\phi} \partial_\tau\phi
  + \frac{1}{|g'|} \partial_x\bar{\phi} \partial_x\phi + |g'|\hat{V} \bar{\phi}\phi\right)dx \nn \\
&=&\frac{1}{2}\int_a^b \left(w(x)\partial_\tau\bar{\phi} \partial_\tau\phi
  + p(x) \partial_x\bar{\phi} \partial_x\phi + q(x) \bar{\phi}\phi\right)dx \ .
  \eea
  Identifying $\psi=\partial_\tau \phi$, and taking $E$ for the square of the norm of the
  vector $u=  (\phi,\psi)$,
  i.e.  prescribing $||u||^2_{_E}:= E$, we recover expression (\ref{e:energy_norm}).
  Considering only the $\phi$-part, this ``energy norm'' is a $H^1$-like norm, so that it takes into
  account the frequency of the mode $\phi$, a most important ingredient
  in our setting, given the role of high-frequency perturbations in
  the ultraviolet instability of QNM overtones. Finally, considering the whole
  $u= (\phi,\psi)$ vector, this
  norm is an $L^2$-norm coming from the energy scalar product $\langle \cdot, \cdot \rangle_{_E}$
  (for $q(x)>0$)
\bea
\label{e:energy_scalar_product_app}
&&\!\!\!\!\!\Big\langle\begin{pmatrix}
  \phi_1 \\
  \psi_1
\end{pmatrix}, \begin{pmatrix}
  \phi_2 \\
  \psi_2
\end{pmatrix}\Big\rangle_{_{E}}  \\
&=&
\frac{1}{2} \int_a^b \!\!\!\left(\! w(x)\bar{\psi}_1 \psi_2 + p(x)  \partial_x\bar{\phi}_1\!\partial_x\phi_2
+  q(x)\bar{\phi}_1 \phi_2 \!\right) dx  \ , \nn
\eea
that coincides with (\ref{e:energy_scalar_product}) upon identification
$q(x) = \tilde{V}$. Note that $\gamma(x)$ plays no role in the energy
scalar product $\langle \cdot, \cdot \rangle_{_E}$.

\subsection{Adjoint operator $L^\dagger$ }
\label{a:adjoint_L}
A very important object in our discussion of QNM spectral instability
and the pseudospectrum construction is the
adjoint $L^\dagger$ of the operator $L$. The definition
of $L^\dagger$ depends on the choice of scalar product and
we shall adopt here the energy scalar product
(\ref{e:energy_scalar_product_app}). The full
construction of the adjoint $L^\dagger$ requires a discussion
of its domain of dependence. This is a delicate question
intimately linked with the boundary and regularity conditions
determining the functional space on which $L$ and $L^\dagger$
are defined. This functional analysis issue will be addressed
elsewhere, and here we focus on the construction
of the so-called ``formal adjoint'', formally satisfying the relation
\bea
\label{e:adj_def}
\Big\langle L^\dagger\begin{pmatrix}
  \phi_1 \\
  \psi_1
\end{pmatrix}, \begin{pmatrix}
  \phi_2 \\
  \psi_2
\end{pmatrix}\Big\rangle_{_{E}}
=
\Big\langle\begin{pmatrix}
  \phi_1 \\
  \psi_1
\end{pmatrix}, L \begin{pmatrix}
  \phi_2 \\
  \psi_2
\end{pmatrix}\Big\rangle_{_{E}} \ ,
\eea
for all $u_1=(\phi_1,\psi_1)$ and $u_2=(\phi_2,\psi_2)$. Taking into account
the definition in Eq. (\ref{e:L_operator}) of the operator $L$, this writes
\begin{widetext}
\bea
\label{e:adj_step1}
\Big\langle L^\dagger\begin{pmatrix}
  \phi_1 \\
  \psi_1
\end{pmatrix}, \begin{pmatrix}
  \phi_2 \\
  \psi_2
\end{pmatrix}\Big\rangle_{_{E}}
&=&
\Big\langle\begin{pmatrix}
  \phi_1 \\
  \psi_1
\end{pmatrix}, \frac{1}{i}\!
\left(
  \begin{array}{c|c}
    0 & 1 \\
    \hline 
   L_1 & L_2
  \end{array}
  \right) \begin{pmatrix}
  \phi_2 \\
  \psi_2
  \end{pmatrix}\Big\rangle_{_{E}} 
  = 
  \Big\langle\begin{pmatrix}
  \phi_1 \\
  \psi_1
  \end{pmatrix},\frac{1}{i}
  \begin{pmatrix}
  \psi_2 \\
  L_1\phi_2 + L_2\psi_2
  \end{pmatrix}\Big\rangle_{_{E}} \nn \\
  &=&
  \Big\langle\begin{pmatrix}
  \phi_1 \\
  \psi_1
  \end{pmatrix},
  \frac{1}{i}\begin{pmatrix}
  \psi_2 \\\displaystyle
  \frac{1}{w(x)}\Big(\partial_x\big(p(x)\partial_x\phi_2\big) - q(x)\phi_2
  +2\gamma(x)\partial_x \psi_2+ \partial_x\gamma(x)\psi_2\Big)
  \end{pmatrix}\Big\rangle_{_{E}} \ ,
  \eea
where we have used the expressions for $L_1$ and $L_2$ in Eq. (\ref{e:L_1-L_2}).
Using the energy scalar product (\ref{e:energy_scalar_product_app}) and integrating by parts
\bea
\label{e:adj_step2}
\Big\langle L^\dagger\begin{pmatrix}
  \phi_1 \\
  \psi_1
\end{pmatrix}, \begin{pmatrix}
  \phi_2 \\
  \psi_2
\end{pmatrix}\Big\rangle_{_{E}}
&=&
  \Big\langle
  \frac{1}{i}\begin{pmatrix}
  \psi_1 \\\displaystyle
  \frac{1}{w(x)}\Big(\partial_x\big(p(x)\partial_x\phi_1\big) - q(x)\phi_1
  +2\gamma(x)\partial_x \psi_1+ \partial_x\gamma(x)\psi_1\Big)
  \end{pmatrix},
  \begin{pmatrix}
  \phi_2 \\
  \psi_2
  \end{pmatrix}\Big\rangle_{_{E}}
  \nn \\
  &+& \frac{1}{i}\Big(2\gamma(b)\bar{\psi}_1(b)\psi_2(b) -2\gamma(a)\bar{\psi}_1(a)\psi_2(a) \Big) \\
  &=&
  \Big\langle
  \frac{1}{i}\left(
  \begin{array}{c|c}
    0 & 1 \\
    \hline 
   L_1 & L_2
  \end{array}
  \right)
  \begin{pmatrix}
  \phi_1 \\
  \psi_1
\end{pmatrix}, 
\begin{pmatrix}
  \phi_2 \\
  \psi_2
  \end{pmatrix}\Big\rangle_{_{E}}
  + w(x) \int_a^b \Big(\overline{\frac{1}{i}2\frac{\gamma(x)}{w(x)}\big(\delta(x-a) - \delta(x-b)\big)\psi_1}\Big)\psi_2 dx \nn \ ,
  \eea
  where we have used $p(a)=p(b)=0$, the real character of $w(x)$, $p(x)$, $q(x)$ and $\gamma(x)$
  and the Dirac-delta $\delta(x)$ distribution to formally evaluate the boundary terms. This allows us to rewrite
\bea
\label{e:adj_step3}
\Big\langle L^\dagger\begin{pmatrix}
  \phi_1 \\
  \psi_1
\end{pmatrix}, \begin{pmatrix}
  \phi_2 \\
  \psi_2
\end{pmatrix}\Big\rangle_{_{E}}
&=&
  \Big\langle
  \frac{1}{i}\left(
  \begin{array}{c|c}
    0 & 1 \\
    \hline 
   L_1 &\displaystyle L_2 + 2\frac{\gamma(x)}{w(x)}\big(\delta(x-a) - \delta(x-b)\big)
  \end{array}
  \right)
  \begin{pmatrix}
  \phi_1 \\
  \psi_1
\end{pmatrix}, 
\begin{pmatrix}
  \phi_2 \\
  \psi_2
  \end{pmatrix}\Big\rangle_{_{E}} \ ,
\eea
\end{widetext}
so that, introducing the operator $L^\partial_2$ as in Eq. (\ref{e:L2_boundary})
\bea
  L^\partial_2 = 2\frac{\gamma(x)}{w(x)}\big(\delta(x-a)-\delta(x-b)\big) \ ,
  \eea
  we can write the formal adjoint in Eqs. (\ref{e:formal_adjoint})
  and (\ref{e:L_dagger})
  \bea
  \label{e:L_dagger_app}
L^\dagger = L + L^{\partial} \ \ , \ \
L^{\partial} =\frac{1}{i}\!
\left(
  \begin{array}{c|c}
    0 & 0 \\
    \hline 
   0 & L^\partial_2
  \end{array}
  \right) \ .
  \eea
  In general $\gamma(x)$ does not vanish at the boundaries, so $L$ is not
  even symmetric and therefore cannot be selfadjoint. Eq. (\ref{e:L_dagger_app}) identifies neatly the
  loss of selfadjointness with such non-vanishing $\gamma(x)$, specifically
  linking spectral instability with a boundary phenomenon, formally cast through the presence of
  the Dirac-delta terms.
  This form also explains the (formal) selfadjoint case $L_2=0$ discussed in
  section \ref{s:pseudospectrum_PT_selfadjointcase}.

More generally, evaluation of adjoints
play a key role in all aspects of our discussion of spectral instability:
i) calculation of conditions numbers $\kappa_n$'s, involving the spectral problem
of the adjoint $L^\dagger$, cf. Eq. (\ref{e:leftright_eigenvectors});
ii) evaluation of the pseudospectrum, involving the calculation
of (generalized) singular values of $R_L(\omega)$ and therefore
the spectral problem of $R^\dagger_L(\omega)R_L(\omega)$, cf.
Eqs. (\ref{e:pseudospectrum_carac_E}) and (\ref{e:pseudospectrum_carac_G});
and iii) the prescription of the norm $||\delta \tilde{V}||_{_E}$ to $\epsilon$
in the exploration of perturbed spectral QNM problems, again involving
the spectral problem of the operator $\delta \tilde{V}^\dagger\delta \tilde{V}$.
Details of the calculation of adjoints in our discretised
approach are given in appendices \ref{a:pseusdospectrum_energy_norm} and \ref{a:Chebyshev_elements}.

\section{Pseudospectrum in the energy norm}
\label{a:pseusdospectrum_energy_norm}
We derive here the relevant expressions for the construction
of pseudospectra in the discretised version of the energy norm.

\subsection{Scalar product and adjoint}
Let us consider a general hermitian-scalar product in $\mathbb{C}^n$
as
\bea
\label{e:matrix_scalar_product_ap}
\langle u, v\rangle_{_G} = (u^*)^i G_{ij}v^j = u^* \cdot G \cdot v \ ,
\eea
with $G$ a positive-definite Hermitian matrix
\bea
\label{e: hermitian-scalar_product}
G^* = G \ \ , \ \  x^* \cdot G \cdot x > 0 \ \ \hbox{if} \ \ x\neq 0 \ ,  
\eea
where ${}^*$ denotes conjugate-transpose, i.e. $u^*=\bar{u}^t$ and $G^*=\bar{G}^t$ 
(we notice that in the problem studied in this work, the Hermitian
positive-definite matrix $G$ is actually a real symmetric positive-definite matrix $G^t = G$,
but we keep the discussion in full generality).
Using (\ref{e:matrix_scalar_product_ap}) and (\ref{e: hermitian-scalar_product})
in the relation
\bea
\langle A^\dagger u, v \rangle_{_{G}} = \langle u, A v \rangle \ ,
\eea
characterising the adjoint $A^\dagger$ of $A$ with respect to the scalar product (\ref{e:matrix_scalar_product_ap}),
we immediately get
\bea
\label{e:matrix_adjoint_app}
A^\dagger = G^{-1}A^*G \ .
\eea
\subsection{Induced matrix norm from a scalar product norm}
The (vector) norm $||\cdot||_{_{G}}$ in $\mathbb{C}^n$ associated with the scalar product $\langle\cdot ,\cdot \rangle_{_{G}}$
in (\ref{e:matrix_scalar_product_ap}), namely
\bea
||v||_{_{G}} = \left(\langle v , v \rangle_{_{G}}\right)^{\frac{1}{2}} \ ,
\eea
induces a matrix norm $||\cdot||_{_{G}}$ in $M_n(\mathbb{C})$ defined as 
 \bea
\!\!\!\!\!\!\!  ||A||_{_{G}} = \max_{||x||=1, x\in \mathbb{C}^n} \left\{ ||Ax||_{_{G}} \right\} \ , \ A \in M_n(\mathbb{C}) \ .
\eea
A more useful characterisation of this  $L^2$ induced matrix norm is given
in terms of the spectral radius $\rho(A^\dagger A)$ of $A^\dagger A$, where
\bea
\label{e:spectral_radius}
\rho(M) = \max_ {\lambda\in\sigma(M)}\left\{|\lambda|\right\} \ .
\eea
Indeed, we can write
\bea
\label{e:step1}
||A||_{_{G}}^2 &=& \left( \max_{||x||=1, x\in \mathbb{C}^n}
\left\{ \left(\langle Ax , Ax \rangle_{_{G}}\right)^{\frac{1}{2}} \right\}\right)^2 \nn \\
&=&  \max_{||x||_{_{G}}=1, x\in \mathbb{C}^n}  \left\{\langle Ax , Ax \rangle_{_{G}}\right\} \nn \\
&=& \max_{||x||_{_{G}}=1, x\in \mathbb{C}^n}  \left\{\langle A^\dagger Ax , x \rangle_{_{G}}\right\} \ .
  \eea
  The rest of the argument essentially follows from Rayleigh-Ritz formula for self-adjoint operators. Explicitly,
  the (self-adjoint) matrix $A^\dagger A$ is unitarily diagonalisable and non-negative definite
  (that is, $\langle x,A^\dagger Ax\rangle_{_{G}}\geq 0, \forall x\in \mathbb{C}^n$),
  so that we can find an orthonormal basis of eigenvectors $\left\{ e_i \right\}$
  \bea
  A^\dagger A e_i = \lambda_i e_i  \ \ , \ \ \langle e_i,e_j \rangle_{_{G}}=\delta_{ij} \ ,
  \eea
  with real non-negative eigenvalues $\lambda_i$ that we order as
  \bea
  0\leq \lambda_1 \leq  \lambda_2 \ldots  \leq \lambda_n \ .
  \eea
  Expanding $\displaystyle x=\sum_i x^i e_i$ for an arbitrary $x\in \mathbb{C}^n$, we write
  \bea
  \!\!\!\!\!\!\!\langle A^\dagger Ax , x \rangle_{_{G}} = \sum_i \lambda_i |x^i|^2 \leq \lambda_n \sum_i |x^i|^2
  = \lambda_n ||x||_{_{G}}^2 \ ,
  \eea
  that we can recast as 
  \bea
  \Big\langle A^\dagger A\frac{x}{||x||_{_{G}}} , \frac{x}{||x||_{_{G}}} \Big\rangle_{_{G}} \leq \lambda_n = \rho(A^\dagger A) \ .
  \eea
  Inserting this in Eq. (\ref{e:step1}), we conclude
  \bea
  ||A||_{_{G}}^2 \leq \rho(A^\dagger A) \ .
  \eea
  To prove that the inequality is actually saturated,  it suffices to show that there exits
  a vector $x$, $||x||_{_{G}}=1$, that realizes the equality, i.e. $||Ax , Ax||^2_{_{G}} = \rho(A^\dagger A)$.
  If we consider $x=e_n$
  \bea
  ||Ae_n||^2_{_{G}} &=& \langle Ae_n , Ae_n \rangle_{_{G}} =  \langle A^\dagger Ae_n , e_n \rangle_{_{G}}
  \nn \\
  &=&  \lambda_n = \rho(A^\dagger A) \ ,
  \eea
  and we can finally conclude
  \bea
  ||A||_{_{G}} = \left(\rho(A^\dagger A)\right)^{\frac{1}{2}} \ .
  \eea

  \subsection{Characterization of the pseudospectrum}
  \label{a:charac_pseudo}
  Given an invertible matrix $A\in M_n(\mathbb{C})$ and
  a non-vanishing eigenvalue $\lambda$, then $1/\lambda$ is an eigenvalue of
  $A^{-1}$ and
  \bea
  \label{e:max_min_lambda}
  \max_{\lambda \in \sigma(A^{-1})} \left\{|\lambda|\right\} = \left(\min_{\lambda \in \sigma(A)} \left\{|\lambda|\right\}\right)^{-1} \ .
  \eea
  Then, for an invertible $M\in M_n(\mathbb{C})$, we can write for the squared
  norm $||\cdot||_{_{G}}$ of  its inverse $M^{-1}$
  \bea
  \label{e:norm_inverse}
  ||M^{-1}||^2_{_{G}} &=& \rho\left((M^{-1})^\dagger M^{-1}\right) = \rho\left(\left(M M^\dagger\right)^{-1}\right)  \\
  &=&  \left(\min_{\lambda \in \sigma(M M^\dagger)} \left\{\lambda\right\}\right)^{-1} =
  \left(\min_{\lambda \in \sigma(M^\dagger M)} \left\{\lambda\right\}\right)^{-1} \ , \nn
  \eea
  where in the passage from the first line to the second we have used (\ref{e:max_min_lambda}) and
  the definition (\ref{e:spectral_radius}) of the spectral radius, whereas in the last equality
  we have used that a matrix $AB$ has the same eigenvalues as the matrix $BA$.

  We consider now the $\epsilon$-pseudospectrum characterisation in Definition 2, namely
  Eq. (\ref{e:pseudospectrum_def2}), applied to the discretised energy norm $||\cdot||_{_{G}}$
  \bea
\label{e:pseudospectrum_def2_m}
\sigma_{_{G}}^\epsilon(A) &=& \{\lambda\in\mathbb{C}:   ||(\lambda \mathrm{Id}- A)^{-1}||_{_{G}}>1/\epsilon\} \ . 
\eea
Using (\ref{e:norm_inverse}), with $M= \lambda \mathrm{Id}- A$, we can write
\bea
||(\lambda \mathrm{Id}- A)^{-1}||_{_{G}}>1/\epsilon \Leftrightarrow
\epsilon > \left(\min_{\lambda \in \sigma(M^\dagger M)} \left\{\lambda\right\}\right)^{\frac{1}{2}} \ .
\eea
 Finally, $\sigma_{_{G}}^\epsilon(A)$
can be written as
\bea
\sigma^\epsilon_{_G}(A) = \{\lambda\in\mathbb{C}:   s_{_G}^\mathrm{min}(\lambda \mathrm{Id}- A)<\epsilon\} \ , 
\eea
where $s_{_G}^\mathrm{min}(M)$ is the minimum of a set of ``generalized singular values'' of $M$,
related to the $\langle \cdot, \cdot \rangle_{_G}$ scalar product
\bea
\label{e:pseudospectrum_carac_G}
s_{_G}^\mathrm{min}(M) := \min \{\sqrt{\lambda}:  \lambda\in \sigma(M^\dagger M) \} \ .
\eea
When choosing the energy scalar product in section \ref{s:numerical_approach}, that is
with $G = G^E$ (see explicit expression in appendix \ref{a:Chebyshev_elements}),
we recover expression (\ref{e:pseudospectrum_carac_E}) for $\sigma^\epsilon_{_E}(A)$.
When using the canonical $L^2$ product we recover  the standard
$\sigma^\epsilon_2(A)$ in (\ref{e:pseudospectrum_carac_L2}), where
\bea
s_{_2}^\mathrm{min}(M) =  \min \{\sqrt{\lambda}:  \lambda\in \sigma(M^* M) \} =: \sigma^{\mathrm{min}} \ ,
\eea
is the smallest of the singular values $\sigma_i(M) = \sqrt{\lambda_i}$, $\lambda_i\in \sigma(M^* M)$,
in the standard singular value decomposition of $M$.

\section{Elements in the Chebyshev discretization}
\label{a:Chebyshev_elements}
\subsection{Chebyshev spectral decomposition}
The Chebyshev's polynomial of order $k$ is given by
\bea
\label{e:Chebyshev_pol}
T_k(x) = \cos\left(k\arccos x\right) \ , \ x\in[-1,1] \ .
\eea
Chebyshev's polynomials provide an orthogonal basis for functions $f\in L^2([-1,1],w(x)dx)$,
with $w(x)=1/\sqrt{1-x^2}$, so that we can write the spectral expansion
\bea
\label{e:Chebyshex_expansion}
f(x) = \frac{c_0}{2} + \sum_{k=1}^\infty c_k T_k(x) \ .
\eea
For sufficiently regular functions $f(x)$, coefficients $c_k$ decay exponentially in $k$.
A $f_N(x)$ approximate of $f(x)$ is obtained by truncating the series to order $N$
\bea
\label{e:f_N_approximate}
f_N(x) = \frac{c_0}{2} + \sum_{k=1}^N c_k T_k(x) \ .
\eea
The function $f$ is therefore approximated by the vector $(c_0, c_1, \ldots, c_N)$ in $\mathbb{C}^n$,
with $n=N+1$. In particular, we can evaluate the integral of $f$ in the interval $[-1,1]$ as
\bea
\label{e:f_N_integral}
\int_{-1}^1  f_N(x) dx = c_0 - \sum_{k=1}^{\lfloor\frac{N}{2}\rfloor}\frac{c_{2k}}{4k^2-1} \ .
\eea

\subsection{Collocation methods: Chebyshev-Lobatto grid}
When dealing with the product of functions, as it is the case in our setting, the description
in terms of spectral coefficients $c_i$'s is not convenient. Instead, one constructs a
Chebyshev's interpolant $f_N(x)$ from the evaluation of $f(x)$ on points $x_i$
\bea
\label{e:f_N_xi}
f_N(x_i) = f(x_i) \ \ , \ \ i\in\{0,1,\ldots, N\} \ ,
\eea
where $x_i\in [-1,1]$ define an appropriately chosen $n$-point quadrature grid.
For concreteness, in the following we focus on the Chebyshev-Lobatto collocation grid including
the interval boundaries $x=\pm 1$, in the spirit of including horizon and null infinity points in
our compactified picture. The Chebyshev-Lobatto $(N+1)$-grid is given by the extrema of $T_N(x)$ (i.e.
the $N-1$ zeros of $T'_N(x)$) together with both extreme points $x_0=1$ and $x_N=-1$, resulting in the values 
\bea
x_i =\cos\left(\frac{\pi i}{N}\right) \ , \ i\in\{0,1,\ldots,N\} \ .
\eea
We can enforce (\ref{e:f_N_xi}) on this grid by constructing a $f_N(x)$ interpolant in the functional form
(\ref{e:f_N_approximate}), with coefficients~\footnote{Note that the resulting associated interpolant $f_N(x)$ does not
exactly coincides with the $N$-degree polynomial truncation from (\ref{e:Chebyshex_expansion}),
since $c_k$'s in (\ref{e:Chebyshex_expansion}) are obtained from the orthogonal projection
of the exact $f$ on the full Chebyshev complete basis.
Both sets of $c_k$'s converge as $N\to\infty$.}
\bea
\label{e:coeffs_Lobatto}
\!\!\!\!\!\!c_i \!=\! \frac{2-\delta_{iN}}{2N}\!\!\!\left[f(x_0) + (-1)^if(x_N) + 2\sum_{j=1}^{N-1} f(x_j) T_i(x_j) \right] \ ,
\eea
with $i\in\{0,1,\ldots,N\}$.
In the construction of our differential operator $L$, the interpolant of the product of
two functions $f$ and $g$ is obtained then by multiplication
on grid points, that is
\bea
\label{e:fg_grid}
(f g)_N(x_i)=f_N(x_i)g_N(x_i) \ . 
\eea
In addition to that, we need an
expression for the interpolant of the derivative $\displaystyle f'_N(x)=\left(\frac{df}{dx}\right)_N\!\!\!\!\!(x)$.
This is determined by
\bea
f'_N(x_i) = \sum_{j=0}^N \mathbb{D}^N_{ij}f_N(x_j) \ ,
\eea
with
\bea
\label{e:CL_derivation_matrix}
\mathbb{D}^N_{ij} =
\left\{
\begin{array}{lcl}
  \displaystyle
  -\frac{2N^2+1}{6} & , & i=j=N \\
  \displaystyle
  \frac{2N^2+1}{6} & , & i=j=0 \\
  \displaystyle
  -\frac{x_j}{2(1-x_j)^2} & , & 0<i=j<N \\
  \displaystyle
  \frac{\alpha_i}{\alpha_j} \frac{(-1)^{i-j}}{x_i-x_j} \ &,& i\neq j 
\end{array}
\right. \ ,
\eea
where
\bea
\alpha_i =
\left\{
\begin{array}{lcl}
  2 \ &,& \ i\in\{0,N\} \\
  1 \ &,& \ i\in\{1,\ldots, N-1\}
\end{array}
\right. \ .
\eea

\subsection{Energy scalar product: Gram matrix $G^{E}$}
Let us first consider the integral
\bea
\label{e:ScalarProd_Int}
I_\mu(f,g) =\int_{-1}^{1} f(x)g(x) d\mu(x) \ ,
\eea
with $d\mu(x) = \mu(x)dx$.
We can get a quadrature approximation $I^N_\mu(f,g)$ to $I_\mu(f,g)$ by using expression (\ref{e:f_N_integral})
for $N$-interpolants  (\ref{e:f_N_approximate}) $f_N$ and $g_N$, combined with the
particular expression (\ref{e:coeffs_Lobatto}) for coefficients in the Chebyshev-Lobatto grid
and the grid multiplication (\ref{e:fg_grid}). We obtain then
\bea
\label{e:ScalarProd_DiscInt}
I^N_\mu(f,g) = f_N^t\cdot C^N_\mu \cdot g_N \ ,
\eea
with $f_N^t=\left(f(x_0),\ldots,f(x_N)\right)^t$, $g_N^t=\left(g(x_0),\ldots,g(x_N)\right)^t$
the $(N+1)$-grid approximates of $f$ and $g$, respectively, and $C^N_\mu$ the diagonal
matrix given by
\bea
\label{e:CN_mu}
(C^N_\mu)_{ij} &=& (C^N_\mu)_{i}\;\delta_{ij} \\
(C^N_\mu)_{i} &=& \frac{2\mu(x_i)}{\alpha_iN} \left(1 - \sum_{k=1}^{\lfloor\frac{N}{2}\rfloor}
T_{2k}(x_i)\frac{2-\delta_{2k,N}}{4k^2-1}\right) \ , \nn
\eea
where we have used $T_0(x)=1$, $T_k(1)=1$ and $T_k(-1)=(-1)^k$.
Then, dropping the
indices $N$, we can
write the discrete version of the scalar product $\langle\cdot,\cdot \rangle_{_E}$ in (\ref{e:energy_scalar_product})
as
\bea
&&\langle u_1,u_2\rangle_{_E} = \Big\langle \begin{pmatrix}
  \phi_1 \\
  \psi_1
  \end{pmatrix},
\begin{pmatrix}
  \phi_2 \\
  \psi_2
  \end{pmatrix}
\Big\rangle_{_E} \\
&=&\frac{1}{2}\Big(\psi_1^* \cdot C_w \cdot\psi_2
+ (\mathbb{D} \phi_1)^*\cdot C_p\cdot \mathbb{D} \phi_1 + \phi_1^* \cdot C_{\tilde{V}_\ell} \cdot \phi_2\Big) \nn \ ,
\eea
that can be rewritten in matrix form as
\bea
\label{e:ScalarProd_WaveEq}
\langle u_1,u_2\rangle_{_E} &=& u_1^*\cdot G^E \cdot u_2 \\ 
&=&(\bar{\phi}_1,\bar{\psi}_1)
\left(
  \begin{array}{c|c}
    G^E_1   & 0 \\
    \hline 
   0 & G^E_2
  \end{array}
  \right)
  \begin{pmatrix}
  \phi_2 \\
  \psi_2
  \end{pmatrix}\ , \nn
  \eea 
  with (here, the matrices $C_{\tilde{V}_\ell}$, $C_p$ and $C_w$ are given by (\ref{e:CN_mu}),
  for the respective functions $\mu(x)=\tilde{V}_\ell(x), p(x), w(x)$)
  \bea
  G^E_1 &=& \frac{1}{2}\left(C_{\tilde{V}_\ell} +\mathbb{D}^t\cdot C_p\cdot \mathbb{D}\right) \nn \\
  G^E_2 &=& \frac{1}{2}C_w \ .
  \eea
  These expressions define the Gram matrix $G^E$ for the discretised version of the
  energy scalar product (\ref{e:energy_scalar_product}), in the basis determined
  from the Chebyshev-Lobatto spectral grid.  

\subsubsection{Grid interpolation}
An important aspect to observe when performing the numerical integration is that Eq.~\eqref{e:f_N_integral} is exact whenever the original function $f(x)$ is a polynomial of order $\leq N$. With this in mind, and assuming that $f(x)$ and $g(x)$ are polynomials, Eq.~\eqref{e:ScalarProd_DiscInt} is exact only for the case where the product $(fg)(x)$ yields polynomials of order $\leq N$. In practical terms, the procedure described above hampers the accuracy of the scalar product's numerical integration whenever the order gets $>N$.

As an illustrative example, take $f(x)=P_{\ell}(x)$ and $g(x)=P_{\ell'}(x)$, with $P_\ell(x)$ the Legendre polynomials. Then the integral \eqref{e:ScalarProd_Int} --- with $\mu(x)=1$ omitted of the expression --- yields $I(f,g) = 2\delta_{\ell, \ell'}/(2\ell +1).$ If we now consider the discrete version $I^N(f,g)$ given by Eq.~\eqref{e:ScalarProd_DiscInt}, one observes that the exact result is obtained only for the cases $\ell + \ell' \leq N$, even though  each individual function $f(x)$ and $g(x)$ is exactly represented for $\ell \leq N$ and $\ell' \leq N$, respectively.

To mitigate this issue, we modify the integration matrix $C^N_\mu$ --- or equivalently the Gram matrix $G^E$ --- by incorporating the following interpolation strategy.

Given an interpolant vector $f_N(x_i)$ associated with a Chebyshev-Lobatto grid $\{x_i\}_{i=0}^{N}$, one can obtain a second interpolant vector $f_{\bar N}(\bar{x}_i)$ associated with another Chebyshev-Lobatto grid  $\{\bar{x}_i\}_{i=0}^{\bar N}$ with a resolution $N\neq\bar N$ via
\bea
f_{\bar N}(\bar{x}_i) =\sum_{i=0}^N {\mathbb I}_{\bar i i} \, f_N(x_i) \ .
\eea
Components ${\mathbb I}_{\bar i i}$ of the interpolation matrix ${\mathbb I}$ are obtained by evaluating Eq.~\eqref{e:f_N_approximate} at the grid $\{\bar{x}_i\}_{i=0}^{\bar N}$, with the coefficients $\{ c_i\}_{i=0}^{N}$ expressed in terms of $f_N(x_i)$ via Eq.~\eqref{e:coeffs_Lobatto}. Then
\bea
\label{e:InterpolationMatrix}
{\mathbb I}_{\bar i i} = \dfrac{1}{\alpha_i N}\left(1 + \sum_{j=1}^N (2-\delta_{j,N}) T_j(\bar{x}_i) T_j(x_i) \right) \ .
\eea
Note that the interpolation matrix ${\mathbb I}$ has size $\bar N \times N$, which reduces to a square matrix only if $\bar N = N$. In this case, Eq.~\eqref{e:InterpolationMatrix} is actually the identity matrix as expected.

Then, for a fixed $N$, we consider the discrete integration \eqref{e:ScalarProd_DiscInt} in terms of a higher resolution $\bar N = 2 N$ and interpolate the expression back to the original resolution $N$. In other words,
defining ${\cal I}^N _\mu(f,g):=I^{\bar N} _\mu(f,g)$, we can consider the grid-interpolated new discrete integration
\bea
{\cal I}^N _\mu(f,g) = f_N^t \cdot {\cal C}^N_\mu \cdot g_N \ ,
\eea
where ${\cal C}^N_\mu = {\mathbb I}^t \cdot C^{\bar N}_\mu \cdot \mathbb I$ or, in terms of its components
\bea
({\cal C}^N_\mu)_{ij} = \sum_{\bar{i}=0}^{\bar N} \sum_{\bar{j}=0}^{\bar N} ({\mathbb I}^t){}_{i \bar i} \,(C^{\bar N}_\mu)_{\bar{i}\bar{j}} \, {\mathbb I}_{\bar j j} \ .
\eea
Going back to the illustrative example where $f(x)=P_{\ell}(x)$ and $g(x)=P_{\ell'}(x)$, we now obtain ${\cal I}^N(f,g) = 2\delta_{\ell, \ell'}/(2\ell +1)$ exactly whenever $\ell, \ell' \leq N$.

In the same way, we grid-interpolate the Gram matrices 
\bea
    {\cal G}^E_1 = {\mathbb I}^t \cdot G^E_1 \cdot \mathbb I,
    \quad {\cal G}^E_2 = {\mathbb I}^t \cdot G^E_2 \cdot \mathbb I \ ,
\eea
that allows to perform the scalar product~\eqref{e:ScalarProd_WaveEq} via
\bea
\label{e:ScalarProd_WaveEq_2}
\langle u_1,u_2\rangle_{_E} &=& u_1^*\cdot {\cal G}^E \cdot u_2 \nn \\ 
&=&(\bar{\phi}_1,\bar{\psi}_1)
\left(
  \begin{array}{c|c}
    {\cal G}^E_1   & 0 \\
    \hline 
   0 & {\cal G}^E_2
  \end{array}
  \right)
  \begin{pmatrix}
  \phi_2 \\
  \psi_2
  \end{pmatrix}.
  \eea

  \section{P\"oschl-Teller QNMs and regularity}
  \label{a:QNM_PT} 
  We give here the derivation of P\"oschl-Teller QNM frequencies (and QNM eigenfunctions in our setting).
  This is done
  for completeness and, more importantly, to illustrate with an explicit example
  the role of regularity in the enforcement of outgoing boundary conditions in the hyperboloidal scheme.

  We start from the Fourier transform in time of the P\"oschl-Teller wave equation in
  Bizo\'n-Mach coordinates, i.e.
Eq. (\ref{e:PT-prehypergeometric})
\bea
\Big((1-x^2)\frac{d^2}{dx^2} - 2(i\omega+1)x\frac{d}{dx} - i\omega(i\omega +1) -1  \Big)\phi =0 \ .
\eea
This equation can be solved in terms of hypergeometric functions.
Making the change $x=1-2z$, it is rewritten as
\bea
&&\Big(z(1-z)\frac{d^2}{dz^2}  \\
&&   + \big((1+i\omega)-2(1+i\omega)z\big)\frac{d}{dz}
 -\big(i\omega(i\omega +1)+1\big)  \Big)\phi =0 \ , \nn
\eea
namely Euler's hypergeometric differential equation
\bea
\Big( z(1-z)\frac{d^2}{dz^2} + \big(c-(a+b+1)z\big)\frac{d}{dz} -ab  \Big)\phi =0 \ ,
\eea
for  the values
\bea
\label{e:abc_PT}
c&=&1+i\omega \nn \\
a&=& \frac{(2i\omega+1) \pm i\sqrt{3}}{2} \\
b&=& (2i\omega+1)-a=\frac{(2i\omega+1) \mp i\sqrt{3}}{2} \nn \ .
\eea
For each choice of $\omega$, this equation admits two linearly independent solutions
that can be built from the Gauss hypergeometric function ${}_2F_1(a,b;c;z)$.
It is only when we enforce some regularity in the solution, that the spectral parameter $\omega$
is discretised and we recover the QNM frequencies. In this particular case, it is
when we truncate the hypergeometric series ${}_2F_1(a,b;c;z)$ to a polynomial,
that we recover P\"oschl-Teller QNM frequencies. Such truncation 
occurs when either $a$ or $b$ is a non-positive integer. From (\ref{e:abc_PT}) we can write
\bea
\omega = \mp \frac{\sqrt{3}}{2} +i\Big(-a +\frac{1}{2}\Big)  = \pm \frac{\sqrt{3}}{2}
+i\Big(-b +\frac{1}{2}\Big) \ .
\eea
Therefore, imposing either $a=-n$ or $b=-n$, with $n\in \mathbb{N}\cup \{0\}$,
we finally get
\bea
\label{e:PT_QNM_appendix}
\omega^\pm_n =  \pm \frac{\sqrt{3}}{2} +i\Big(n +\frac{1}{2}\Big) \ .
\eea
Choosing the $a=-n$ version, the corresponding eigenvectors can be written as Jacobi polynomials
$P_n^{(\alpha,\beta)}(x)$, defined as
\bea
\label{e:JacobiPol}
P_n^{(\alpha,\beta)}(x) = \frac{(\alpha+1)_n}{n!}{}_2F_1(-n,1+\alpha+\beta;\alpha+1;\frac{1-x}{2}) \ ,
\eea
with $(y)_n$ the Pochhammer symbol (i.e. $\displaystyle (y)_n=\prod_{k=0}^{n-1} (y-k)$).
Inserting,  for a given $n\in\mathbb{N}\cup \{0\}$, the values (\ref{e:abc_PT}) and
(\ref{e:PT_QNM_appendix}) into ${}_2F_1(a,b;c;z)$
we get, upon comparison with  (\ref{e:JacobiPol})
\bea
\alpha=\beta=i\omega_n \ ,
\eea
so that P\"oschl-Teller QNM eigenfunctions write, in Bizo\'n-Mach coordinates,
as
\bea
\label{e:phi_n_appendix}
\phi^{\pm}_n(x) = P_n^{(i\omega^\pm_n, i\omega^\pm_n)}(x) \ , \ x \in [-1,1] \ .
\eea

\bibliographystyle{spmpsci}
\bibliography{Biblio}

\end{document}